\documentclass[11pt]{article} 

\usepackage[utf8]{inputenc}
\usepackage{multirow}
\usepackage{amsfonts}
\usepackage{amsmath}
\usepackage{amssymb}
\usepackage{amsthm}
\usepackage{marvosym}
\usepackage{fontawesome5}
\usepackage{booktabs, array}

\newcolumntype{M}{>{$}c<{$}}  
\usepackage[export]{adjustbox}
\usepackage[english]{babel}

\usepackage{tikz}
\usetikzlibrary{shapes.misc,decorations.markings,arrows,decorations.pathmorphing,backgrounds,matrix,patterns,positioning,decorations.pathreplacing,calc}

\newcommand{\canadagoose}{\vcenter{\hbox{$\mathord{\includegraphics[width=2.5ex]{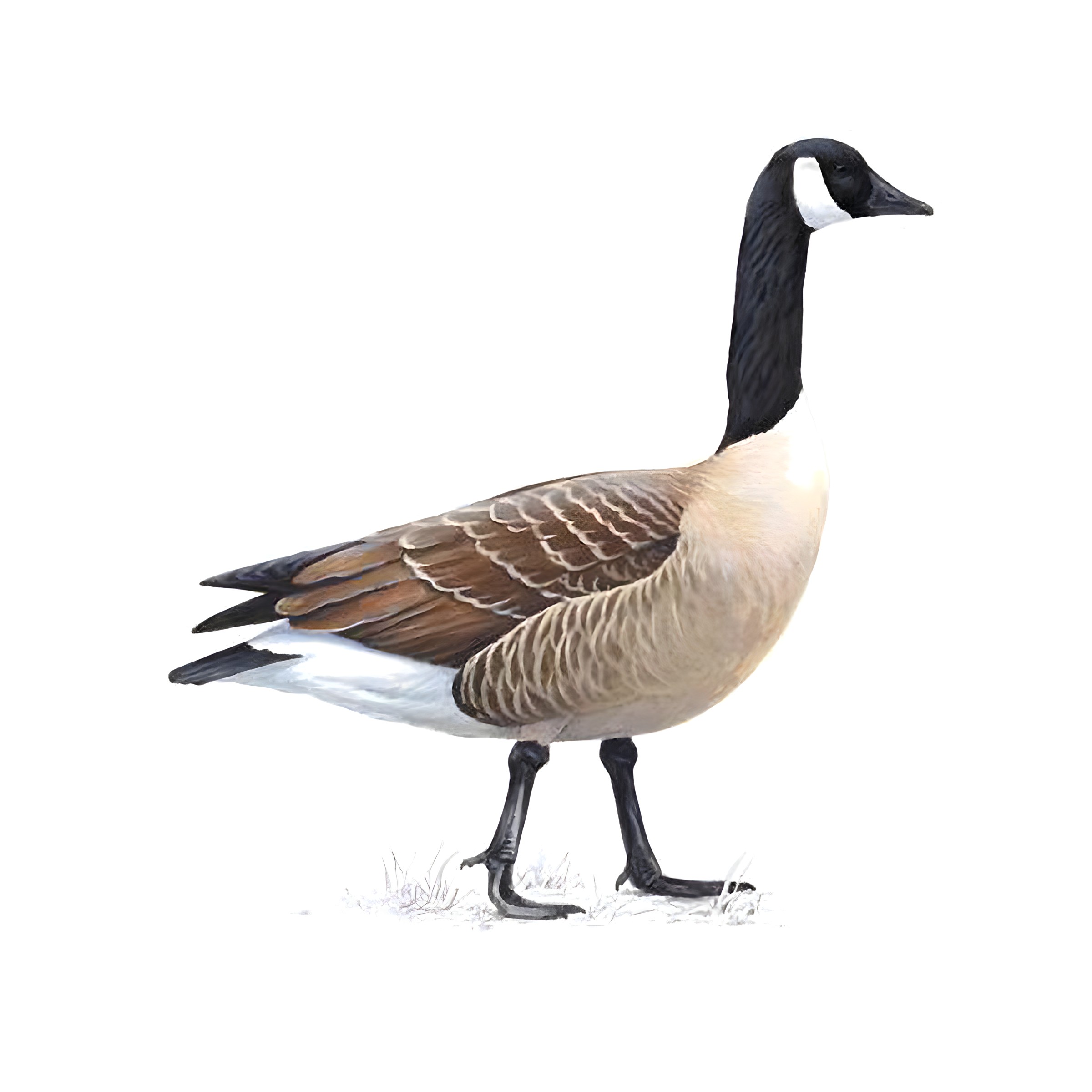}}$}}}

\makeatletter
\usepackage{comment}
\usepackage{float}
\usepackage{caption}
\usepackage{xcolor}
\definecolor{dgreen}{rgb}{0,0.5,0}
\definecolor{darkblue}{rgb}{0,0,0.6}
\definecolor{lblue}{rgb}{0.5,0.5,1.0}
\definecolor{purple}{rgb}{0.4,.2,0.7}
\definecolor{ggray}{rgb}{0.85,.85,0.85}
\definecolor{dpink}{rgb}{1,0.5,0.5}
\def\red{\color{red}}

\usepackage[margin = 2.5cm]{geometry}
\usepackage{graphicx}
\usepackage[hyperfootnotes = false, colorlinks = true, linkcolor = blue, citecolor = purple]{hyperref}
\usepackage{subcaption}
\usepackage{blkarray}
\def\tr{\mathrm{tr}}

\newcommand{\bes}{\begin{equation} \begin{split} }	
	\newcommand{\ees}{\end{split} \end{equation} }
	
	\renewcommand{\i}{\mathrm{i}}

	\renewcommand{\O}{\mathcal{O}}

	\newcommand{\evs}[1]{\langle #1 \rangle} %
    \newcommand{\gev}[1]{\evs{\Omega| #1 |\Omega}}
	
	\newcommand{\inv}{^{-1}}
	
	\newcommand{\hf}{\tfrac{1}{2}}
	\newcommand{\qrt}{\frac{1}{4}}
	\usepackage{physics}

\usepackage{braket}
\usepackage{cite}
\usepackage{enumitem}
\usepackage{mathrsfs}
\usepackage{setspace}
\usepackage{titlesec}
\usepackage{makecell}
\usepackage{wasysym}

\graphicspath{{./figures/}}

\definecolor{darkgreen}{rgb}{0,0.5,0}
\newcommand{\M}{\mathcal{M}}
\newcommand{\N}{\mathcal{N}}

\renewcommand{\d}{\mathrm{d}}
\newcommand{\TreeSymbol}{%
\tikz[scale=0.1,baseline=-.1ex]{
  \draw[fill=brown!60!black] (0,-0.3) rectangle (0.4,0.4);
  \draw[fill=green!70!black] (-0.9,0.4)--(0.2,1.8)--(1.3,0.4)--cycle;
  \draw[fill=green!70!black] (-0.7,1.2)--(0.2,2.5)--(1.1,1.2)--cycle;
}}

\newcommand{\cheese}[1][.3em]{%
  \tikz[baseline=-0.5ex, x=#1, y=#1, rotate=-25]{ %
    \coordinate (A) at (0,0);
    \coordinate (B) at (2,0.4);
    \coordinate (C) at (0,1.2);
    \coordinate (S) at (0.8,0.5);
    \coordinate (A2) at ($(A)+(S)$);
    \coordinate (B2) at ($(B)+(S)$);
    \coordinate (C2) at ($(C)+(S)$);

    \fill[black, opacity=0.12] (1.0,-0.12) ellipse (1.05 and 0.16);

    \begin{scope}
      \clip (A)--(B)--(C)--cycle;
      \fill[yellow!87!orange] (A)--(B)--(C)--cycle;
      \path[fill=orange!60!brown]
        (A)--(B)--($(B)!0.14!(C)$)--($(A)!0.14!(C)$)--cycle;
      \foreach \x/\y/\r in {0.65/0.36/0.14, 1.25/0.70/0.18, 0.32/0.86/0.10}{
        \fill[orange!40!brown, opacity=0.9] (\x,\y) circle (\r);
        \fill[white,opacity=0.4] (\x-\r*0.35,\y+\r*0.35) circle (\r*0.45);
      }
    \end{scope}

    \begin{scope}
      \clip (A2)--(B2)--(C2)--cycle;
      \fill[yellow!95!white] (A2)--(B2)--(C2)--cycle;
      \foreach \x/\y/\r in {1.20/1.00/0.12, 0.95/1.28/0.09}{
        \fill[orange!35!brown, opacity=0.8] (\x,\y) circle (\r);
        \fill[white,opacity=0.55] (\x-\r*0.35,\y+\r*0.35) circle (\r*0.40);
      }
    \end{scope}

    \fill[yellow!82!orange] (B)--(B2)--(C2)--(C)--cycle;

    \draw[line width=0.0em, draw=orange!40!brown!70!white, opacity=0.6, line join=round]
      (A)--(B)--(B2)--(C2)--(C)--(A)--(C);
  }%
}

\begin{document}

\thispagestyle{empty}
\begin{center}
    ~\vspace{5mm}

     {\LARGE \bf Bootstrapping Euclidean Two-point Correlators}

   \vspace{0.5in}
     
   {Minjae Cho$^\text{\faPizzaSlice}$, Barak Gabai$^{\cheese}$, Henry W. Lin$^{\TreeSymbol,{\color{orange} \text{\CleaningP}}}$, Jessica Yeh$^{\TreeSymbol}$, Zechuan Zheng$^{\canadagoose,\text{\Coffeecup}}$  }

    \vspace{0.5in}

   ~
   \\
   {$^\text{\faPizzaSlice}$ Leinweber Institute for Theoretical Physics, University of Chicago, Chicago, IL 60637, USA}\\
    {$^{\cheese}$ Laboratory for Theoretical Fundamental Physics, Institute of Physics, \\
    École Polytechnique Fédérale de Lausanne (EPFL), CH-1015 Lausanne, Switzerland}  \\
    {$^{\TreeSymbol}$ Leinweber Institute for Theoretical Physics, Stanford University, Stanford, CA 94305, USA}  \\
   {$^{\color{orange} \text{\CleaningP}}$ Jadwin Hall, Princeton University, Princeton, NJ 08540, USA}   \\
    {$^{\canadagoose}$ Perimeter Institute for Theoretical Physics, Waterloo, ON N2L 2Y5, Canada}\\
    {$^\text{\Coffeecup}$ Laboratoire de Physique de l\textquoteright Ecole normale sup\'erieure, ENS, Universit\'e
PSL,  \\
CNRS, Sorbonne Universit\'e, Universit\'e Paris Cit\'e, F-75005 Paris, France}
    \vspace{0.5in}

    \vspace{0.5in}
    
\end{center}

\vspace{0.5in}

\begin{abstract}
We develop a bootstrap approach to Euclidean two-point correlators, in the thermal or ground state of quantum mechanical systems. We formulate the problem of bounding the two-point correlator as a semidefinite programming problem, subject to the constraints of reflection positivity, the Heisenberg equations of motion, and the Kubo–Martin–Schwinger condition or ground-state positivity. In the dual formulation, the Heisenberg equations of motion become ``inequalities of motion'' on the Lagrange multipliers that enforce the constraints. This enables us to derive rigorous bounds on continuous-time two-point correlators using a finite-dimensional semidefinite or polynomial matrix program. We illustrate this method by bootstrapping the two-point correlators of the ungauged one-matrix quantum mechanics, from which we extract the spectrum and matrix elements of the low-lying adjoint states. Along the way, we provide a new derivation of the energy–entropy balance inequality and establish a connection between the high-temperature two-point correlator bootstrap and the matrix integral bootstrap.

\end{abstract}

\vspace{1in}

\pagebreak

\setcounter{tocdepth}{3}

\tableofcontents

\section{Introduction}

The quantum mechanical bootstrap is a non-perturbative method that can be used to study strongly coupled and/or large $N$ quantum systems \cite{Anderson:2016rcw, Han:2020bkb}. It enforces positivity of the inner product, the Heisenberg equations of motion, and kinematical constraints to produce nontrivial bounds on physical observables. Until recently, the quantum mechanical bootstrap had been primarily applied to time-independent quantities such as one-point functions in thermal equilibrium \cite{Fawzi:2023fpg, Cho:2024kxn} or energy eigenstates \cite{Han:2020bkb, LinZheng1, LinZheng2}.\footnote{See, however, \cite{Nancarrow:2022wdr, Lin:2024vvg} for bootstrapping matrix elements of operators, \cite{Anderson:2016rcw, Kazakov:2022xuh, Cho:2022lcj, Kazakov:2024ool} for bootstrap bounds on lattice systems (where one of the dimensions can be viewed as a discrete Euclidean time), and \cite{Cho:2025dgc} for bootstrapping nonequilibrium statistical mechanical systems.}

The use of positivity and semidefinite constraints in quantum mechanics has earlier precursors in the reduced-density-matrix literature. In particular, variational constraints on reduced density matrices were studied as early as 1955 \cite{PhysRev.100.1579}, and semidefinite programming was later used to constrain ground states of quantum chemistry systems \cite{PhysRevA.57.4219,10.1063/1.1360199} and Hubbard-type systems \cite{PhysRevLett.108.200404}. These developments contain several ingredients of what is now called the quantum mechanical bootstrap, although they predate modern bootstrap terminology and software.

While these quantities are ``dynamical'' in the sense that they are not determined by kinematics or symmetries alone, they represent only a small subset of physically relevant observables. Many important physical questions concern how correlators behave as the temporal or spatial separation between operators varies, with prominent examples including the gap, response, thermalization, and chaos. At the same time, systems exhibiting such rich behavior often pose computational challenges due to strong coupling dynamics and the large number of degrees of freedom. It is therefore desirable to develop a bootstrap method that efficiently constrains multi-point correlators.

As a first step in this direction, we propose in this work a bootstrap approach to constraining Euclidean two-point correlators in thermal equilibrium
\begin{equation}\label{eqn:euclid2ptgen}
\langle {\cal O}_1(\tau) {\cal O}_2(0)\rangle_\beta,
\end{equation}
where ${\cal O}_{1,2}$ are operators of interest, $\tau$ is their Euclidean time separation, and $\langle\cdots\rangle_\beta$ denotes the thermal expectation value at inverse temperature $\beta$. In particular, this includes the ground-state case $\beta^{-1}=0$. The bootstrap ingredients we employ for such Euclidean two-point correlators are reflection positivity, the Heisenberg equations of motion, and either the Kubo–Martin–Schwinger (KMS) condition \cite{doi:10.1143/JPSJ.12.570,PhysRev.115.1342} for finite $\beta$ or ground-state positivity for $\beta\to \infty$, all of which are convex constraints on the space of Euclidean two-point correlators. The KMS condition, in particular, expresses the periodicity of Euclidean two-point correlators by definition and thus provides natural bootstrap constraints on their space.

A main obstacle in this \textit{primal} bootstrap problem is that Euclidean two-point correlators are functions of $\tau$, making the space of bootstrap variables infinite-dimensional even when ${\cal O}_{1,2}$ in (\ref{eqn:euclid2ptgen}) are drawn from a finite set of operators. A well-known approach to such problems in optimal control theory \cite{2007math......3377L,2022arXiv221115652H,holtorf2024bounds} is to consider the \textit{dual} problem, for which any feasible solution provides a rigorous bound on the primal objective—here, the Euclidean two-point correlator—by the weak duality theorem. We may therefore adopt a finite-dimensional ansatz for the dual-feasible variables, resulting in a standard semidefinite programming (SDP) problem that yields rigorous bootstrap bounds on the Euclidean two-point correlators.

This work is largely motivated by a recent study by Lawrence, McPeak, and Neil (LMN) \cite{Lawrence:2024mnj}, who formulated a quantum mechanical bootstrap problem to bound the Lorentzian time evolution of one-point functions $\langle {\cal O}\rangle_{\rho(t)}$ for a density matrix $\rho(t)$ evolving under the von Neumann equation, given some knowledge of the initial values $\langle {\cal O}\rangle_{\rho(t=0)}$.\footnote{We review the LMN bootstrap method in Appendix \ref{app:lawrence}.} LMN also implemented the dual formulation to derive bootstrap bounds on time-dependent expectation values $\langle {\cal O}\rangle_{\rho(t)}$, extending previous quantum mechanical bootstrap methods that considered only time-independent observables. Our work can be regarded as an extension of LMN to Euclidean two-point correlators, which may eventually be generalized to constrain multi-point correlators with generic complex time separations.

Time-dependent quantum constraints can also be incorporated at the operator level using differential non-commutative polynomial optimization (DNPO), which embeds dynamics into SDP hierarchies \cite{Araujo:2023vnf,Araujo:2024cpe}, although it does not directly treat multi-time correlators as bootstrap variables.

To illustrate our method, we apply it to the one-matrix quantum mechanics (1-MQM) with the Hamiltonian\footnote{We use the convention $\tr \mathcal{O}= \frac{1}{N} \Tr \mathcal{O} = \frac{1}{N} \sum_a O_{aa}$. We also define $P = P_\text{canonical}/\sqrt{N}$ and $X = X_\text{canonical}/\sqrt{N}$ so that $[X_{ab},P_{cd}]=\frac{\i}{N} \delta_{ad}\delta_{bc}$. With this convention, $V$ is a polynomial whose coefficients are independent of $N$ in the 't Hooft limit. For an introduction to MQM, we refer the readers to \cite{Klebanov:1991qa}.}
\begin{equation} \label{eq:HamiltonianMQM}
    H= N^2 \tr (\hf P^2 +V(X)).
\end{equation}

This system possesses a $U(N)$ symmetry under which $X$ and $P$ transform in the adjoint representation. We study the ``ungauged'' model, where all representations of $U(N)$ are included in the Hilbert space. Unlike the ``gauged’’ model, whose Hilbert space is restricted to $U(N)$ singlets, this ungauged model is not analytically solvable even at large $N$ and therefore serves as a good testing ground for the bootstrap method.

Specifically, we obtain highly precise lower and upper bounds on $\langle\tr X(\tau)X(0)\rangle_\beta$ at both zero and nonzero temperatures. At zero temperature (the ground state), these bounds allow us to extract properties of low-lying adjoint energy eigenstates, showing nice agreement with expected results from the Marchesini-Onofri equation \cite{Marchesini:1979yq}. At nonzero temperature, our bootstrap bounds are more precise than Monte Carlo results over a range of Euclidean time separations $\tau$.

This paper is organized as follows. In Section \ref{2ptSetup} we set up the bootstrap problem and derive the dual SDP. In Section \ref{analyticBds}, we derive some analytic bounds using the two-point bootstrap. This includes a novel derivation of the energy-entropy balance (EEB) inequalities and a derivation of the integral/statistical-mechanics bootstrap from the high-temperature limit of the two-point correlator bootstrap. In Section \ref{numericalBds}, we apply the two-point correlator bootstrap to the ungauged MQM and present the numerical results. %
We discuss the physics of the adjoint gap and some future directions in Section \ref{Discussion}. The Appendices collect numerous technical details about the implementation of the bootstrap (\ref{app:implementation}, \ref{app:b-spline},  \ref{app:PMP}, \ref{app:extension}, \ref{app:montecarlo}), an improvement of the time-dependent bootstrap of LMN \ref{app:lawrence}, a review of some analytic results \ref{app: analytic} for the 1-MQM, and some novel results about the one-point function bootstrap of the MQM \ref{app:1ptMQM}.

\section{Bootstrapping two-point correlators \label{2ptSetup}}
In this section, we introduce the bootstrap constraints that Euclidean two-point correlators must obey and describe how to systematically implement them using SDP.

\subsection{Ground state}
We begin with the ground state. Variables of the bootstrap problem are given by the following matrix of two-point correlators
\begin{align}
\mathcal{M}_{ij}(\tau) =	\bra{\Omega} \bar{\mathcal{O}}_i(\tau) \mathcal{O}_j(0) \ket{\Omega} ,
\end{align}
where $|\Omega\rangle$ is the ground state, $\O_i$ provides a basis of all operators, $\bar{\O}_i(\tau) = e^{\tau H} \bar{\O}_i e^{-\tau H}$, and $\bar{\O}_i$ is the adjoint of the operator $\O_i$. Time-translation invariance of the ground state implies that we may write
\begin{align}
\mathcal{M}_{ij}(\tau) = \ev{\Omega|\bar{\mathcal{O}}_i(\tau/2) \mathcal{O}_j(-\tau/2)|\Omega}.
\end{align}
Hence, the first bootstrap constraint is reflection positivity:
\begin{align}
\mathcal{M}(\tau) \succeq 0, \quad \forall \tau \ge 0 .
\end{align}
The time-evolution of two-point correlators is dictated by the Heisenberg equations of motion. We can expand the commutator of $H$ with any operator in terms of the basis of operators ${\O}_i$:
\begin{align}
    [H ,\O_i] =  \O_k D_{ki}, \quad [H,\bar{\O}_i] = -[H,\O_i]^\dagger = - \bar{D}_{ki} \bar{\O}_k = - (D^\dagger)_{ik} \bar{\O}_k 
\end{align}
We then arrive at the second bootstrap constraints:
\begin{align}
\partial_\tau \mathcal{M}_{ij} 
&=  \bra{\Omega} [H,\bar{\mathcal{O}}_i(\tau)] \mathcal{O}_j(0) \ket{\Omega} = -\bra{\Omega} \bar{\mathcal{O}}_i(\tau) [H, \mathcal{O}_j(0)] \ket{\Omega}\nonumber  \\
&= - \mathcal{M}_{ik} D_{kj}\\ 
0 &= D^\dagger \mathcal{M}- \mathcal{M} D 
\label{Nline2}%
\end{align}
In the last line, we used time-translation invariance to derive a linear constraint on the matrix $\mathcal{M}$.

All of the constraints above apply to correlators in any stationary state. Next, we introduce a bootstrap constraint that applies specifically to the ground state. For that purpose, we define another matrix, which we refer to as the ``ground-state positivity'' matrix, following \cite{Araki:1977px, Fawzi:2023fpg, Cho:2024kxn}, whose positivity constitutes the third bootstrap constraint:
\begin{align}
\mathcal{N}_{ij} &=   \bra{\Omega} [\bar{\mathcal{O}}_i(\tau),H] \mathcal{O}_j(0) \ket{\Omega} \nonumber\\
 &=   \bra{\Omega} \bar{\mathcal{O}}_i(\tau/2) [H,\mathcal{O}_j(-\tau/2)] \ket{\Omega} \label{Nline1}\\
\mathcal{N}(\tau) &\succeq 0. \label{ground_state_positivity}
\end{align}
Here we have used the fact that the ground state has lower energy than the state $\mathcal{O}_j (-\tau/2) \ket{\Omega}$. This matrix is related to $\mathcal{M}$ via
\begin{align}
\mathcal{N}(\tau) &= -\partial_\tau \mathcal{M} (\tau)= \M(\tau) D %
\end{align}

In this work, we consider quantum systems with an anti-unitary time-reversal symmetry $\mathcal{T}$ that leaves the ground state invariant. Furthermore, since $\mathcal{T}^2 = 1$, we can organize all operators into $\mathcal{T}$-even or $\mathcal{T}$-odd operators. We then multiply all the $\mathcal{T}$-odd operators by $\i$ so that all elements of this operator basis satisfy $\mathcal{T} \O_i \mathcal{T}\inv = \O_i$. Then,
\begin{align}
 \overline{\mathcal{M}_{ij}(\tau)} =  \overline{\ev{ \O_i(\tau/2) \Omega , \O_j(\tau/2) \Omega}} =   \ev{ \mathcal{T}\O_i(\tau/2) \Omega , \mathcal{T}\O_j(\tau/2) \Omega} \nonumber\\
  =    \ev{ \mathcal{T}\O_i(\tau/2) \mathcal{T}\inv  \Omega , \mathcal{T}\O_j(\tau/2)\mathcal{T}\inv  \Omega} = \mathcal{M}_{ij} (\tau)
\end{align}
Together with Hermiticity of $\mathcal{M}$, we conclude that for time-reversal invariant systems $\mathcal{M}$ is actually real and symmetric.
Time-reversal invariance is not essential to the bootstrap construction. It allows us to choose a basis in which $\mathcal{M}(\tau)$ and the corresponding SDP data are real. When time-reversal symmetry is broken, $\mathcal{M}(\tau)$ is instead a complex Hermitian matrix, and the same positivity, equations-of-motion, and duality arguments continue to apply. For solvers accepting only real data, each complex Hermitian constraint can be represented by its standard real symmetric block form. We impose time-reversal invariance below only to simplify the implementation.

If the quantum system of interest possesses symmetries, they provide further bootstrap constraints. Suppose we have a symmetry that leaves the ground state invariant, %
\begin{equation*}
    \bra{\Omega} U (g) \bar\O_i(\tau) U(g\inv ) \O_j(0)  \ket\Omega = \bra{\Omega} \bar\O_i(\tau) U(g\inv ) \O_j(0)  U (g)  \ket\Omega.
\end{equation*} 
This leads to
\begin{align}
  U(g\inv) \mathcal{O}_j U(g) &=  \O_k R(g)_{kj}, \quad U(g) \bar{\O}_i U(g\inv) = (U(g) \O_i U(g\inv))^\dagger =  \bar{\O}_k \bar{R}(g\inv)_{ki}\\
    {R}^\dagger(g\inv) \M &= \M R(g)
\end{align}
In our application the symmetry group will be $\mathbb{Z}_2$. For a continuous symmetry generated by $C_A$ we may instead write
\begin{align}
    \bra{\Omega} [C_A,\bar\O_i(\tau)] \O_j(0)  \ket\Omega = - \bra{\Omega} \bar\O_i(\tau) [C_A, \O_j(0)]    \ket\Omega.
\end{align}

Collecting all the bootstrap constraints we described above, we arrive at the primal bootstrap problem where the objective is a Euclidean two-point correlator at time $\tau=T$:
\begin{equation}\label{primalMN}
\begin{split}
\operatorname{minimize } & \operatorname{Tr} \widehat{O} \M(T) \\
\text { subject to } & \M(\tau) \succeq 0\\
& \N(\tau)\succeq0,  \\
& \left(D^\dagger - \partial_\tau \right) \M(\tau)=0 \\
& \N(\tau )- \M(\tau)D =0\\
& D^\dagger \M(\tau) - \M(\tau) D = 0\\
& R^\dagger(g\inv) \M(\tau) - \M(\tau) R(g) = 0
\end{split}
\end{equation}
Here we have encoded the objective function into the matrix $\widehat{O}_{ij}$. For example, if we want to derive a lower bound on $\ev{ \bar{\phi}(\tau=T) \phi(0)}$ for some operator $\phi$, we should expand $\phi$ in our basis of operators $\phi = f_i \mathcal{O}_i$ and then $\widehat{O}_{ij} = \bar{f}_i f_j$. By taking $\widehat{O} \to - \widehat{O}$, one can also try to obtain an upper bound. Furthermore, for simplicity, we assume the knowledge of ${\cal M}(0)$, even though the following discussion straightforwardly generalizes to the case where ${\cal M}(0)$ is not given (see Appendix \ref{sec:gsbootnoinitPD}). 

Importantly, all of the constraints in (\ref{primalMN}) are convex in the variables ${\cal M}(\tau)$ and $\N(\tau)$, and thus, (\ref{primalMN}) is an SDP problem. However, these variables are functions of $\tau$, making the problem infinite-dimensional even if we truncate to a finite-dimensional operator basis. In order to make the problem finite-dimensional while still producing rigorous bounds on the objective function, we now turn to the dual problem.

To this end, we introduce the following action\footnote{In the optimization literature this is sometimes referred to as a Lagrangian functional.}: %
\begin{align}
	I &= \Tr  \bigg\{\widehat{O} \M(T)\nonumber\\ &+ \int_0^T    \left[ \lambda_A D^\dagger - D \lambda_A  +\hf (D \lambda_G + \lambda_G D^\dagger )+   \hf (\lambda_D D^\dagger+ D \lambda_D) - \lambda_D \partial_\tau   -\Lambda_\M \right] \mathcal{M}(\tau) \d \tau \nonumber\\
    & + \int_0^T  \left[ -\lambda_G  - \Lambda_\N\right] \mathcal{N}(\tau)  \, \d \tau \bigg \}\\
    &= \Tr  \bigg\{ \int_0^T    \left[ \lambda_A D^\dagger   - D \lambda_A+\hf (D \lambda_G + \lambda_G D^\dagger )  + \hf (\lambda_D D^\dagger+ D \lambda_D)   + \partial_\tau\lambda_D     -\Lambda_\M \right] \mathcal{M}(\tau)  \, \d \tau \nonumber\\
    &\quad + \int_0^T \left[ -\lambda_G - \Lambda_\N\right] \mathcal{N}(\tau) \, \d \tau + \lambda_D(0)  \M(0) + \left[ \widehat{O} -\lambda_D(T)  \right]  \M(T) \bigg \}\nonumber
\end{align}
Here $\vec{\lambda} = \{\lambda_A(\tau), \lambda_D(\tau), \lambda_G(\tau)\}$ are Lagrange multipliers. $\lambda_A$ is anti-Hermitian because $D^\dagger \M - \M D$ is anti-Hermitian, while $\lambda_D$ and $\lambda_G$ are Hermitian matrices.\footnote{Since $\M$ and $\N$ are Hermitian, the corresponding Lagrange multipliers should be Hermitian.} We have also included $\vec{\Lambda} = \{ \Lambda_\M(\tau), \Lambda_\N (\tau) \}$ which are Hermitian, positive-semidefinite matrices that impose the positivity of $\M$ and $\N$. For simplicity, we have ignored any possible symmetries, but the generalization is straightforward. %
Varying with respect to $\M(\tau)$ and also the boundary value $\M(T)$, we arrive at the {\it dual problem}: %
\begin{equation}\label{SDP:MNboot}
\boxed{    \begin{split}
        \text { maximize } &  \operatorname{Tr} \lambda_D(0) \M(0) \\
\text { subject to } & \widehat{O}=\lambda_D(T)   \\
&  \hspace{-15pt}\lambda_A(\tau) D^\dagger - D \lambda_A(\tau)    + \hf (D \lambda_G(\tau) +\lambda_G(\tau) D^\dagger + D \lambda_D(\tau) + \lambda_D(\tau) D^\dagger )  + \partial_\tau \lambda_D(\tau)     \succeq 0\\
&\hspace{-15pt}-\lambda_G(\tau)  \succeq 0, \quad \forall \tau \in [0,T]
\end{split}}
\end{equation}
The dual problem has the interesting property that the Heisenberg {\it equations of motion have been converted into an inequality}, as opposed to an equality. This means that we do not have to solve the equations of motion; if we find any functions $\vec{\lambda}(\tau) = \{\lambda_A(\tau), \lambda_G(\tau), \lambda_D(\tau) \}$ which satisfy the inequalities in \eqref{SDP:MNboot}, we immediately get a lower bound on $\Tr \widehat{O} \M(T)$.  Note that the primal objective is encoded as a constraint in the dual problem.

We now introduce the first truncation scheme regarding the space of operators. For a word $\tilde\O$ consisting of $n$ number of $X$'s and $m$ number of $P$'s (e.g. $X^2P^{m-4}X^{n-2}P^4$), we define its weight $\ell({\tilde\O})$ to be $n+2m$. Then, we define a level $L$ for any given operator $\O$ that is a sum of words consisting of $X$ and $P$ to be the maximal weight of its words. 

The asymmetric assignment $\ell(X)=1$ and $\ell(P)=2$ is adapted to the quartic Hamiltonian. Indeed, the Heisenberg equations have the schematic form $[H,X]\sim P$ and $[H,P]\sim X+X^3$. With this assignment, the highest-weight term in either commutator raises the weight by one.

The operator basis $\{\O_i\}$ we choose to construct the matrix $\mathcal{M}$ is organized in terms of its level. Truncation of $\M$ to level $L$ defines a finite-dimensional submatrix of $\M$ that is built out of basis operators $\{\O\}_i$ of levels up to $L/2$. For the matrix $\N$, note that the commutator $[H,\O]$ has a weight bigger than that of $\O$ by 1. Therefore, truncation of $\N$ to level $L+1$, constructed by operator basis of levels up to $L/2$, produces a finite-dimensional submatrix of $\N$ whose matrix elements are expectation values of operators with levels up to $L+1$. In this work, when we declare that ground-state bootstrap is performed at level $L$, we have used truncation of $\M$ to level $L$ and truncation of $\N$ to level $L+1$. Similarly, for thermal bootstrap at level $L$, we have used truncation of $\M$ to level $L$. {In practice, the nontrivial step in this truncation is the computation of $D$ and the construction of linearly independent bases for the matrices $\lambda_{D/G}$ and $\lambda_A D^\dagger - D \lambda_A$. A concrete algorithm for performing this task is given in Appendix~\ref{appendix:finitedimdiscussion}. While not guaranteed to be complete, it systematically reproduces the majority of the independent constraints up to the corresponding level in all cases examined.}%

Primal bootstrap constraints up to level $L$ still produce rigorous bounds on the physical observables of interest. As explained above, one can formulate the dual of such level-$L$ primal problem that takes the form \eqref{SDP:MNboot} where the dimension of the matrices are finite. However, such a dual optimization problem is still infinite-dimensional because the variables $\vec{\lambda}(\tau)$ are functions of $\tau$. 
A simple way to proceed is to test positivity on a finite grid of times $\{\tau_i\}$. A naive fixed-grid check is not rigorous, since positivity might be violated between grid points. However, by refining the grid one can hope to converge on reliable bounds.\footnote{This limitation is not intrinsic to grid-based or barrier-function methods. For a finite-dimensional function ansatz, adaptive refinement combined with certified derivative bounds, interval arithmetic, or controlled subdivision can establish positivity between sampled points. Likewise, adaptive quadrature can evaluate a barrier function and its gradient with controlled error \cite{Lawrence:2024spectral}.} Note that this method is similar to the barrier function method adopted in \cite{Lawrence:2024mnj}; see Appendix \ref{app:lawrence} for more on this. 

We will instead opt for a rigorous approach, based on the weak duality theorem in SDP, which implies that \textit{any} feasible solution of (\ref{SDP:MNboot}) provides a rigorous lower bound on the primal objective $\operatorname{Tr} \widehat{O} \M(T)$ in (\ref{primalMN}). Therefore, we can render the dual problem (\ref{SDP:MNboot}) finite-dimensional by expanding $\vec{\lambda}(\tau)$ in a finite-dimensional basis ansatz and searching for the optimal feasible solution within this restricted function space. The resulting feasible solution still yields a rigorous lower bound on the primal objective.

To ensure that the positivity constraints are satisfied, we expand the functions in a basis consisting of positive functions, which need not be orthonormal. Let this basis be denoted by ${\cal B}_{\cal I} = \{\phi_I(\tau)\}_{I \in {\cal I}}$, where ${\cal I}$ is a finite index set, and each $\phi_I(\tau) \geq 0$ for all $I \in {\cal I}$ and $\tau \in [0,T]$. A sufficient condition for the feasibility of (\ref{SDP:MNboot}) is that, once the positivity constraints are expressed in this basis, the resulting \textit{coefficient} matrices are all positive semidefinite. Expanding in this basis,
\begin{equation} \label{finiteBasis}
    [\lambda_A(\tau)]_{ij}=\sum_{I\in\cal I}a_{ij}^I \phi_I(\tau),~~[\lambda_D^{}(\tau)]_{ij}=\sum_{I\in\cal I}d^{I}_{ij}\phi_I(\tau),~~[\lambda_G^{}(\tau)]_{ij}=\sum_{I\in\cal I}g^{I}_{ij}\phi_I(\tau).
\end{equation}
A subtlety is the following: since the positivity constraints involve the derivative term $\frac{\d}{\d \tau} \lambda_D^{(k)}(\tau)$, we must ensure that $d^I_{ij} \neq 0$ only for those $I \in \cal I$ for which $\frac{\d}{\d \tau}\phi_I(\tau)$ is well-defined\footnote{For example, with linear splines, the derivative can be discontinuous.}. To this end, we define the maximal subset ${\cal J} \subseteq \cal I$ such that $\frac{\d}{\d \tau}\phi_I(\tau)$ is well-defined for all $I \in \cal J$, and is also closed under differentiation:
\begin{equation} \label{dDJI}
    \frac{\d}{\d \tau}\phi_J(\tau)=\sum_{I\in{\cal I}}\mathcal{D}_{JI}\phi_I(\tau), ~~~\forall J\in\cal J.
\end{equation}  
So once we identify this subset, we simply restrict $d^I = 0$ if $I \notin \cal J$.
Such a basis of positive functions can be constructed from polynomial-based functions. In this work, we choose to use the clamped B-spline basis, whose details are provided in Appendix~\ref{app:b-spline}.

To summarize, the final {\it finite-dimensional} dual problem is given by
\begin{equation}\label{SDP:finiteDim1}
\boxed{    \begin{split}
        \text { maximize } & \sum_{I\in\cal J}\phi_I(0) \operatorname{Tr} d^I \M(0) \\
\text { subject to } & \widehat{O} =\sum_{I\in\cal J}d^I\phi_I(T)   \\
& a^I D^\dagger -   D a^I  + \hf ( D g^I + g^I D^\dagger + D d^I + d^I D^\dagger) +\sum_{J\in\cal J}\mathcal{D}_{JI} d^J \succeq 0,~~~\forall I\in\cal I\\
&-g^I \succeq 0,~~~\forall I\in\cal I,
\end{split}}
\end{equation}
where here we have implicitly assumed that $d^I = 0$ if $I \notin \mathcal{J}$.
The variables of the SDP problem are the matrix elements of $a^I, d^J,$ and $g^I$. The positivity constraints in (\ref{SDP:finiteDim1}) can be efficiently implemented using standard SDP solvers. In this work, we mostly used \texttt{MOSEK} \cite{MOSEK}.

We also remark that feasible solutions of the dual problem \eqref{SDP:MNboot} can be found using the polynomials instead of splines, as we explain in Appendix \ref{app:PMP}.
The basic idea is that we take a polynomial ansatz for the Lagrange multipliers
\begin{align} \label{polySDP}
    \lambda_A(\tau) =  \sum_{j=0}^d a_j \tau^j, \quad \lambda_D(\tau) =  \sum_{j=0}^d d_j \tau^j, \quad \lambda_G(\tau) =  \sum_{j=0}^d g_j \tau^j, \quad \tau = \frac{T y}{1+y}.
\end{align}
We have mapped $\tau \to y$ in such a way that $\tau \in [0,T]$ gets mapped to $y \in [0, \infty)$. In \eqref{polySDP}, $\{ a_j, d_j, g_j\} $ are  matrices, so the resulting dual problem will involve a positivity constraint on a matrix of polynomials\footnote{Note that in the $y$ variables, we have a rational ansatz $\lambda_A = (1+y)^{-d} \sum_j T^j a_j y^j (1+y)^{d-j}$. However, the overall positive factor of $(1+y)^{-d}$ essentially factors out of the dual constraints, see Appendix \ref{app:PMP}.}.
In Appendix \ref{app:PMP}, we show that the dual problem has the precise form of a polynomial matrix program (PMP), which can be efficiently solved by \texttt{SDPB} \cite{Simmons-Duffin:2015qma,Landry:2019qug}, which we also used in this work. An advantage of the PMP approach is that a smaller number of parameters can be used to achieve similar numerical bounds, see Figure \ref{splineConvergence}. In other words, the PMP method appears to converge faster for the physical problems we considered; a more thorough comparison of the computational cost is left to future work.

\subsection{Thermal state}
\def\T{\mathcal{T}}
A variant of the previous problem is the Euclidean two-point correlator at finite temperature:
\begin{align}
	\M_{ij}(\tau) = %
    \frac{1}{Z} \Tr[ e^{-\beta H} \bar{\mathcal{O}}_i(\tau) \mathcal{O}_j(0)]
    = \ev{\bar{\mathcal{O}}_i(\tau) \mathcal{O}_j(0)} , \quad \M(\tau) \succeq 0
\end{align}
We have the same constraints on $\mathcal{M}$ from the reflection positivity, the Heisenberg equations of motion $\partial_\tau \mathcal{M} = - \mathcal{M} D$, and time-translation invariance $D^\dagger \mathcal{M} = \mathcal{M} D$.
However, instead of imposing the ground-state positivity condition, we now impose the KMS condition:
\begin{align} \label{kms}
	\mathcal{M}_{ij}(\beta) &= {\mathcal{M}}_{\bar{j} \bar{i}}(0)=\langle\O_j\bar\O_i \rangle%
\end{align}
This follows from $ Z(\beta)	\mathcal{M}_{ij}(\beta) = \Tr \bar{\O}_i e^{-\beta H} \O_j = \Tr e^{-\beta H} \O_j \bar{\O}_i   
$.
Note that the RHS of \eqref{kms} is a linear function of $\mathcal{M}(0)$: we can define a matrix $\T$ such that $\O_i = \T_{ij} \O^\dagger_j$.

We first consider the case where $\M(0)$ is not known. There are still linear relations between different matrix elements $\M_{ij}(0)$ that take the form $\operatorname{Tr}B^{(i)}\M(0)=b^{(i)}$ for some matrices and constants $B^{(i)}$ and $b^{(i)}$. The corresponding primal problem is given by the following SDP problem: %
\begin{equation}\label{primalThermal}
\begin{split}
\operatorname{minimize } & \operatorname{Tr} \widehat{O} \M(T) \\
\text { subject to } & \M(\tau) \succeq 0\\
& \left(D^\dagger - \partial_\tau \right) \M(\tau)=0 \\
& D^\dagger \M(\tau) - \M(\tau) D = 0\\
& R^\dagger(g\inv) \M(\tau) - \M(\tau) R(g) = 0\\
&\mathcal{M}(\beta) = \mathcal{T} \mathcal{M}^T(0) \mathcal{T}^\dagger %
\\
& \Tr B^{(i)} \mathcal{M}(0) = b^{(i)}
\end{split}
\end{equation}

The derivation of the dual problem follows essentially the same steps as before. The action is given by
\begin{align}
    I &= \Tr  \bigg\{\widehat{O} \M(T) + \lambda_E (\M(\beta) - \T \mathcal{M}^T(0) \T^\dagger]) \nonumber\\ &+ \int_0^\beta  \d \tau \left[ \lambda_A D^\dagger - D \lambda_A  +   \hf (\lambda_D D^\dagger+ D \lambda_D) -  \lambda_D\partial_\tau     -\Lambda_\M \right] \mathcal{M}(\tau)   \bigg\}\\
    &+ \lambda_{B,i} \left( \Tr  (B^{(i)} \M(0)) - b^{(i)} \right). \nonumber%
\end{align}
We may think of $\widehat{O}$ as the strength of a $\delta$-function source\footnote{We may consider the more general problem of minimize $\int \d \tau J(\tau) \M(\tau)$. Then one would obtain the constraint
\begin{align}
 & j(\tau) + \lambda_A D^\dagger - D \lambda_A    - \hf (- D \lambda_D - \lambda_D D^\dagger )  + \partial_\tau \lambda_D     \succeq 0.
\end{align}
Then we may specialize to the case where $j = \widehat{O} \delta(\tau-T)$. Integrating over the discontinuity gives $\widehat{O} + \lambda_D(T+\epsilon) - \lambda_D(T-\epsilon) \succeq 0$.
} at time $\tau = T$. Therefore the Lagrange multiplier $\lambda_D$ may have a discontinuity at $\tau = T$. The dual SDP problem is given by 
\begin{equation}\label{dualsdp:thermalNoInit}
\begin{split}
\text { maximize } & -\lambda_{B,i} b^{(i)}\\
\text { subject to } & \widehat{O}+\lambda_D(T+\epsilon)-\lambda_D(T-\epsilon) \succeq  0   \\
&  \lambda_A D^\dagger - D \lambda_A    - \hf (%
- D \lambda_D - \lambda_D D^\dagger )  + \partial_\tau \lambda_D     \succeq 0\\
&\lambda_D(0) + \lambda_{B,i} B^i - [\T^\dagger \lambda_E \T]^T \succeq 0\\
&-\lambda_D(\beta) + \lambda_E \succeq 0
\end{split}
\end{equation}

We can also consider a variant of the problem where $\M(0)$ is already known. We can then solve for $\M(\beta)$ using KMS, so we can treat $\M(\beta)$ as known. We therefore get a simpler action %
\begin{align}
    I &= \Tr  \bigg\{\widehat{O} \M(T)  + \int_0^\beta  \d \tau \left[ \lambda_A D^\dagger - D \lambda_A  +   \hf (\lambda_D D^\dagger+ D \lambda_D) - \lambda_D \partial_\tau    -\Lambda_\M \right] \mathcal{M}(\tau)  \bigg\}.\nonumber \\
    &= \Tr  \bigg\{\int_0^\beta  \d \tau \left[ \lambda_A D^\dagger - D \lambda_A  +   \hf (\lambda_D D^\dagger+ D \lambda_D) + \partial_\tau \lambda_D   -\Lambda_\M \right] \mathcal{M}(\tau)   \\
    &\quad + (\widehat{O}+\lambda_D(T+\epsilon) - \lambda_D(T-\epsilon)) \M(T) +  \lambda_D(0) \M(0)  - \lambda_D(\beta) \M(\beta)    \bigg\}\nonumber
\end{align}
We arrive at the dual problem:
\begin{equation}\label{SDP:dualThermal}
\boxed{    \begin{split}
        \text { maximize } &  \operatorname{Tr} \lambda_D(0) \M(0) - \lambda_D(\beta) \M(\beta) \\
\text { subject to } & \widehat{O}+\lambda_D(T+\epsilon)-\lambda_D(T-\epsilon) \succeq  0   \\
&  \lambda_A D^\dagger - D \lambda_A    + \hf ( D \lambda_D + \lambda_D D^\dagger )  + \partial_\tau \lambda_D     \succeq 0,\, \forall \tau \in [0,T]
\end{split}}
\end{equation}

Making the dual problem finite-dimensional can be done analogously to (\ref{SDP:finiteDim1}). We introduce finite-dimensional bases of positive functions ${\cal B}_{{\cal I}_L} = \{\phi_{I_L}(\tau)\}_{I_L\in{\cal I}_L}$ and ${\cal B}_{{\cal I}_R} = \{\phi_{I_R}(\tau)\}_{I_R\in{\cal I}_R}$ over $\tau \in [0,T]$ and $\tau \in [T,\beta]$, respectively, where ${\cal I}_L$ and ${\cal I}_R$ are finite index sets, and
\begin{align}
    {}&\phi_{I_L}(\tau)\geq0,~~~\forall I_L\in{\cal I}_L,~\forall \tau\in[0,T],\\
    &\phi_{I_R}(\tau)\geq0,~~~\forall I_R\in{\cal I}_R,~\forall \tau\in[T,\beta].
\end{align}
We also define the maximal subsets ${\cal J}_L \subseteq {\cal I}_L$ and ${\cal J}_R \subseteq {\cal I}_R$ such that ${\d\over \d\tau}\phi_{I_{L}}(\tau)$ and ${\d\over \d\tau}\phi_{I_{R}}(\tau)$ are well-defined for all $I_L \in {\cal J}_L$ and $I_R \in {\cal J}_R$, and further satisfy the closure:
\begin{align}
    {}&\frac{\d}{\d \tau}\phi_{J_L}(\tau)=\sum_{I_L\in{\cal I}_L}U^L_{J_LI_L}\phi_{I_L}(\tau), ~~~\forall J_L\in{\cal J}_L,
    \\
    &\frac{\d}{\d \tau}\phi_{J_R}(\tau)=\sum_{I_R\in{\cal I}_R}U^R_{J_RI_R}\phi_{I_R}(\tau), ~~~\forall J_R\in{\cal J}_R.
\end{align}
Such bases can, for example, be constructed explicitly from clamped B-splines.

We can now expand $\lambda_A^{(i)}(\tau)$ and $\lambda_D^{(k)}(\tau)$ as
\begin{align}
    {}&\lambda_A^{(i)}(\tau)=\theta(T-\tau)\sum_{I_L\in{\cal I}_L}a^{(i)}_{I_L}\phi_{I_L}(\tau)+\theta(\tau-T)\sum_{I_R\in{\cal I}_R}a^{(i)}_{I_R}\phi_{I_R}(\tau),
    \\
    &\lambda_D^{(k)}(\tau)=\theta(T-\tau)\sum_{I_L\in{\cal J}_L}q^{(k)}_{I_L}\phi_{I_L}(\tau)+\theta(\tau-T)\sum_{I_R\in{\cal J}_R}q^{(k)}_{I_R}\phi_{I_R}(\tau).
\end{align}
Then, the finite-dimensional dual problem becomes
\begin{align}\label{dualthermalSDPspline}
\text { maximize } & \sum_{I_L\in{\cal J}_L}q^{(k)}_{I_L}\phi_{I_L}(0) \operatorname{Tr} C^{(k)} \M(0) - \sum_{I_R\in{\cal J}_R}q^{(k)}_{I_R}\phi_{I_R}(\beta) \operatorname{Tr} C^{(k)} \M(\beta) \\
\text { subject to } 
&  \O + C^{(k)} \left[\sum_{I_R\in{\cal J}_R}q^{(k)}_{I_R}\phi_{I_R}(T) -\sum_{I_L\in{\cal J}_L}q^{(k)}_{I_L}\phi_{I_L}(T)\right] \succeq 0\\
& A^{(i)}a^{(i)}_{I_L}+\Theta_{{\cal J}_L}(I_L)D^{(k)}q^{(k)}_{I_L}+C^{(k)}\sum_{J_L\in{\cal J}_L}U^L_{J_L I_L}q^{(k)}_{J_L} \succeq 0,~~~\forall I_L\in{\cal I}_L\\
& A^{(i)}a^{(i)}_{I_R}+\Theta_{{\cal J}_R}(I_R)D^{(k)}q^{(k)}_{I_R}+C^{(k)}\sum_{J_R\in{\cal J}_R}U^R_{J_R I_R}q^{(k)}_{J_R} \succeq 0,~~~\forall I_R\in{\cal I}_R
\end{align}

In bootstrapping the thermal/ground state correlators, there are minor variants of the method depending on the degree to which the one-point functions (two-point correlators $\ev{\O_1(\tau)\O_2(0)}$ with $\tau = 0$) are known. For the 1-MQM, the zero-time limit of the adjoint two-point correlators are single-trace expectation values. At zero temperature, these expectation values are known analytically at infinite $N$ \cite{Brezin:1977sv}, see Appendix \ref{app: analytic}. At finite temperature, the non-singlet states contribute and thus the single-trace expectation values are unknown. In principle, the two-point correlator bootstrap can still be used without any knowledge of the zero-time correlators (and in fact the two-point bootstrap does constrain one-point functions, as demonstrated in Figure \ref{fig:thermalEEB}). However, in practice, it may be easier to first bootstrap the one-point functions and then feed the partial knowledge gained from the one-point function bootstrap into the two-point correlator bootstrap. 
\begin{table}[H]
\centering
\begin{tabular}{|l|p{6.5cm}|l|}
\hline
\textbf{Zero-time input} & \textbf{Description} & \textbf{Reference} \\
\hline
Full knowledge & Exact knowledge of zero-time correlators (e.g.\ from large $N$ solution). & App.~\ref{app:gswithinitialdata}, \ref{app:withInitThermal} \\[3pt]
\hline
Partial knowledge & Assumes bounds derived from one-point function bootstrap (ground state/EEB). & App.~\ref{app:initialIneqThermal} \\[3pt]
\hline
No knowledge & combined bootstrap & App.~\ref{sec:gsbootnoinitPD}, \ref{app:noInitThermal} \\[3pt]
\hline
\end{tabular}
\caption{Summary of possible zero-time input assumptions and corresponding sections.}
\label{tab:zerotime_inputs}
\end{table}

\section{Analytic results \label{analyticBds}}
In this section, we present analytic results that can be derived from Euclidean two-point correlator bootstrap described in the previous section.

\subsection{An exponential bound on the two-point correlator\label{universalBd}}

Let us consider the $\M$ matrix with operators $\O_i\in\{\mathbf{1}, \mathcal{O}, [\O,H] \}$, leading to
\begin{equation} \label{pos3}
    \mathcal{M}(\tau) =
\begin{pmatrix}
1 & \langle \O \rangle & 0 \\
\langle \bar\O \rangle & G(\tau) & G'(\tau) \\
0 &G'(\tau) & G''(\tau)
\end{pmatrix}\succeq 0
\end{equation}
Here we have defined $G(\tau) = \ev{\bar\O(\tau) \O(0)}$. By the Schur Complement, this is equivalent to 
\begin{equation}
\label{eq:toLogConv}
    \begin{bmatrix}
G(\tau) & G'(\tau) \\
G'(\tau) &G''(\tau)
\end{bmatrix}-\begin{bmatrix}
\ev{\bar\O} \\
0
\end{bmatrix}
\begin{bmatrix}
\langle \O \rangle &
0
\end{bmatrix}\succeq0
\end{equation}
This is simply a condition on the connected two-point correlator $G_c(\tau) =  \ev{\bar\O(\tau) \O(0)}-  \ev{\bar\O}\ev{\O}$:
\begin{equation} \label{mcpos}
    \mathcal{M}_c(\tau) =
\begin{pmatrix}
	G_c(\tau) &  G_c'(\tau)  \\
 G_c'(\tau)	& G_c''(\tau) \end{pmatrix}\succeq0\\
\implies (\log G_c(\tau))'' \ge 0.
\end{equation}
In other words, the connected two-point correlator is {\it log-convex}. The function that saturates this inequality takes the exponential form. Assuming $\tau>0$, we obtain
\begin{align}\label{universal-lower-bound}
    &\log G_c(\tau)\geq \log G_c(0)+\tau \frac{G_c'(0^+)}{G_c(0)}\Rightarrow G_c(\tau)\geq G_c(0) \exp \left(-\mu \tau  \right), \quad  \mu = \frac{-G'_c(0^+)}{G_c(0)}.
\end{align}

\subsubsection{Applications}
Now we can apply this to a non-relativistic particle in a potential
\begin{equation}
    H=\frac{1}{2} p^2+ V(x).
\end{equation}
We choose $\mathcal{O} = x$. Then $G_c(0)=\langle x(0)^2\rangle-\langle x(0)\rangle^2$  and $G_c'(0^+)=-\frac{1}{2}$ so $\mu = \frac{1}{2 G_c(0)}$.
We will compare this analytic result to the numeric bounds in Figure \ref{fig:euclideanThermalFig} later.

This result can easily be generalized to (multi-)MQM. 
\begin{align}\label{eqn:adjointgapuniversal}
    \ev{\tr X(\tau) X(0)} \ge \ev{\tr X^2} e^{-\mu \tau}, \quad \mu = \frac{1}{2{\ev{\tr X^2}}}
\end{align}
This bound implies an upper bound\footnote{An interesting {\it lower} bound on the adjoint gap in MQM was derived in \cite{Gross:1990ub}, see their equation (4.16). } on the adjoint gap $\Delta_\text{adj} \le \mu$. An alternative derivation of this inequality was given in \cite{Lin:2025srf}. This bound will be explained in more detail and further refined in section \ref{gap2}.

\subsubsection{Generalization to the thermal state}
Note that for the thermal state, the two-point correlator should be symmetric about $\tau =\beta/2$, and we have two bounds that are simultaneously valid,
\begin{align}\label{eqn:univExpThermal}
  \frac{G_c(\tau)}{G_c(0)}  \ge   \text{max} \left\{ \exp \left(- \mu \tau \right), \exp \left(- \mu (\beta-\tau) \right) \right\}.
\end{align}

\subsubsection{Saturation of the bound by the harmonic oscillator}
Let us comment that the ground state bound \eqref{mcpos} is saturated by the simple harmonic oscillator. (Or more generally, $N$ copies of the harmonic oscillator.) In the basis $\{x,\i p\}$, we can write down the $\mathcal{N}$ matrix:
\begin{align}
\mathcal{N}(\tau) =
\begin{pmatrix}
-G'(\tau) &  -G(\tau) \\
 -G(\tau) & -G'(\tau)
\end{pmatrix}
\succeq 0,
\end{align}
where we have used the equations of motion. This gives $G(\tau)^2-G'(\tau)^2 \leq0$. The previous inequality \eqref{mcpos} can be simplified for the simple harmonic oscillator by using $G''(\tau) = G(\tau)$, which yields $G^2+G'(\tau)^2 \ge 0$. So we conclude that %
\begin{equation}
G'(\tau) = \pm  G(\tau)
\end{equation}
One-point function bootstrap leads to $\langle x(0)^2\rangle = {1\over2}$. Together with the decay conditions at $\tau=\pm\infty$, we arrive at the known answer
\begin{equation}
 G(\tau) = {1\over2}e^{-|\tau|}.
\end{equation}
which is simply $1/\omega^2$ in the frequency domain.

\subsection{Derivation of the Energy-Entropy balance inequality}

As we have seen from \eqref{mcpos}, the two-point correlator $G$ is log-convex\footnote{If $G_c$ is log-convex, then $G = G_c + \text{constant}$ is log-convex for positive constant. Hence both $G$ and $G_c$ are log-convex.}. Let us apply this to the thermal two-point correlator $G = \ev{\bar{O}(\tau) O(0)}_\beta$ where $\tau \in [0,\beta]$. Then
\begin{align}
    \log \frac{G(\beta)}{G(0)} \ge \beta \frac{G'(0)}{G(0)}. \label{logconvex}
\end{align}
Now the boundary values are
\begin{align}
G(0) = \ev{\bar{\O}\O}_\beta, \quad G'(0) = -  \ev{\bar{\O} [H, \O] }_\beta \\ 
    G(\beta) = \tr \{ \rho e^{ \beta H} \bar{\O} e^{-\beta H} \O \} =  \tr   \{ \O \bar{\O} \rho\}   =  \ev{\O \bar{\O}}_\beta.
\end{align}
In the last line we used the KMS condition. Plugging into \eqref{logconvex}, we arrive at the Energy-Entropy balance (EEB) inequality \cite{Araki:1977px,cmp/1103900930,Bratteli:1996xq}\footnote{See Appendix A of \cite{Cho:2024owx} for a physicist-friendly version of the proof.}:
\begin{align}\label{eqn:EEBineq}
    \log \frac{\ev{\bar{\O}\O}_\beta}{\ev{\O \bar{\O}}_\beta } \le \beta \frac{  \ev{\bar{\O} [H, \O] }_\beta}{\ev{\bar{\O} \O }_\beta }.
\end{align}
This inequality was used recently to bootstrap thermal one-point functions \cite{Fawzi:2023fpg, Cho:2024kxn}.

\begin{figure}[H]
    \centering
\includegraphics[width=.75\linewidth]%
{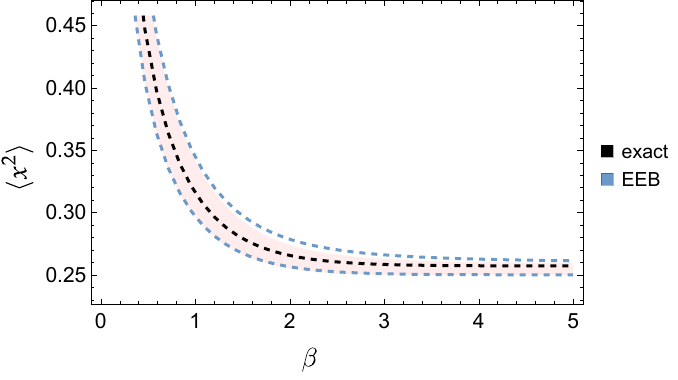}
    \caption{Comparison of the allowed region (in {\color{dpink} pink}) for the level 8 two-point correlator bootstrap (\ref{dualsdp:thermalNoInit}) using the KMS condition with the level 7 one-point function bootstrap using the EEB inequality \cite{Fawzi:2023fpg, Cho:2024kxn} for the anharmonic oscillator $V(x) = \frac12 x^2 + x^4$. We also show the ``exact'' value from Hamiltonian truncation (in black).
    }
    \label{fig:thermalEEB}
\end{figure}

For Hamiltonians with quartic potential where we define $\ell(x)=1$ and $\ell(p)=2$ similarly to the MQM case, the EEB inequality follows from considering operators of level $\ell(\O)+1$. We present the comparison between the two results for the anharmonic oscillator $V(x) = \frac12 x^2 + \frac14 x^4$ in Figure \ref{fig:thermalEEB}, where the two-point correlator bootstrap (\ref{dualsdp:thermalNoInit}), which assumes no knowledge of the initial condition, is used to obtain bootstrap bounds on the one-point function $\langle x^2\rangle$\footnote{Note that in \eqref{primalThermal}, we can choose some $\widehat{O}$ that picks out a one-point function.}.

\subsection{High-temperature bootstrap}
Here we show that the two-point correlator bootstrap imposing the KMS condition at high temperatures $(\beta\rightarrow0)$ reduces to the bootstrap of an integral. Consider the classical statistical mechanics of a particle in a potential:
\begin{align}
    Z = \int \d x \,  e^{-\beta V(x)}.
\end{align}
We can bootstrap moments of the Boltzmann distribution using the Schwinger-Dyson equation \cite{Lin:2020mme}:
\begin{align}
    n \ev{x^{n-1}} = \beta \ev{x^n V'(x)} \label{SD1particle}
\end{align}
This classical statistical mechanics problem is expected to be the high-temperature limit of the quantum mechanical problem with the Hamiltonian $H={p^2\over2}+V(x)$. %

The KMS condition implies
\begin{equation}
\langle\mathcal O_{1}(\beta)\,\mathcal O_{2}(0)\rangle=\langle\mathcal O_{2}(0) \,\mathcal O_{1}(0)\rangle .
\end{equation}
For small $\beta$,
\begin{equation}
\mathcal O_{1}(\beta)=\mathcal O_{1}(0)+\beta[H,\mathcal O_{1}(0)]+O(\beta^{2}),
\end{equation}
leading to
\begin{equation}\label{eqn:KMShighT}
\langle[\mathcal O_{2},\mathcal O_{1}]\rangle
=\beta\,\big\langle[H,\mathcal O_{1}]\,\mathcal O_{2}\big\rangle+O(\beta^{2}).
\end{equation}
Consider $\mathcal O_{1}=p$ and $\mathcal O_{2}=x^{n}$. Using $[x^{n},p]=\i n x^{\,n-1}$ and $[H,p]=iV'(x)$, we have
\begin{equation}
n\,\langle x^{\,n-1}\rangle=\beta\,\langle V'(x)\,x^{n}\rangle .
\end{equation}
which is precisely \eqref{SD1particle}.

\begin{figure}[H]
    \centering
    \includegraphics[width=0.7\linewidth]{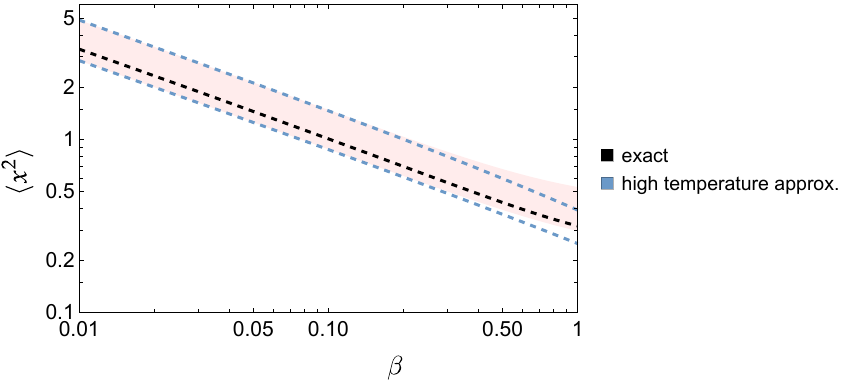}
    \caption{Comparison of the integral bootstrap (in dashed blue) with the two-point correlator bootstrap (in {\color{dpink} pink}, uses \eqref{dualsdp:thermalNoInit}) for the anharmonic oscillator $V(x) = \frac12 x^2 + x^4$. This shows good agreement at high temperature, where a bosonic quantum system reduces to classical statistical mechanics. Both bootstrap results are obtained at level 6.}%
    \label{fig:thermalX2smallbetaL6}
\end{figure}

We can then compare the two-point correlator bootstrap with the integral bootstrap at the same level. Let's consider a level $\ell$ integral bootstrap where we consider operators up to $x^\ell$. For the two-point bootstrap to contain the same equation, we need the operator $\O_2$ to be $x^{\ell-3}$ because of the $x^3$ term in $V'(x)$ in \eqref{SD1particle}. Since we include operators up to level $L/2$ in the level $L$ two-point bootstrap, they contain the same information if $L/2 = \ell-3$. For example, at high temperatures a level $L=6$ two-point correlator bootstrap (\ref{dualsdp:thermalNoInit}) is equivalent to a $\ell=6$ integral bootstrap, as shown in Figure \ref{fig:thermalX2smallbetaL6}.

\subsubsection{Generalization to MQM}
For the case of MQM, we expect the quantum system to reduce to the matrix integral at high temperatures. For the ungauged model, we expect to get the one-matrix model. This matrix model can be bootstrapped using the loop equations, together with positivity of the Hankel matrix $\M$  \cite{Lin:2020mme}. For the matrix integral defined by the partition function
\begin{equation}
    Z=\int \d X \, e^{-\beta N\tr V(X)},
\end{equation}
the loop equations are given by
\begin{equation}\label{eqn:matrixIntegralSD}
    \sum_{k=1}^n\langle \tr X^{k-1} \tr X^{n-k}\rangle=\beta\langle \tr \,X^n V'(X)\rangle.
\end{equation}
For the ungauged MQM with the Hamiltonian \eqref{eq:HamiltonianMQM},
we take ${\cal O}_1=P_{ab}$ and ${\cal O}_2=X^n_{ba}$ in (\ref{eqn:KMShighT}), after which we sum over the indices $a,b$. Using $\sum_{a,b}[X^n_{ba},P_{ab}]=\i N\sum_{k=1}^n\tr X^{k-1}\tr X^{n-k}$ and $[H,P_{ab}]=\i N\partial_{X_{ba}}V(X)$, we obtain
\begin{equation}
    \sum_{k=1}^n\langle \tr X^{k-1} \tr X^{n-k}\rangle=\beta\langle X^n_{ab}\partial_{X_{ba}}V(X)\rangle,
\end{equation}
in agreement with (\ref{eqn:matrixIntegralSD}). Note that for the gauged MQM, the matrix integral obtained in the high temperature limit will involve an additional matrix corresponding to the gauge field. This proof can be easily generalized for multi-MQM. For MQM that involves fermionic matrices (such as Banks-Fischler-Shenker-Susskind (BFSS) matrix model \cite{Banks:1996vh}), the fermions will be suppressed in the high temperature limit.

\section{Bootstrap bounds \label{numericalBds}}
We now turn to the numerical bootstrap bounds on the Euclidean two-point correlators. Dual SDP formulations described in section \ref{2ptSetup} produce rigorous and sometimes highly tight bounds, which allow us to extract further observables such as spectrum and matrix elements.

\subsection{Warm-up: anharmonic oscillator}
\begin{figure}[H]
    \centering
\includegraphics[width=.75\linewidth]{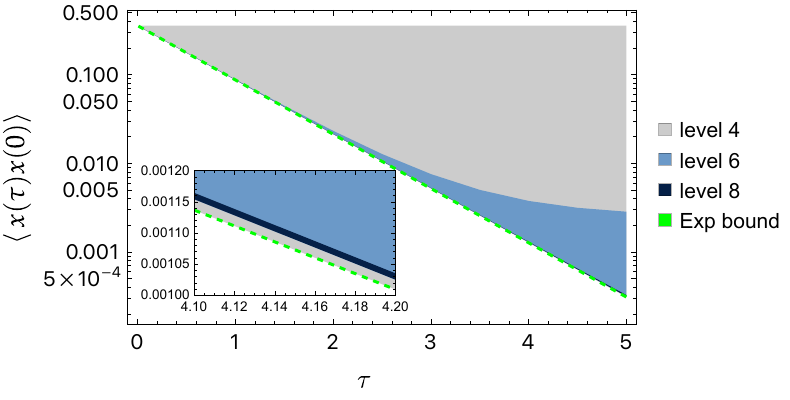}
    \caption{The ground-state Euclidean two-point correlator of the anharmonic oscillator with $H=\frac{1}{2} p^2+ \frac{1}{2} x^2 + \frac{1}{4} x^4$. The shaded regions indicate the bootstrap allowed regions at level-$\{4,6,8\}$. Note that level $L$ means $\M$ with level $L$ and $\N$ with level $L+1$. The green dashed line indicates the exponential lower bound \eqref{universal-lower-bound}. This plot is obtained by feeding the zero-time data at $\tau=0$ from Hamiltonian truncation into the dual problem \eqref{SDP:MNbootPractical}, discretized using the PMP formulation described in Appendix~\ref{app:PMP} with $d=16$. The bounds at different levels are also reported in Table  \ref{tab:anharmonic-levels}.
    }
    \label{fig:euclideanGroundFig}
\end{figure}

As a toy problem, we consider bootstrapping the two-point correlator of the anharmonic oscillator
\begin{align}
	H = \hf p^2 + \hf x^2 + g x^4.
\end{align}
To bootstrap the ground-state two-point correlator up to Euclidean time $\tau=5$, we consider the dual problem \eqref{SDP:MNboot} whose practical implementation is discussed in Appendix~\ref{app:gswithinitialdata}. We use the PMP form of the problem (Appendix \ref{app:PMP}) and use \texttt{SDPB} to obtain numerical bounds. The zero-time data are inputted from Hamiltonian truncation. Figure \ref{fig:euclideanGroundFig} shows the upper and lower bounds we obtained at different levels. The bounds become looser as $\tau$ increases. We list the results at the largest time $\tau=5$ in Table~\ref{tab:anharmonic-levels}. Note that the numerical lower bounds at level 4 saturate the universal lower bounds in \eqref{universal-lower-bound} as the highest-weight operator in \eqref{mcpos} is exactly level 4.

\begin{table}[]
    \centering
    \begin{tabular}{c|c|c|c}
         & lower & upper & difference\\
         \hline
         \text{level 4}& 0.0003092033 & 0.3548402512 & 0.3545310478 \\ 
         \text{level 6}& 0.0003151403 &  0.0028321050& 0.0025169648\\ 
         \text{level 8}& 0.0003151607 & 0.0003256455 & \( 1.04848 \times 10^{-5}\) \\ 
         \text{level 10}& 0.000315160730 & 0.000315201914 & $4.1184\times 10^{-8}$ \\ 
    \end{tabular}
    \caption{The bootstrap results for $\gev {x(\tau\!=\! 5)x(0)}$ in the anharmonic oscillator with potential $V(x)=x^2/2+x^4/4$ at levels \( \{4,6,8,10\}\), using the PMP formulation with $d=16$. The exact value from Hamiltonian truncation is~$\approx 0.000315160732$.}
    \label{tab:anharmonic-levels}
\end{table}

\begin{figure}
    \centering
\includegraphics[width=1\linewidth]{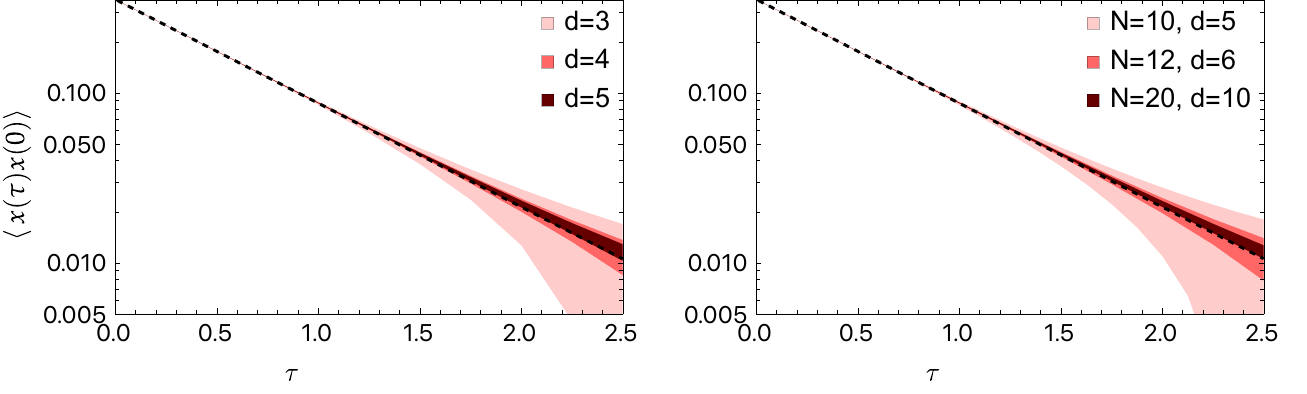}
    \caption{The lower and upper bounds converge as the parameters of either the polynomial or spline basis increase (such as the degrees of the polynomial used in the Lagrange multipliers). On the left, we show the convergence in the polynomial basis where $\mathsf{d}$ is the degree of the polynomial. On the right, we show the convergence in the B-spline parameters, which include the number of nodes $\mathsf{N}$ and the degree $\mathsf{d}$ (see Appendix \ref{app:b-spline}). We show the results in the case of the ground state of an anharmonic oscillator with $V(x) = \hf x^2 + \qrt x^4$.}
    \label{splineConvergence}
\end{figure}

At a fixed level, the bootstrap bounds converge as we increase the number of parameters in either the polynomial or the spline basis, which means the increased dimension of the search space. We study this convergence for both the clamped B-spline basis and PMP in Figure \ref{splineConvergence}. This convergence, along with the agreement between two different bases, provides a strong evidence that we reached the optimum in the dual problem at a fixed level.

\begin{figure}[H]
        \centering
\includegraphics[width=1\linewidth]
{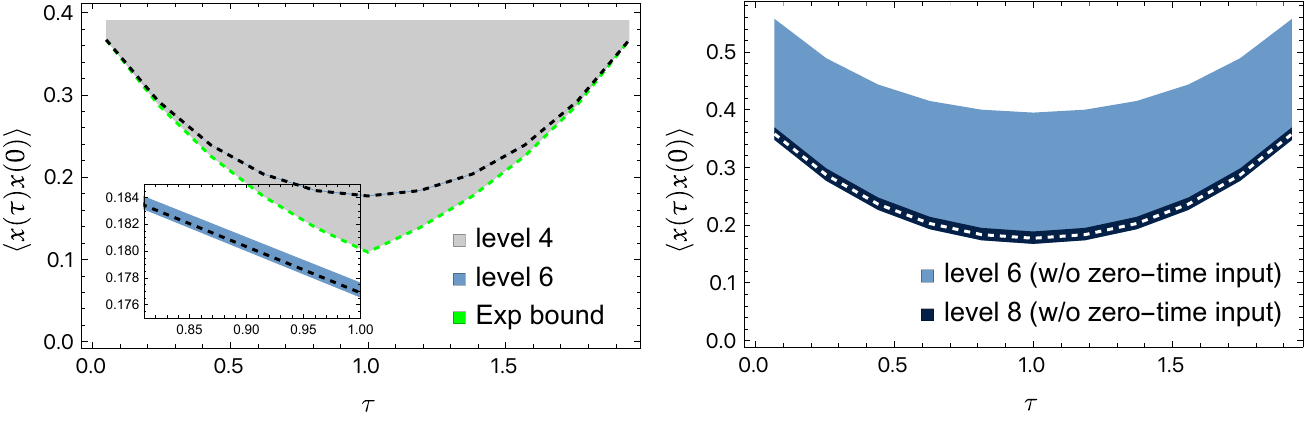}
        \caption{Bootstrap bounds on the Euclidean two-point correlator in the thermal state of anharmonic oscillator with $H=\frac{1}{2} x^2+ \frac{1}{2} p^2 + \frac{1}{4} x^4$ at $\beta=2$. {\it Left}: Bootstrap results using the zero-time input obtained from Hamiltonian truncation, as stated in \eqref{SDP:dualThermal} and Appendix \ref{app:withInitThermal}. {\it Right}: Bootstrap results without the zero-time input, as stated in \eqref{dualsdp:thermalNoInit} and Appendix \ref{app:noInitThermal}. The shaded regions indicate the bootstrap allowed regions at different levels. The green dashed line indicates the exponential lower bound \eqref{eqn:univExpThermal}. Both figures are obtained with the B-spline basis.
        }
    \label{fig:euclideanThermalFig}
\end{figure}

We also consider the bootstrap bounds on the thermal two-point correlators. In Figure \ref{fig:euclideanThermalFig}, we present bootstrap bounds obtained from \eqref{dualthermalSDPspline} using clamped B-splines, at $\beta=2$. The left figure assumes the knowledge of the zero-time data obtained by Hamiltonian truncation as discussed in \eqref{SDP:dualThermal} and Appendix \ref{app:withInitThermal}, while the right figure does not assume any zero-time data as discussed in \eqref{dualsdp:thermalNoInit} and Appendix \ref{app:noInitThermal}. We observe that, \textit{a priori}, zero-time input are not required to obtain rigorous bounds that converge as the level increases. Nonetheless, the inclusion of zero-time input strengthens the bounds at any given level.

\subsection{Matrix Quantum mechanics}
In this section, we study the 1-MQM with Hamiltonian \eqref{eq:HamiltonianMQM}. We will take
\begin{align}
    V(X) = \hf X^2 + g X^4
\end{align}
The $U(N)$ symmetry generators are given by
\begin{align} \label{gaugeGen}
    C = \i [X,P] + 1.
\end{align}
The moments (one-point functions) of this system obey various identities implied by %
the cyclicity of the trace, the gauge constraints for singlet states, time-reflection symmetry, and translation symmetry. We provide more details in Appendix ~\ref{app:1ptMQM}. In particular, time-reversal invariance implies that if we define $\Pi = -\i P$, all correlators involving $\Pi$ and $X$ are real. For example, $\ev{\tr X \Pi} = - \ev{\tr \Pi X} = \frac{1}{2} $. 

Using the $U(N)$-invariance of the ground state $C \ket{\Omega} = 0 $, we obtain
\begin{align}
    \ev{\Omega|\tr (\O_i(\tau) \O_j(0) C(0))|\Omega} &= \ev{\Omega|\tr (C(\tau) \O_i(\tau) \O_j(0) )|\Omega} = 0 \label{gaugeinv1} %
    \end{align}
We also have a constraint when $C$ is sandwiched between two operators:   
    \begin{align}
     \ev{\Omega| \O_{i,a_1 a_2}(\tau) C_{a_2,a_3}(\tau) \O_{j,a_3,a_1}(0)|\Omega}  &= \ev{\Omega| \O_{i,a_1 a_2}(\tau) C_{a_2,a_3}(0) \O_{j,a_3,a_1}(0)|\Omega} \nonumber\\
     &=
      \ev{\Omega| \O_{i,a_1 a_2}(\tau) [C_{a_2,a_3} ,\O_{j,a_3,a_1}(0)]|\Omega}\\
      &=\ev{\Omega| \tr \O_{i} (\tau) \O_{j}(0)|\Omega} - \ev{\Omega|\tr \O_{i}|\Omega} \ev{\Omega|\tr \O_j|\Omega}. \nonumber%
\end{align}
Here we have used that the action of the generator on an adjoint operator is
\begin{align}
[C_{ab},\, \mathcal{O}_{cd}] = \frac{1}{N}
    \left( \mathcal{O}_{ad}\delta_{bc} - \mathcal{O}_{cb}\delta_{ad} \right),
\end{align}
Note that the gauge constraint and cyclicity of the zero-time correlators are the only constraints where we have used large $N$. Curiously large $N$ factorization enters essentially in the zero-time data, see Appendix \ref{app:1ptMQM}. %

We can also derive weaker conditions for the two-point correlator in the thermal state of the ungauged model. We use that the thermal density matrix $\rho_\beta$ obeys $[\rho_\beta, C_{ab}] = 0$:
\begin{align}
    \ev{\tr (\O_i(\tau) \O_j(0) C(0))}_\beta = \Tr_\mathcal{H} \rho_\beta \O_{i,a_1 a_2} (\tau) \O_{j,a_2 a_3} C(0)_{a_3 a_1}\nonumber\\
    = \Tr_\mathcal{H} \rho_\beta C(0)_{a_3 a_1}  \O_{i,a_1 a_2} (\tau) \O_{j,a_2 a_3} \\
       = \Tr_\mathcal{H} \rho_\beta C(\tau)_{a_3 a_1}  \O_{i,a_1 a_2} (\tau) \O_{j,a_2 a_3}\nonumber
\end{align}
The last line follows from the fact that $C$ commutes with the Hamiltonian.

\subsection{Two-point correlator in the ground state}
We first present the bootstrap bounds on the Euclidean two-point correlators in the ground state of the 1-MQM. Note that even though the potential is unbounded from below for $g<0$, tunneling effects are exponentially suppressed at large $N$, allowing for normalizable ground states as long as $g \geq g_c = -{\sqrt{2}\over6\pi} \approx -0.075026$. In Appendix~\ref{app:1ptMQM}, we show that all ground-state one-point functions can be expressed in terms of $\gev{\tr X^m}$. In turn, these quantities can be obtained analytically by mapping the 1-MQM to non-interacting fermions~\cite{Brezin:1977sv}. Hence, we implement the dual problem that assumes knowledge of the zero-time data \eqref{SDP:MNboot}, in the PMP formulation.

\begin{figure}[h]
    \centering
\includegraphics[width=1\linewidth]{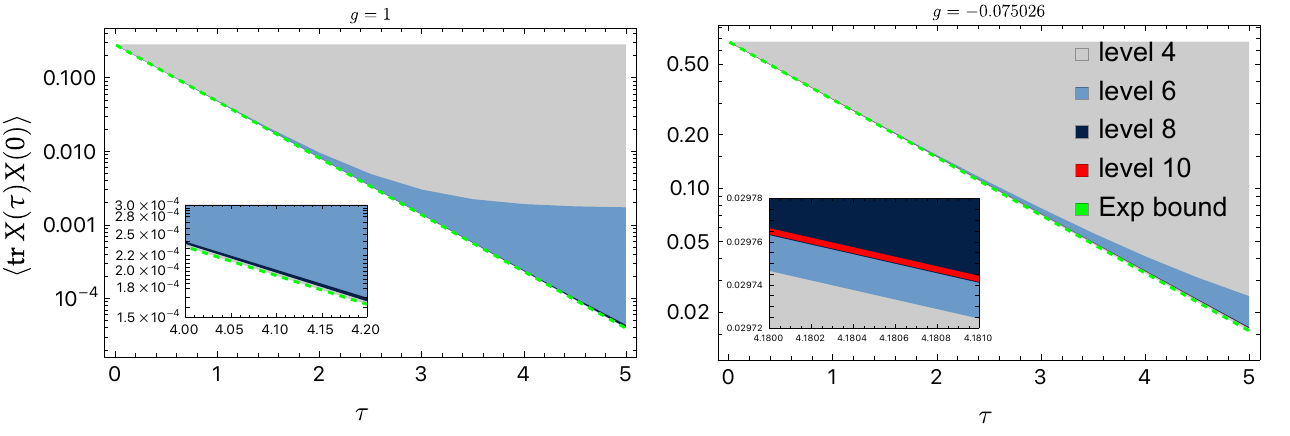}

    \caption{Euclidean two-point correlator $\ev{\tr X(\tau) X(0)}$ in the ground state of the 1-MQM with $V= \frac{1}{2} \Tr X^2+   g \Tr X^4$ at $g=1$ (left) and $g=-0.075026\approx g_c$ (right). The shaded region indicates the allowed bootstrap region at level-$\{4,6,8,10\}$. The green dashed line indicates the universal lower bound in \eqref{universal-lower-bound}. Bounds are obtained from the PMP formulation using \texttt{SDPB} with $d=16$ (Appendix \ref{app:PMP}). The level 8 allowed region is barely visible in the main figures; in the inset panel, we show a zoomed in version where the level 8 (and level 10 in the right figure) regions can be more clearly seen.  }
    \label{fig:euclideanGroundMQM}
\end{figure}

\begin{figure}[h]
    \centering
\includegraphics[width=0.7\linewidth]{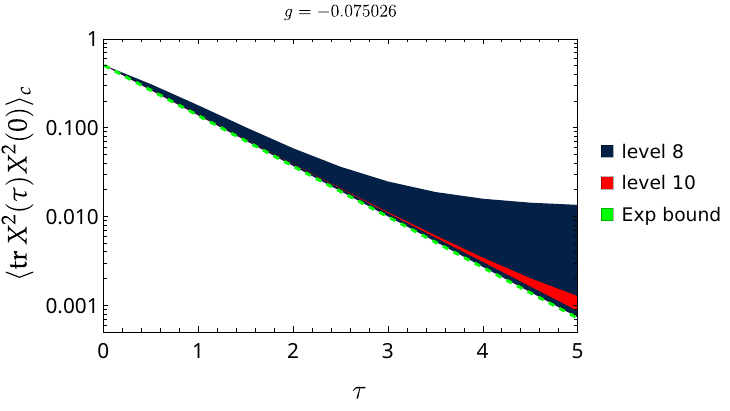}
    \caption{Euclidean two-point correlator $\ev{\tr X^2(\tau) X^2(0)}_c$ in the ground state of the 1-MQM with $V= \frac{1}{2}  X^2+   g X^4$ at $g=-0.075026\approx g_c$. The shaded region indicates the allowed bootstrap region at level 8 and level {\red 10}. The green dashed line indicates the universal lower bound in \eqref{universal-lower-bound}. We bootstrap the two-point correlator using analytically obtained zero-time expectation values as input (see Appendix~\ref{app:1ptMQM}). This plot is obtained with the polynomial formulation of the dual problem and \texttt{SDPB} with $d=16$ and $d=8$ for level 8 and 10, respectively (Appendix \ref{app:PMP}).}
    \label{fig:euclideanGroundMQMX2}
\end{figure}

We show the resulting bootstrap bounds in Figure~\ref{fig:euclideanGroundMQM} for two couplings,\footnote{We also considered the purely quartic potential $V = \tr X^4$ and obtained similar results. For the 
quartic potential, the level 4 bound on the adjoint gap is somewhat worse.} $g=1$ and $g= -0.075026$.  We observe that the bootstrap bounds at a fixed level tighten as $g\rightarrow g_c$. The results again satisfy the universal lower bound in \eqref{universal-lower-bound}. We also show the result for the connected Euclidean two-point correlator $\gev{ \tr X^2 (\tau) X^2 (0)}_c$ in Figure \ref{fig:euclideanGroundMQMX2}. At the same level, bounds on $\gev{ \tr X^2 (\tau) X^2 (0)}_c$ are looser compared to those on $\gev{ \tr X (\tau) X(0)}_c$as expected, since it’s a correlator of higher-level operators.%

\subsection{Bounds on the adjoint gap}\label{gap2}
If we view the $U(N)$ symmetry as a global symmetry, we can discuss the energy gaps for different irreducible representations of $U(N)$. The adjoint gap is the lowest energy in the adjoint representation of $U(N)$ minus the singlet ground state energy. All other non-trivial irreps will have larger gaps.
When the model also has a $\mathbb{Z}_2$ symmetry that commutes with $U(N)$, one may define separate adjoint gaps in the even and odd parity sectors.
In this section, we discuss how to improve the analytic upper bound on the adjoint gap \eqref{eqn:adjointgapuniversal}. Consider an operator $\O_{ab}$ in the adjoint representation of $U(N)$. Its ground-state two-point connected correlator allows for the following spectral decomposition
\begin{align}\label{eqn:spectralsum}
    G_c(\tau)=\ev{\Omega|\bar{\O}_{ba}(\tau)\O_{ba}(0)|\Omega}_c=\sum_n |\bra{\Omega}\bar{\O}_{ba}\ket{n,ab}|^2 e^{-\Delta_n \tau}, \quad \Delta_n = E_n - E_0,
\end{align}
where $|n,ab\rangle$ denotes an orthonormal basis of energy eigenstates with energies $E_n$ in the adjoint sector (with $ab$ labeling the adjoint indices), ordered such that $E_1\leq E_2\leq\cdots$. For notational convenience, we introduce the shorthand $\langle\Omega|\O|n\rangle=\bra{\Omega}\O_{ba}\ket{n,ab}$. Terms in the sum are manifestly positive and subject to $\mathbb{Z}_2$-selection rules. At large $\tau$, terms with the smallest $\Delta_n$ dominate. Then, log-convexity of $G_c(\tau)$ discussed in \eqref{universal-lower-bound} implies
\begin{equation}
    \exp(-\Delta_{\text{gap}}\tau)\geq{G_c(\tau)\over G_c(0)}\geq \exp\left(-\frac{G_c'(0)}{G_c(0)}\tau\right),
\end{equation}
leading to $\Delta_{\text{gap}} \leq -\frac{G_c'(0)}{G_c(0)}$, where $\Delta_{\text{gap}}$ corresponds to $\Delta_1$ or $\Delta_2$ depending on the parity of the operator in $G_c$.

The lowest-level bound for the $\mathbb Z_2$-odd sector is obtained by choosing $\O=X$:\footnote{Similarly for the anharmonic oscillator, we obtain $ \Delta_{1} \leq {1}/({2\ev{x^2}})$.}%
\begin{equation}
    \Delta_{1} \leq \frac{1}{2\ev{\tr X^2}}. \label{bound1odd}
\end{equation}
 For the lowest-level bound in the $\mathbb Z_2$-even sector, we choose $\O=X^2$ and use that the ground state is $U(N)$ invariant $\ev{\Omega|{\O_{ab}|\Omega}} = \delta_{ab} \ev{\Omega|\tr \O|\Omega}$:\footnote{For the anharmonic oscillator, we obtain \begin{align}\label{delta2}
  \Delta_{2} \leq  -\frac{ \ev{ [H,x^2]x^2}}{\ev{x^4}-\ev{x^2}^2} = \frac{2\ev{x^2 }}{\ev{x^4}-\ev{x^2}^2}.
\end{align}}
\begin{align}
    \gev{\tr \O(\tau) \O(0)}_c &= \frac{1}{N} \gev{\O_{ij}(\tau) \O_{ji}(0)} - \frac{1}{N} \gev{\O_{ij}} \gev{\O_{ji}}\nonumber\\
    &= \gev{\tr \O(\tau) \O(0)} - \gev{\tr \O} \gev{\tr \O}.%
\end{align}
Using the constraints including large $N$ factorization\footnote{Notice that \eqref{delta2ad} differs from \eqref{delta2} by a factor of 2. The adjoint gap bound uses large $N$ factorization, so we cannot simply set $N=1$ to recover \eqref{delta2} from \eqref{delta2ad} by setting $N=1$. In particular, we used $\ev{\tr \Pi X^3} = \ev{\tr X^3 \Pi} - 2 \ev{\tr X^2} - \ev{\tr X \tr X}$ and used large $N$ to drop the last term. }, this gives%
\begin{align} \label{delta2ad}
  \Delta_{2} \leq   \frac{\gev{\tr X^2 }}{\gev{\tr X^4}-\gev{\tr X^2}^2}.
\end{align}
More generally, we can improve the bounds by considering a superposition of operators $\mathcal{O} = \alpha_i (\O_i - \gev{\O_i}) $. Then according to \eqref{universal-lower-bound} we have %
\begin{align}
G_c(\tau) \ge G_c(0) e^{-\mu \tau}, \quad \mu = -\frac{ \bar\alpha_{i}\widetilde{\N}_{ij}(0) \alpha_j  }{\bar\alpha_{i} \widetilde\M_{ij}(0) \alpha_j }  %
\end{align}
Here, $\widetilde{\M}$ and $\widetilde{\N}$ are the connected correlator versions of ${\M}$ and ${\N}$. This implies that we can improve the bound on the gap by considering the more general bound:
\begin{align}
   \Delta_\text{gap} \le  \mu = \frac{ \bar\alpha_{i}\widetilde\N_{ij}(0) \alpha_j  }{\bar\alpha_{i} \widetilde\M_{ij}(0) \alpha_j }  
\end{align}
We can rewrite this inequality as 
\begin{align}
   \Delta_\text{gap}  \bar\alpha_{i} \widetilde\M_{ij}(0) \alpha_j      \le  { \bar\alpha_{i}\widetilde{\N}_{ij}(0) \alpha_j  }, \quad  \forall \alpha  %
\end{align}
Since this inequality is true for all choices of $\alpha$, this is equivalent to finding the largest value of $\Delta$ such that 
\begin{align} \label{bootstrapGapcondition}
       \widetilde\N(0) \succeq \Delta \widetilde\M(0).
\end{align}
This formulation of the gap problem was first derived in \cite{ZhengInPrep} without using the time-dependent bootstrap formalism. Since we are maximizing over $\Delta$, the resulting estimate is an upper bound on the gap.

\begin{figure}[H]
    \centering
    \includegraphics[width=0.46\linewidth]{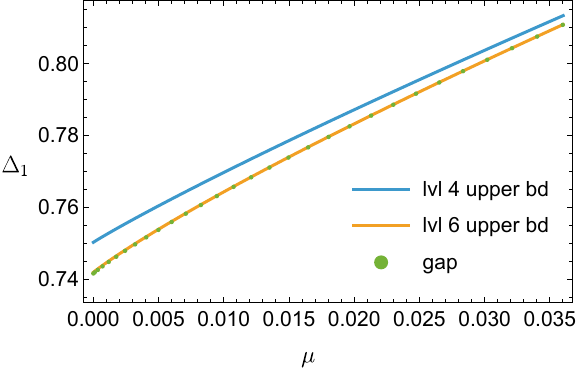} \hspace{0.5cm}
        \includegraphics[width=0.46\linewidth]{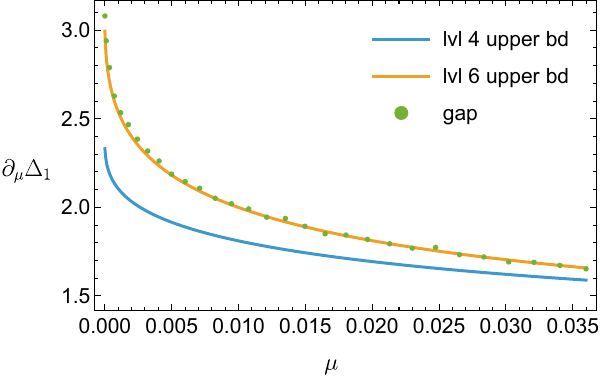}
    \caption{Adjoint gap near criticality. On the right, we see clear evidence of the non-analytic behavior of the gap as $\mu \to 0$. The green dots are obtained by numerically solving the Marchesini-Onofri equation \cite{Marchesini:1979yq}, see Appendix \ref{app: analytic}. We checked that the bootstrap results at level 8 agree well with these green dots, see Table \ref{tab:adj-spectrum}. }
    \label{fig:MQM_adjgapCritical}
\end{figure}

For the level 6 bound on the $\mathbb{Z}_2$-odd adjoint gap in MQM, we choose $\O = \alpha_X X + \alpha_\Pi \Pi$. %
Then,~\eqref{bootstrapGapcondition} gives:
\begin{align}\label{gapEq2}
   \begin{pmatrix}
        \hf  & -\ev{\tr \Pi^2} \\
        -\ev{\tr \Pi^2} & -\ev{\tr \Pi V'(X)}
    \end{pmatrix}\succeq  \Delta_\text{gap} \begin{pmatrix}
        \ev{\tr X^2} & \hf  \\
        \hf  & -\ev{\tr \Pi^2}\ .
    \end{pmatrix} 
\end{align}
Combining this with the virial theorem, $ \ev{\tr \Pi^2} = -\ev{\tr X^2} - 4 g \ev{\tr X^4}$ and the identity $\ev{\tr \Pi X^3} = -\ev{\tr X^2}$, we obtain%
\begin{align}
   \begin{pmatrix}
        \hf  & t_2 +4gt_4  \\
        t_2 +4gt_4  & \hf + 4g t_2
    \end{pmatrix}- \Delta_\text{gap} \begin{pmatrix}
      t_2 & \hf  \\
        \hf  &  t_2 +4gt_4
    \end{pmatrix} \succeq 0 
\end{align}
where we introduced the shorthand $t_k = \gev{\tr X^k}$. This implies the bound
\begin{align} \label{bound2odd}
\Delta_\text{gap} \leq \frac{-2 \sqrt{\frac{1}{4} \left(64 g^2 t_4^2+8 g t_2 (4 t_4-1)+4 t_2^2-1\right) (4 t_2 (4 g t_4+t_2)-1)+\left(4 g t_2^2-2 g t_4\right)^2}+8 g t_2^2-4 g t_4}{4 t_2 (4 g t_4+t_2)-1}
\end{align}
We can compute $t_2(g)$ and $t_4(g)$ analytically, see Appendix \ref{app: analytic}. We show the resulting bounds for the gap in Figure \ref{fig:MQM_adjgapCritical}. Here we parameterize the coupling by $\mu(g)$ which is the difference in the Fermi level from the top of the potential $\mu(g) = \mu_c - \mu_F(g)$, see \eqref{muf}. This parameterization makes the universal properties of the system more apparent.

\begin{figure}[H]
    \centering
    \includegraphics[width=0.46\linewidth]{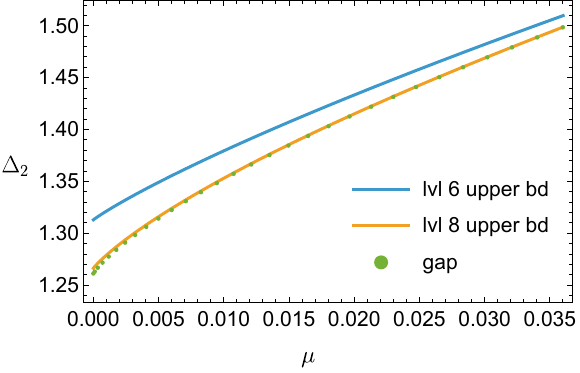} \hspace{0.5cm}
        \includegraphics[width=0.46\linewidth]{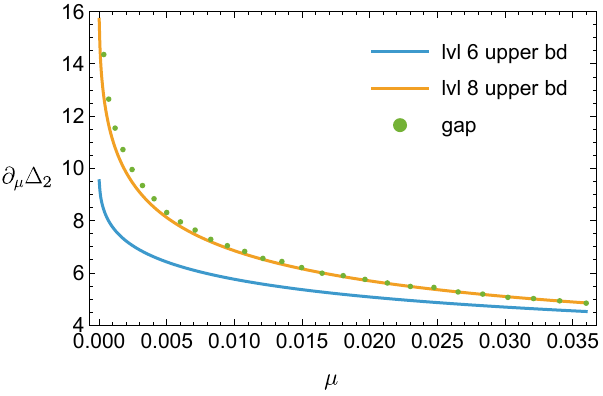}
    \caption{$\mathbb{Z}_2$-even adjoint gap near criticality. As with the $\mathbb{Z}_2$-odd adjoint gap, we see clear evidence of the non-analytic behavior as $\mu \to 0$. We observe that the behavior of the level 6 and 8 bounds on the $\mathbb{Z}_2$-even gap are qualitatively similar to the level 4 and 6 bounds on the $\mathbb{Z}_2$-odd gap. }
    \label{fig:MQM_adjgapCriticalEven}
\end{figure}

\begin{figure}[H]
    \centering
    \includegraphics[width=0.46\linewidth]{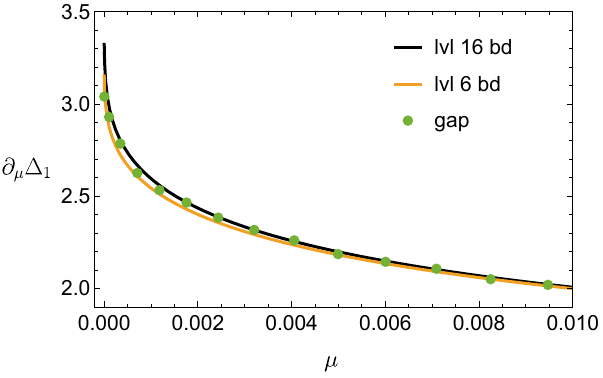} \hspace{0.5cm}
        \includegraphics[width=0.46\linewidth]{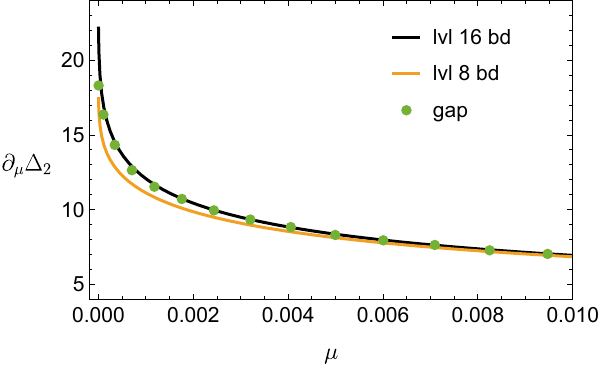}
    \caption{Derivative of the adjoint gap near criticality. Here we show the derivative of the level 16 bound (level 17 for $\N$), which we believe is more precise than our numerical solution to the Marchesini-Onofri equation. On the left, we show the $\mathbb{Z}_2$-odd gap, whereas on the right we show the $\mathbb{Z}_2$-even result. The derivative appears to diverge as $\mu \to 0$.}
    \label{fig:MQM_adjgapCritical2}
\end{figure}

For the level 8 bound on the $\mathbb{Z}_2$-even adjoint gap, we choose $\O = \alpha_1 X^2 + \alpha_2 X\Pi$.
Then \eqref{bootstrapGapcondition} gives:
\begin{align}
   &\begin{pmatrix}
        \ev{\tr X^3 \Pi} + \ev{\tr X^2 \Pi X} & \ev{\tr X^2 \Pi^2} + \ev{\tr X^4}+4g \ev{\tr X^6} \\
         \ev{\tr X^2 \Pi^2} + \ev{\tr X^4}+4g \ev{\tr X^6} & - \ev{\tr \Pi X \Pi^2} - \ev{\tr \Pi X^3}-4g \ev{\tr \Pi X^5}
    \end{pmatrix}\nonumber\\
    &\succeq  \Delta_\text{gap} \begin{pmatrix}
        \ev{\tr X^4} - \ev{\tr X^2}^2 & \ev{\tr X^3 \Pi} - \ev{\tr X^2} \ev{\tr X\Pi} \\
         \ev{\tr X^3 \Pi}  - \ev{\tr X^2} \ev{\tr X\Pi} & - \ev{\tr \Pi X^2 \Pi} +  \ev{\tr \Pi X} \ev{\tr X\Pi}
    \end{pmatrix} 
\end{align}
In our conventions, the matrix $\N$ includes the inner product of the identity operator, whereas the matrix $\widetilde\N$ does not include the identity. Similarly, $\widetilde\M_{ij} = \M_{ij} - \M_{1,\mathbf{1}} \M_{j,\mathbf{1}} $.
Using the relations $\ev{\tr \Pi X^3} = -\ev{\tr X^2}$, $\ev{\tr X^2 \Pi ^2} = -\frac{1}{3} t_4 -\frac{4}{3}g t_6+ \frac{1}{6}$, $\ev{\tr \Pi X^5} = t_4+\hf t_2^2$, $\ev{\tr \Pi X \Pi^2}=0$, and $\ev{\tr \Pi X^2 \Pi}= -\frac{1}{3} t_4 -\frac{4}{3}g t_6- \frac{1}{3}$,
\begin{align}
   \begin{pmatrix}
        t_2  & \frac{1}{6}-\frac{t_4}{3}-\frac{4}{3} g t_6 +t_4+4g t_6\\
        \frac{1}{6}-\frac{t_4}{3}-\frac{4}{3} g t_6 +t_4+4g t_6  & t_2 +4g (t_4+\hf t_2^2)
    \end{pmatrix}- \Delta_\text{gap} \begin{pmatrix}
      t_4 -t_2^2& \hf t_2 \\
        \hf t_2  & \frac{1}{3}+\frac{t_4}{3}+\frac{4}{3} g t_6 - \qrt
    \end{pmatrix} \succeq 0 
\end{align}
We arrive at the bound
\begin{align}\label{bound2even}
\Delta_{\mathrm{gap}}
\le
\frac{3}{24 t_2^2 S_1 - 6 t_4 S_2}\left[
 T
+
{12 \sqrt{\,R^2 + \tfrac{1}{108} Q U\,}}
\right]
\end{align}
where
\begin{align}
&S_1 = 1 + t_4 + 4 g t_6, \qquad 
S_2 = 1 + 4 t_4 + 16 g t_6, \\[4pt]
& T = t_2 (1 + 12 t_2^2 + 24 g t_2^3 - 8 t_4
+ 24 g t_2 t_4+ 16 g t_6) - 48 g t_4^2 \\ %
&R = t_2^3 + 2 g t_2^4 + 2 g t_2^2 t_4 - 4 g t_4^2 
+ \tfrac{1}{12} t_2 (1 - 8 t_4 + 16 g t_6), \\[4pt]
&Q = 4 t_2^2 S_1 - t_4 S_2, \qquad 
U = 36 t_2^2 + 72 g t_2^3 + 144 g t_2 t_4 - S_2^2. %
\end{align}

With the help of some analytic results from Appendix \ref{app: analytic} and \ref{app:1ptMQM}, we can also go to much higher level numerically using a standard SDP solver. We simply try to maximize $\Delta$ in \eqref{bootstrapGapcondition}. Alternatively, since $\N$ and $\M$ are known matrices, we can use \eqref{rewrite}, to be discussed below.

\begin{figure}[H]
    \centering
    \includegraphics[width=0.46\linewidth]{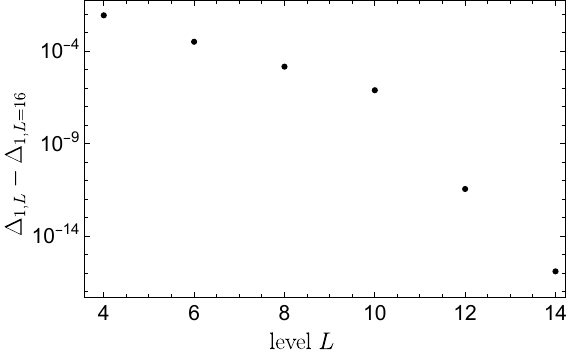} \hspace{0.5cm}
        \includegraphics[width=0.46\linewidth]{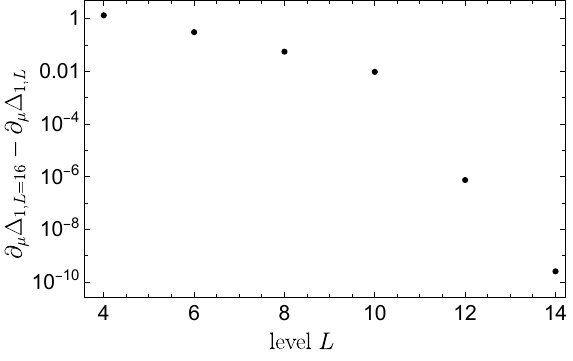}
    \caption{Convergence of the estimates for the gap $\Delta_1$ and the gap derivative $\partial_\mu \Delta$ as function of level $L$ of the bounds. {\it Left}: the convergence of the gap $\Delta_1$ evaluated at $g=g_c$. {\it Right}: convergence of the derivative $\partial_\mu \Delta$ near the critical point $g-g_c = \hf \delta$ where $\delta = 10^{-150}$. This was achieved using arbitrary precision arithmetic. }
    \label{fig:MQM_adjgapConvergence}
\end{figure}

We see clear evidence that the gap is non-analytic as $g \to g_c$ or $\mu \to 0$. To estimate the non-analytic behavior, one can consider the simple bound $\Delta_1 \le 1 /(2 t_2)$. It can be seen that $t_2 \sim \mu^2 \log \mu + \text{analytic}$, where the analytic terms includes a constant as $\mu \to 0$.
Then $ \frac{1}{2 t_2} \sim  C_2 \mu^2 \log \mu + \text{analytic}.$ Of course, the level 4 bound is not saturated by the true gap; nevertheless the higher level bounds rapidly converge at large $L$, see Figure \ref{fig:MQM_adjgapConvergence}. 
We thus conjecture that the adjoint gap $\Delta_\text{gap}$ has a non-analytic behavior $\propto C \mu^k \log \mu$ where $ 1 \le k \le 2$.  An interesting problem is to compute $C$ and $k$, either directly from the matrix model or by understanding the string dual.  

\subsubsection{Extremal functional method \label{sec: extremal}} 
The above results give completely rigorous upper bounds on the gap. We observe that these bounds rapidly converge, suggesting that the ``true'' gap is close to saturating the upper bound at sufficiently large level. This assumption is strongly supported by a different approach to solving the gap numerically, using the Marchesini-Onofri equation \cite{Marchesini:1979yq, Casartelli:1979iz} discussed in Appendix \ref{app: analytic}.
Given the rapid convergence, it is also interesting to consider the ``extremal eigenvector'' that saturates the bootstrap bound. This is a null eigenvector of the matrix $\widetilde\N - \Delta_c \widetilde\M$, where $\Delta_c$ is the optimal value of $\Delta$ such that $\widetilde\N - \Delta_c \widetilde\M$ is positive semi-definite.

In our simple bootstrap problem, $\N$ and $\M$ are known matrices (whereas in more general bootstrap problems one would have to consider the extremal values of $\N$ and $\M$), so we can consider
the eigenvalue equation:
\begin{align} \label{rewrite}
    \M^{-1/2} \tilde\N \M^{-1/2} \vec{v}_a  = \Delta_a \vec{v}_a
\end{align}
We have chosen to conjugate $\N$ by $\M^{-1/2}$ so that the LHS is a Hermitian operator to ensure that the eigenvectors are orthogonal. (To obtain precise estimates, we found it necessary to use arbitrary precision arithmetic as $\M$ contains many eigenvectors with numerically small eigenvalues.)
We expect the eigenvalues $\{ \Delta_a\} $ to give the energies above the ground state (in the particular symmetry sector of interest). Although this is not rigorous, we expect that the eigenvectors $\vec{v}$ to be reasonable approximations to the actual adjoint eigenvectors (and we expect that the estimates should converge when we take the max level of the $\M$ and $\N$ matrices to be large). For example, we may compute 
\begin{align}
    |\bra{\Omega} \O \ket{n}|^2 \approx |\vec{e}_\O  \M^{1/2}  \vec{v}_a|^2
\end{align}
where $\vec{e}_\O$ is the vector representing the adjoint state $\bra{\Omega} \O_{ab}$. In Table \ref{tab:adj-spectrum}, we show the results for the first and second $\mathbb{Z}_2$-odd excited states $n=1$, $n=3$ for $\O = X$ and $\O = X^3$. For the first $\mathbb{Z}_2$-even state $n=2$, we show the result for $\O = X^2$ and $\O =X^4$. In all cases, we see reasonable agreement with the Marchesini-Onofri equation. 

The 1-MQM is sufficiently simple that one can compute $\M$ and $\N$ analytically; however, for more general quantum systems one could still use an extreme value of $\M$ and $\N$, e.g., if there is a ``kink'' in the allowed bootstrap region. It would be interesting to explore this more generally. Note that in the large $N$ context, this is closely related to a variational method where the wavefunction is parameterized in terms of large $N$ ``words.'' This is more efficient than a traditional variational problem where one would try to give an explicit description of the wavefunction in terms of an exponentially large Hilbert space.

\subsection{Extracting properties of the adjoint sector}\label{sec:fitting}
With the dual formulation, we have obtained bootstrap bounds on the Euclidean two-point correlator in the MQM ground state. With this data in hand, we can extract some physical properties of the system in the adjoint sector by comparing against the spectral decomposition \eqref{eqn:spectralsum}. The bootstrap bounds on the {adjoint gap} from \S \ref{gap2}\footnote{Here we are contrasting the bootstrap bounds from \S \ref{gap2} that did not explicitly involve time separation with the results on the time-dependent correlator.} provide strong evidence that the adjoint energy levels and {matrix elements} can be extracted from the 2-pt correlator bootstrap. Note that, in principle, the ground state two-point correlator contains much more information.%

In this subsection, we extract the first few adjoint energy levels $\Delta_n$ and matrix elements by fitting to the late-time behavior of our bootstrap bounds for the Euclidean two-point correlator $\gev{ \tr X(\tau) X(0)}$ or $\gev{\tr X^2(\tau)X^2(0)}$. %
Then, we compare the obtained values against those acquired independently by solving the Marchesini-Onofri (MO) equation \cite{Marchesini:1979yq} (see Appendix \ref{app:longstring}).
In the following, we discuss the various fitting strategies, with the corresponding estimates presented in Table~\ref{tab:adj-spectrum}. %
All fits were obtained using the level-8 bootstrap lower bounds from Figures~\ref{fig:euclideanGroundMQM} and~\ref{fig:euclideanGroundMQMX2}, restricted to the interval $\tau \in [2.5, 5]$. The lower bounds were chosen because they appear to converge faster. Indeed, they lie closer to the exact values in the anharmonic oscillator case (see, e.g., Figure~\ref{fig:euclideanGroundFig}).

For the $\mathbb{Z}_2$-odd sector, let $G(\tau)=\gev{ \tr X(\tau) X(0)}$. To obtain a leading order estimate of $\Delta_1$, we approximate $G(\tau)$ at late times by,
\begin{equation}
    \widetilde{G}(\tau) = a_1 e^{-\Delta_1 \tau} \ .
\end{equation}
Here $a_n = |\langle\Omega|X|n\rangle|^2$. Then, we either eliminate one unknown by imposing the zero-time condition,
\begin{equation}
    a_1 = \widetilde{G}(0) = G(0)=\gev{\tr X^2},
\end{equation}
and perform a one-parameter fit to determine $\Delta_1$, or directly perform a two-parameter fit for both $a_1$ and $\Delta_1$. In either case, we retain only the value of $\Delta_1$ from these fits and they account for the results for $\Delta_1$ in Table \ref{tab:adj-spectrum}. %

\begin{figure}
    \centering
    \includegraphics[width=1\linewidth]{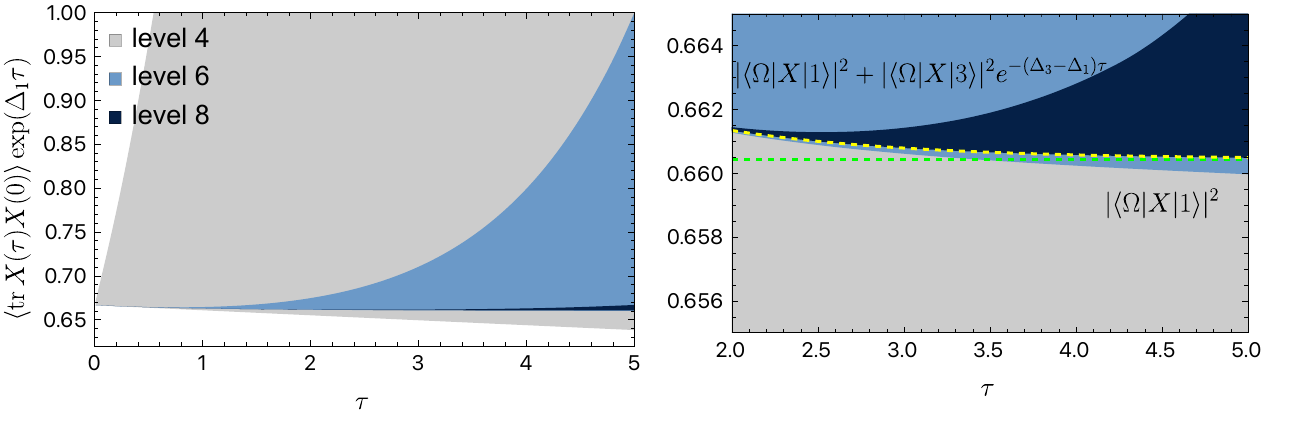}
    \caption{Euclidean two-point correlator in the ground state of the 1-MQM at $g=-0.075026$, close to the critical coupling $g_c$, multiplied by the adjoint gap factor $e^{\Delta_{1}\tau}$ obtained from the Marchesini-Onofri equation. This quantity approaches the first matrix element $|\langle\Omega|X|1\rangle|^2$ as $\tau$ increases (green dashed line). The correction from the next subleading exponential at smaller $\tau$ (yellow dashed line). Both dashed lines were obtained by fitting  $|\langle\Omega|X|1\rangle|^2 ,|\langle\Omega|X|3\rangle|^2$, and $\Delta_3$ to the lower bound while assuming the value of $\Delta_1$. See the main text for details about the fit.}
    \label{fig:MQM_adjgap}
\end{figure}

To extract more information, we include the next energy level in the ansatz,
\begin{equation}
    \widetilde{G}(\tau) = a_1 e^{-\Delta_1 \tau} +  a_3 e^{-\Delta_3 \tau} \ .   
\end{equation}
By incorporating the zero-time data $G(0) = a_1 + a_3$ and potentially $G'(0) = \langle \mathrm{tr} PX   \rangle=  - \Delta_1 a_1 - \Delta_3 a_3$, we can reduce the number of unknowns to two or one. We then perform a one- or two-parameter fitting respectively. These fits account for the results for $a_1$, $a_3$ and $\Delta_3$ in Table \ref{tab:adj-spectrum}.
The results of the 2-parameter fit are also plotted in the RHS of Figure \ref{fig:MQM_adjgap}.

As an alternative approach, we also attempted to fit $a_3$ and $\Delta_3$ without using the zero-time input, instead incorporating the estimates of $a_1$ and $\Delta_1$ obtained from the extremal functional method (see section \ref{sec: extremal}). We found that the resulting estimates ($a_3 = 0.0075539$ and $\Delta_3 = 1.80309$) deviate further from the expected values.

For the $\mathbb{Z}_2$-even sector, a similar procedure can be applied. Let, 
\begin{equation}
    {G}(\tau)=\gev{ \tr X^2(\tau) X^2(0)},    
\end{equation}
and take the ansatz,
\begin{equation}
    \widetilde{G}(\tau)=\gev{\tr X^2}^2 + a_2 e^{-\Delta_2 \tau},    
\end{equation}
which we expect to hold at large~$\tau$. As %
an additional approximation, we again impose the zero-time condition by setting $a_2 = \gev{ \tr X^4} - \gev{\tr X^2}^2$ and subsequently perform a one- or two-parameter fit to extract $a_2$ and $\Delta_2$.

We observe that the estimates for $\Delta_1$ and $|\langle\Omega|X|1\rangle|^2$ are in excellent agreement with the values obtained from the Marchesini–Onofri equation and the level-16 estimates discussed in Section~\ref{gap2}. In contrast, the estimates for the $\mathbb{Z}_2$-even sector deviate by a few percent from the Marchesini–Onofri predictions. Similarly, the estimates for $\Delta_3$ and $a_3$ are less robust with the current level-10 data. We observe a $\sim 10\%$ change in the extracted values while varying the fitting strategy. We expect these estimates to improve at higher levels and/or by considering correlators of the form $\langle \mathrm{tr}(\bar{\mathcal{O}}(\tau)\mathcal{O}(0))\rangle$ with more general operators, such as $\mathcal{O} = \alpha_{X^2}X^2 + \alpha_{XP}XP + \alpha_P P^2 + \cdots $, {in a similar fashion to the discussion in \S \ref{gap2}.}

\begin{table}[]
    \centering
    \begin{tabular}{c|c|c|c|c|c|c}
        &\text{1-param fit} & \text{2-param fit} & \text{MO} & \text{bound (1)} & \text{bound (2)} & \text{bound (3)} \\
        \hline
        $\Delta_1$ &0.74436& 0.74199 & 0.74158 & 0.75027 & 0.74190 & 0.741573662448591\\
        \hline
        $\Delta_2$ &1.28163& 1.26660 & 1.26130 & 1.31299 &1.26636 & 1.26122 \\
        \hline
        $\Delta_3$ &1.71370 & 1.69548  &1.68218 &  --- &  --- & 1.68192** \\
    \end{tabular}\\
    \vspace{2cm}
        \begin{tabular}{c|c|c|c|c}
         &  \text{1-param fit}  &\text{2-param fit} & \text{MO} & \text{extremal fct} \\
        \hline
       $|\langle\Omega|X|1\rangle|^2$ &0.660469& 0.660449 & 0.660453 &  0.660455 \\ \hline
       $|\langle\Omega|X^3|1\rangle|^2$ & --- & ---  & 1.48535 &  1.48543 \\ \hline
        $|\langle\Omega|X^2|2\rangle|^2$ &0.50756 & 0.487051 & 0.476621 & 0.476603\\ \hline
         $|\langle\Omega|X^4|2\rangle|^2$ & --- & ---  & 2.77206 &  2.77227 \\ \hline
         $|\langle\Omega|X|3\rangle|^2$ &0.00595684& 0.00597743 &0.00574156 & 0.00574315 \\ \hline
    \end{tabular}
    \caption{Bootstrap estimates for the low-lying spectrum and corresponding matrix elements of the 1-MQM model at $g=-0.075026$. The first two columns show estimates from fitting a simple functional form to the bootstrap lower bounds on $G_c(\tau)$ at level 10, as explained in Section \ref{sec:fitting}.
    We compare the estimates to the results from solving the Marchesini-Onofri (MO) equation in the third column (see Appendix \ref{app:longstring}). The upper bound (1) is given by the analytic expression \eqref{bound1odd} and \eqref{delta2ad}, and (2) is given by \eqref{bound2odd} and \eqref{bound2even}. The upper bound (3) is obtained by numerically solving for $\Delta$ at level 16. The stars$^{**}$ indicate that (unlike the $\Delta_1, \Delta_2$ bounds) the $\Delta_3$ estimate is not a rigorous upper bound, see  subsection \ref{sec: extremal}. }
    \label{tab:adj-spectrum}
\end{table}

\subsection{Thermal two-point correlator}

 \begin{figure}[H]
     \centering
 \includegraphics[width=\linewidth]%
 {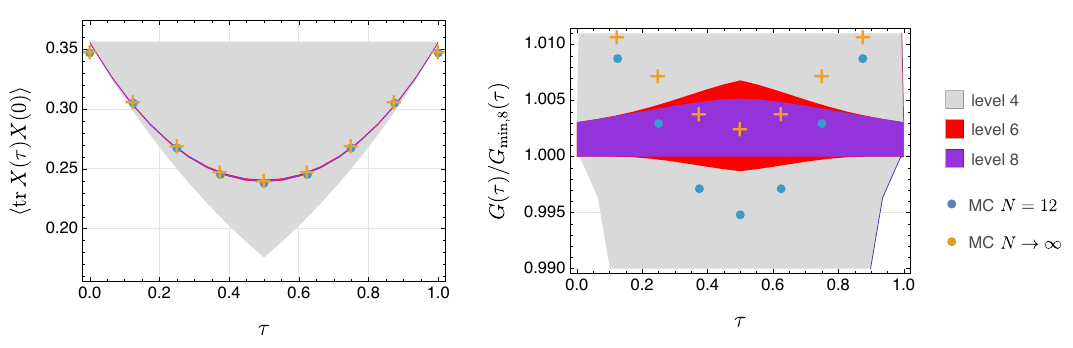}
     \caption{Bootstrap bounds on the thermal two-point correlator $\langle\operatorname{tr}X(\tau)X(0)\rangle$ for the ungauged 1-MQM with $\beta=1$ and $g=1$, at levels 4, {\red 6}, and \textcolor{purple}{8}. For the zero-time conditions, we use inequalities obtained from the level 14 bootstrap of \cite{Cho:2024kxn} based on the EEB inequalities. We also compare to Monte Carlo (MC) data, see Appendix \ref{app:montecarlo} for details about the Monte Carlo simulation. The uncertainty in the MC results is completely dominated by systematic errors associated with the large $L$ (number of lattice sites) and large $N$ extrapolation. For clarity, we also plot on the right the ratio of $G(\tau)=\langle\operatorname{tr}X(\tau)X(0)\rangle$ obtained from bootstrap and MC to the level 8 bootstrap lower bound $G_{\text{min,8}}(\tau)$.}
     \label{fig:thermal_2pt}
 \end{figure}

\begin{figure}[H]
    \centering
\includegraphics[width=.7\linewidth]{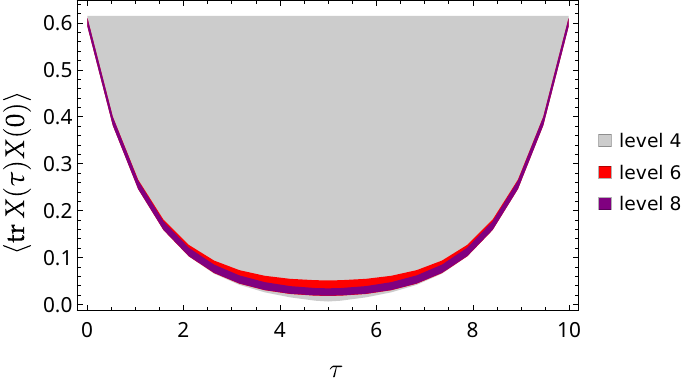}
    \caption{Bootstrap bounds on the thermal two-point correlator $\langle\operatorname{tr}X(\tau)X(0)\rangle$ for the ungauged 1-MQM with $\beta=10$ and $g=-0.06$, at levels 4, {\red 6}, and \textcolor{purple}{8}. For the zero-time conditions, we use inequalities obtained from the level 14 bootstrap of \cite{Cho:2024kxn} based on the EEB inequalities.}
    \label{fig:thermalMQM-negativeg}
\end{figure}
Bootstrapping Euclidean two-point correlators at nonzero temperatures for matrix quantum mechanical systems, {without assuming knowledge of the zero-time correlators}, involves a technical subtlety due to large $N$ factorization. The zero-time matrix ${\cal M}(0)$ has matrix elements that generally involve quadratic terms in the expectation values once large $N$ factorization is imposed. This arises from the cyclicity of the trace, as can be seen from the following example:
\begin{equation}
    \tr P^2X P^3 = \tr P^5 X+\i \tr P^4+ \i \left(\tr P^2\right)^2.
\end{equation}
Therefore, the matrix elements of ${\cal M}(0)$ contain quadratic expressions such as $\langle\tr P^2\rangle^2$, which render the bootstrap problem non-convex. One may explicitly scan over a sufficient number of single-trace expectation values to restore convexity, as illustrated in \cite{LinZheng1,Lin:2025srf}, or instead employ a convex relaxation of the non-convex problem \cite{Kazakov:2021lel}. The former approach, which scans over possible values of a few single-trace expectation values, leads to a feasibility problem for the bootstrap constraints. This is discussed in Appendix \ref{app:noInitThermal}, where the final dual formulation takes the form of (\ref{eqn:noinitthermalInfeas}).

In this work, we adopt a variant of the approach in (\ref{SDP:dualThermal}). In (\ref{SDP:dualThermal}), we assumed knowledge of the exact values of ${\cal M}(0)$. Instead, we consider the case where only rigorous \textit{bounds} on the matrix elements of ${\cal M}(0)$ are known, taking the form
\begin{equation}\label{eq:boundsOnM0}
l^{(p)}\leq\operatorname{Tr}J^{(p)}{\cal M}(0)\leq u^{(p)},
\end{equation}
for suitable indices $p$, matrices $J^{(p)}$, and lower and upper bounds $l^{(p)}$ and $u^{(p)}$, respectively. Such inequality constraints are manifestly linear and hence convex. They can be obtained using the EEB inequalities (\ref{eqn:EEBineq}) as bootstrap constraints, as studied in \cite{Cho:2024kxn} for the case of ungauged matrix quantum mechanical systems. 
In \cite{Cho:2024kxn}, the non-convexity arising from large $N$ factorization was handled via a combination of the inclusion of factorized double-trace equations of motion,
\begin{equation}
    \braket{\tr(\mathcal{O}_1)[H,\tr(\mathcal{O}_2)]} = \braket{\tr(\mathcal{O}_1)}\braket{[H,\tr(\mathcal{O}_2)]} = 0,
\end{equation}
and a crude (but still rigorous) convex relaxation of the remaining variables, by declaring that expressions quadratic in expectation values are treated as new independent variables—for example, by defining $\langle\tr P^2\rangle^2$ as a new bootstrap variable $w_{P^2}$. Even such a simple relaxation produced rigorous and precise bounds on ${\cal M}(0)$ in \cite{Cho:2024kxn}.

Specifically, we use bounds on ${\cal M}(0)$ obtained from the $(m,k)=(3,3)$ bootstrap defined in \cite{Cho:2024kxn} at level 14.\footnote{See sections 3.1 and 3.2 of \cite{Cho:2024kxn} for notation and further details. Instead of their $L$ which stands for the maximal word length, we used the level defined in this work for the finite truncation of the bootstrap problem.} The practical implementation of the thermal Euclidean two-point bootstrap using these bounds on ${\cal M}(0)$ is summarized in Appendix \ref{app:initialIneqThermal}. We present the corresponding bounds on $\langle\tr X(\tau)X(0)\rangle$, together with Monte Carlo (MC) results discussed in Appendix \ref{app:montecarlo}. More detailed comparisons presented in Appendix \ref{app:montecarlo} show that bootstrap bounds are more precise than uncertainties of MC results due to large lattice number and large $N$ extrapolations (see Figure \ref{fig:monteCarloNextrap}).

We also present similar bounds for $g=-0.06$ in Figure \ref{fig:thermalMQM-negativeg}. It should be noted that, unlike the $g>0$ case where a thermal state exists at any temperature, for $g_c\leq g<0$ thermal states exist only up to a critical temperature $T_*(g)$, since the potential is unbounded. In \cite{Cho:2024kxn}, upper bounds on $T_*(g)$ were obtained by bootstrapping the EEB inequalities. We chose $T=0.1$ at $g=-0.06$ in Figure \ref{fig:thermalMQM-negativeg}, which lies well below the upper bound $T_*(g=-0.06)=0.31$ reported in \cite{Cho:2024kxn}. We remark that MC simulations are not applicable for $g<0$ because nonzero tunneling effects at any finite $N$ render the system unstable. In this regard, the results presented in Figure \ref{fig:thermalMQM-negativeg} demonstrate the power of the bootstrap formulation in addressing physics defined intrinsically at large $N$, which is otherwise difficult to study.

\section{Discussion \label{Discussion}}
In this work, we have demonstrated that the bootstrap provides a powerful framework for studying Euclidean two-point correlators. There are several interesting directions for further studies, which we discuss below.

For thermal two-point correlators, the KMS condition imposes boundary conditions at $\tau = 0$ and $\tau = \beta$. Given the analyticity of the Heisenberg equations of motion in $\tau$, it is possible to analytically continue the correlator beyond $\tau = \beta$. In principle, the bootstrap formulation introduced in this work is capable of performing such an analytic continuation, since it effectively time-evolves ``inequalities of motion," which are themselves analytic. In practice, the thermal two-point correlator is expected to grow as $\tau$ grows beyond $\beta$, implying that the basis of dual Lagrange multipliers should be simultaneously enlarged to obtain meaningful bootstrap bounds. It would be particularly interesting to observe the presence of poles/cuts at $\tau=\beta_0>\beta$, which are expected due to thermalization and chaos \cite{Dodelson:2025rng, Lin:2023trc}.\footnote{We thank Matthew Dodelson for explaining to us the interpretation of poles in Euclidean time beyond $\beta$.}

Relatedly, it would be interesting to explore the Lorentzian (or, more generally, complex-time) bootstrap for two-point correlators. This would be a generalization of \cite{Lawrence:2024mnj}. The Lorentzian-time thermal two-point correlator could potentially be used to extract the quasi-normal modes (QNMs), for example by fitting the late-time behavior of the correlator. In the BFSS context, there exists a prediction for the QNM at low temperatures in the ’t~Hooft regime, derived from the Type~IIA black hole~\cite{Biggs:2023sqw}, although in general little is known. We envision that the bootstrap will serve as a boundary-based technique for computing these QNMs in the planar limit (see also~\cite{Dodelson:2024atp, Dodelson:2025rng}).

Further future directions include $n$-point or out-of-time-ordered correlators and retarded Green's function. The latter, in particular, possesses well-known analytic properties and manifest positivity, which may provide additional input for the bootstrap.\footnote{See \cite{Chowdhury:2025dlx} for a relevant discussion in the context of transport.} In general, our bootstrap approach extends straightforwardly to generic multi-point correlators, provided they possess some notion of positivity, from which additional physical information can be extracted.

Our main goal in this work was to set up the formalism for the two-point correlator bootstrap and to calibrate it on the simplest large $N$ quantum system, the 1-matrix model. To this end, we used very modest computational resources; we believe it should be possible to scale this computation up to obtain bounds at least up to level $L \sim 14$, and to study more complicated matrix models (see e.g. \cite{Lin:2025srf}) where the quantum system is strongly believed to be chaotic at large $N$.

A curiosity that arose in bootstrapping the adjoint gap is that it remains finite as $g \to g_c$, e.g., as $\mu \to 0$ the gap does not diverge. Indeed, just from the level 4 analytic bound on the adjoint gap
$ \Delta_\text{gap} \le \frac{1}{2 \ev{\tr X^2}} $
implies that the gap remains finite, since $\ev{\tr X^2}$ is finite in the double scaling limit. This is naively in tension with \cite{Gross:1990md}  if one interprets their equations to mean that the adjoint gap diverges like $\sim \log \mu$; the resolution\footnote{We thank Xi Yin and Igor Klebanov for useful conversations about the adjoint gap.} is that the typical adjoint state has a logarithmic energy divergence (that is accurately captured by the WKB analysis of \cite{Gross:1990md}) but there are some relatively rare states with energies that are much lower. The existence of rare  low energy states (``short'' long strings) does not contradict the long string picture of \cite{Maldacena:2005hi}; see 3.20 in \cite{Maldacena:2005hi} for the adjoint density of states, which is highly suppressed at large negative renormalized energies.
It would be interesting to understand whether the leading non-analyticity of this gap as $\mu \to 0$ has an interpretation in the $c=1$ string theory.

On the mathematical side of our bootstrap formulation, there are important questions regarding the convergence of bootstrap bounds as the level $L$ increases. Given the rapid numerical convergence observed in this work, it is plausible that one can prove the asymptotic convergence of the bootstrap bounds to the physical values, in the spirit of the Lasserre hierarchy in polynomial optimization \cite{10.1007/3-540-45535-3_23, doi:10.1137/S1052623400366802}. In particular, it is interesting to ask whether reflection positivity—rather than the positivity of density matrices—is sufficient to guarantee such convergence. For the case of bootstrapping the statistical Ising model on the lattice, it has been shown that positivity of the Gibbs state is necessary for convergence, and that reflection positivity alone does not suffice, at least in one dimension \cite{Cho:2023ulr}. Another related problem is to establish the \textit{strong} duality of the primal–dual formulations introduced in this work, which asserts that the dual optimum coincides with the primal optimum. This property is necessary for ensuring that the bounds obtained from the dual formulations converge to the physical values. Such strong duality theorems have been proven in problems closely related to those discussed here,\footnote{See, e.g., \cite{holtorf2024bounds} for a summary of strong duality theorems in similar contexts.} and may thus extend to the present setting.

Finally, it would be interesting to apply the time-dependent bootstrap to lattice systems (e.g., spin chains or Hamiltonian lattice gauge theories) and to other large-$N$ quantum systems, particularly those suffering from a sign problem that renders Monte Carlo methods challenging.

\section*{Acknowledgments}
We thank Mathew Dodelson, David Simmons-Duffin, Davide Gaiotto,  Andrea L. Guerrieri, Luca Iliesiu, Raghu Mahajan, Harish Murali, Igor Klebanov, Pedro Vieira and Xi Yin for useful discussions. The work of MC is supported by Clay C\'ordova's Sloan Research Fellowship from the Sloan Foundation. BG is supported by Simons Foundation grant \#994310
(Simons Collaboration on Confinement and QCD Strings). HL is supported by a Bloch Fellowship and by NSF grant PHY-2310429.
JY is partially supported by National Science Foundation awards No. 2440805 and 2016245. Z.Z. is supported by Simons Foundation grant \#994308 for the Simons Collaboration on Confinement and QCD Strings. The work of MC and BG was performed in part at Aspen Center for Physics, which is supported by National Science Foundation grant PHY-2210452.
We also acknowledge the workshop ``Solving Holographic Theories'' at Imperial College London, where this collaboration was initiated.

\pagebreak

\appendix

\section{Practical Implementation}\label{app:implementation}
In practice, it is convenient to choose a basis in which to represent the constraints in~\eqref{primalMN}. In fact, this is the approach originally proposed in~\cite{Lawrence:2024mnj}. In the following, we describe the practical implementations for the different cases considered in this work.

\subsection{Finite-dimensional operator basis truncation}\label{appendix:finitedimdiscussion}
We first present more details about the level $L$ truncation of the operator basis introduced in section \ref{2ptSetup}. {For a given $L$, we define the spaces of (anti-)Hermitian truncated matrices $ \mathcal{H}_L$ ($\bar{\mathcal{H}}_L$) that include only operators of weight less or equal to $L$,
\begin{align}
  \left(
\begin{array}{cc}
 * & 0 \\
 0 & 0 \\
\end{array}
\right),
\end{align}
Next, we define the linear operators,
\begin{equation}
    \mathcal{D}_\pm[M] = D M \pm M D^{\dagger}\ .
\end{equation}
We can now re-write \eqref{Nline2} as follows. For any anti-hermitian matrix $M$, $\Tr[\mathcal{D}_-[M] \cdot \mathcal{M}]=0$. The space of such constraints is hence the large $L$ limit of the image of $\mathcal{D}_-$ acting on $\bar{\mathcal{H}}_{L}$, which we denote $\mathcal{D}_-(\bar{\mathcal{H}}_{L})$. Note that it is a subspace of $\mathcal{H}_{L+1}$. Clearly, all matrices $A \in \mathcal{D}_-(\bar{\mathcal{H}}_{L})$ satisfy $\Tr[A \cdot \mathcal{M}]=0$. From this we learn that we strictly relax the constraint imposed by $\lambda_A$ by replacing,\footnote{We have observed empirically that this often gives most of the constraints that include all operators up to truncation level $L$. However, in some cases such as $L=8$ for the anharmonic oscillator there are a few equations of motion constraints that are not generated in this way. Explicitly, they are linear combination of higher level constraints that accidentally truncate to level $L$. %
}
\begin{equation}
    \lambda_A D^\dagger - D \lambda_A \to \sum_j \lambda_A^{(j)} A^{L}_j\ ,    
\end{equation}
where $\{A^{L}_j\}$ is a basis of $\mathcal{D}_-(\bar{\mathcal{H}}_{L-1})$, and $\lambda_A^{(j)}$ are numbers. The larger $L$ is, more of the constraints are imposed.

Now we turn to $\lambda_D$ and $\lambda_G$. They both live in
\begin{equation}
\mathcal{H}_\infty/\mathcal{D}_-(\bar{\mathcal{H}}_{\infty})\ ,
\end{equation}
where the subscript $\infty$ is a shorthand for the $L \to \infty$ limit. To truncate to a finite value of $L$, we also need to take $\lambda_{D/G}$ s.t., $\mathcal{D}_+ (\lambda_{D/G}) \in \mathcal{H}_L$. Hence, the truncated $\lambda_{D/G}$ live in,
\begin{equation}
    \mathcal{C}_L \equiv \{ C \in \mathcal{H}_L \ |\  \exists A \in \mathcal{D}_-(\bar{\mathcal{H}}_{L})\ , \ \ \text{s.t.} \ \ \mathcal{D}_+(C) + A \in \mathcal{H}_L\}\ .
\end{equation}
In practice, we then perform the replacement,
\begin{equation}
    \lambda_{D/G} \to \sum_j \lambda_{D/G}^{(j)} C^{L}_j\ ,    
\end{equation}
where $\{C^{L}_j\}$ is a basis of $\mathcal{C}_L$. As before, the larger $L$ is the more of the constraints are imposed. This procedure is equivalent to only imposing a subset of the constraints in the primal problem.
}

\subsection{Ground state with zero-time input}\label{app:gswithinitialdata}
Consider bootstrapping the time evolution in the ground state in the case where we input the zero-time data ${\cal M}(0)$. We can write the primal problem as:
\begin{equation}
\label{eq:probGSInit}
\begin{split}
\operatorname{minimize } & \operatorname{Tr} \widehat{O} \M(T) \\
\text { subject to } & \M(\tau) \succeq 0\\
& \N(\tau)\succeq 0,  \\
& \operatorname{Tr} A^{(i)} \M(\tau)= 0 \\
& \operatorname{Tr}\left(D^{(k)}-C^{(k)} \frac{\d}{\d \tau}\right) \M(\tau)=0 \\
& \operatorname{Tr}\left(G^{(j)}\M(\tau)-H^{(j)}\N(\tau)\right)=0.\\
\end{split}
\end{equation}
The first two positivity constraints are imposed for $\tau \in [0,T]$. Note that a more general condition is $\operatorname{Tr} A^{(i)} \M(\tau)= a^{(i)}$ for some constants $a^{(i)}$, but we can always set $a^{(i)}=0$ by rearranging the equations and using the fact that $\M_{11}=1$. Moreover, the zero-time input has $\M_{11}(0)=1$ and the time-evolution demands that $\frac{d \M_{11}(\tau)}{d\tau}=0$, this together implies $\M_{11}(\tau)=1$ and we can thus set all $a^{(i)}$ to be zero.

Let us now derive the dual problem by introducing the action and the Lagrange multipliers:
\begin{align}
	I &= \Tr  \bigg\{\widehat{O} \M(T) + \int_0^T \d \tau   \left[ \lambda_A^{(i)} A^{(i)} + \lambda_G^{(j)} G^{(j)} + \lambda_D^{(k)} \left(D^{(k)}-C^{(k)} \partial_\tau \right)  -\Lambda_\M \right] \mathcal{M}(\tau) \nonumber\\
    &\quad - \int \d \tau \left[ \lambda_G^{(j)} H^{(j)} + \Lambda_\N\right] \mathcal{N}(\tau) \bigg \}\nonumber\\
    &= \Tr  \bigg\{ \int_0^T \d \tau   \left[ \lambda_A^{(i)} A^{(i)} + \lambda_G^{(j)} G^{(j)} + \left(D^{(k)}+C^{(k)} \frac{\d}{\d \tau}\right) \lambda_D^{(k)} -\Lambda_\M \right] \mathcal{M}(\tau) \\
    &\quad - \int \d \tau \left[ \lambda_G^{(j)} H^{(j)} + \Lambda_\N\right] \mathcal{N}(\tau)+ \lambda_D^{(k)}(0) C^{(k)} \M(0) + \left[ \widehat{O} -\lambda_D^{(k)}(T) C^{(k)} \right]  \M(T) \bigg \} \nonumber
\end{align}
Varying with respect to $\M(\tau)$ and also the boundary value $\M(T)$, we arrive at the {\it dual problem}:

\begin{equation}\label{SDP:MNbootPractical}
\boxed{    \begin{split}
        \text { maximize } & \lambda_D^{(k)}(0) \operatorname{Tr} C^{(k)} \M(0) \\
\text { subject to } & \widehat{O}- \lambda_D^{(k)}(T) C^{(k)} = 0 \\
& \lambda_A^{(i)}(\tau) A^{(i)}+\lambda_G^{(j)}(\tau)G^{(j)}+\left(D^{(k)}+C^{(k)} \partial_\tau \right) \lambda_D^{(k)}(\tau) \succeq 0\\
&-\lambda_G^{(j)}(\tau)H^{(j)}\succeq 0, \quad \forall \tau \in [0,T]
\end{split}}
\end{equation}

\subsection{Ground state without zero-time input}\label{sec:gsbootnoinitPD}
Even when $\mathcal{M}(0)$ is not known, one can still perform the Euclidean two-point correlator bootstrap. In such a case, $\M(0)$ and $\N(0)$ are also treated as primal bootstrap variables, both of which are required to be positive semidefinite. Furthermore, the operator algebra and Hamiltonian commutators provide linear relations between the matrix elements of $\mathcal{M}(0)$ and $\mathcal{N}(0)$, taking the form,
\begin{equation}
    \operatorname{Tr} B^{(j)}\M(0)=b^{(j)} \quad,\quad \text{and} \quad,\quad \operatorname{Tr}\left(E^{(l)}\M(0)\right)-\operatorname{Tr}\left(F^{(l)}\N(0)\right)=0
\end{equation}
The primal problem is then written as
\begin{equation}
\begin{split}
\operatorname{minimize} & \operatorname{Tr} \widehat{O} \M(T) \\
\text { subject to } & \M(\tau) \succeq 0,~~ \M(0) \succeq 0, \\
& \N(\tau)\succeq0,~~\N(0) \succeq 0,  \\
& \operatorname{Tr} A^{(i)} \M(\tau)= 0 \\
& \operatorname{Tr} B^{(j)} \M(0)= b^{(j)} \\
& \operatorname{Tr}\left(D^{(k)}-C^{(k)} \frac{\d}{\d \tau}\right) \M(\tau)=0 \\
& \operatorname{Tr}\left(E^{(l)}\M(0)\right)-\operatorname{Tr}\left(F^{(l)}\N(0)\right)=0\\
& \operatorname{Tr}\left(G^{(n)}\M(\tau)\right)-\operatorname{Tr}\left(H^{(n)}\N(\tau)\right)=0\\
\end{split}
\end{equation}
Denoting the Lagrange multipliers collectively by $\vec{\lambda}$, the Lagrangian is
\begin{equation}
\begin{split}
	&L[\M(\tau), \N(\tau),\vec{\lambda}(\tau)]\\
    &=\operatorname{Tr} (\widehat{O} \M(T))-\operatorname{Tr}(\Lambda_{\M 0} \M(0))+\lambda_{B}^{(j)}(\operatorname{Tr} (B^{(j)} \M(0))- b^{(j)})\\
    &+\lambda_E^{(l)}\left(\operatorname{Tr}\left(E^{(l)}\M(0)\right)-\operatorname{Tr}\left(F^{(l)}N(0)\right)\right)-\operatorname{Tr}\left(\Lambda_{\N 0} \N(0)\right)-\operatorname{Tr}\left(\Lambda_{\N T} \N(T)\right)   \\
    &+\int_0^T\, d\tau\operatorname{Tr}\left[\left(\lambda_A^{(i)}(\tau)A^{(i)}+\lambda_G^{(n)}(\tau)G^{(n)}+\lambda_D^{(k)}(\tau)\left(D^{(k)}-C^{(k)}\frac{\d}{\d \tau}\right)-\Lambda_\M(\tau)    \right)  \M(\tau)\right]\\
    &-\int_0^Td\tau\operatorname{Tr}\left[\left(\lambda_G^{(n)}(\tau)H^{(n)}+\Lambda_\N(\tau)\right)\N(\tau)\right]\\
    &=\operatorname{Tr}\left( \left(\widehat{O}-\lambda^{(k)}_D(T) C^{(k)}\right) \M(T)\right)\\    &+\operatorname{Tr}\left(\left(\lambda^{(k)}_D(0) C^{(k)}+\lambda_{B}^{(j)}B^{(j)}+\lambda_E^{(l)}E^{(l)}-\Lambda_{\M 0}\right)\M(0)\right)\\
    &-\operatorname{Tr}\left(\left(\Lambda_{\N 0}+\lambda_E^{(l)}F^{(l)}\right)\N(0)\right)  -\lambda_{B}^{(j)} b^{(j)}\\
    &+\int_0^T d\tau  \left[ \lambda_A^{(i)}(\tau) A^{(i)}+\lambda_G^{(n)}(\tau)G^{(n)}+ \left(D^{(k)}+C^{(k)} \frac{\d}{\d \tau}\right) \lambda_D^{(k)} -\Lambda_\M(\tau)\right] \M(\tau)\\
    &-\int_0^T d\tau\operatorname{Tr}\left[\left(\lambda_G^{(n)}(\tau)H^{(n)}+ \Lambda_\N(\tau)\right)\N(\tau)\right]
\end{split}
\end{equation}

The dual problem is then given by

\begin{equation}\label{dual0T}
\boxed{\begin{split}
        \text { maximize } & -\lambda_{B}^{(j)} b^{(j)}\\
\text { subject to } & \widehat{O}-\lambda_D^{(k)}(T) C^{(k)}=0 \\
& \lambda^{(k)}_D(0) C^{(k)}+\lambda_{B}^{(i)}B^{(i)}+\lambda_E^{(l)}E^{(l)}\succeq0 \\
& -\lambda_E^{(l)}F^{(l)}\succeq0\\
& \lambda_A^{(i)}(\tau) A^{(i)}+\lambda_G^{(n)}(\tau)G^{(n)}+\left(D^{(k)}+C^{(k)} \frac{\d}{\d \tau}\right) \lambda_D^{(k)} \succeq 0\\
&-\lambda_G^{(n)}(\tau)H^{(n)}\succeq0
    \end{split}}
\end{equation}
Note that we can also impose $\mathcal{M}(T) \succeq 0$ in the primal problem and convert the first constraint in the dual problem into an inequality. In practice, we obtain the same numerical results either way.

\subsection{Thermal state with zero-time input}\label{app:withInitThermal}
In this variant of the bootstrap problem, we assume that the thermal one-point functions $\mathcal{M}(0)$ (and, by the KMS condition, also $\mathcal{M}(\beta)$) are known. The primal problem for the thermal Euclidean two-point correlator bootstrap is given by
\begin{equation}
\begin{split}
\operatorname{minimize} & \operatorname{Tr} \widehat{O} \M(T) \\
\text { subject to } & \M(\tau) \succeq 0 \\
& \operatorname{Tr} A^{(i)} \M(\tau)= 0 \\
& \operatorname{Tr}\left(D^{(k)}-C^{(k)} \frac{\d}{\d \tau}\right) \M(\tau)=0. 
\end{split}
\end{equation}
Here, the positivity of $\M(\tau)$ is imposed for $\tau \in (0,\beta)$. An analogous derivation through the action leads to the dual problem
\begin{equation}
\label{thermaldualM0}
\boxed{\begin{split}
\text { maximize } & \lambda_D^{(k)}(0) \operatorname{Tr} C^{(k)} \M(0) - \lambda_D^{(k)}(\beta) \operatorname{Tr} C^{(k)} \M(\beta) \\
\text { subject to } 
&  \widehat{O}+ C^{(k)} \left[\lambda_D^{(k)}(T+\epsilon) -\lambda_D^{(k)}(T-\epsilon)\right] \succeq 0\\
& \lambda_A^{(i)}(\tau) A^{(i)}+\left(D^{(k)}+C^{(k)} \frac{\d}{\d \tau}\right) \lambda_D^{(k)}(\tau) \succeq 0 .\end{split}}
\end{equation}

\subsection{Thermal state without zero-time input}\label{app:noInitThermal}

In the case where $\M(0)$ (and also $\M(\beta)$) is unknown, we can use the KMS condition in the primal problem: 
\begin{equation}\label{thermal_no_init_primal}
\begin{split}
\operatorname{minimize} & \operatorname{Tr} \widehat{O} \M(T) \\
\text { subject to } & \M(\tau) \succeq 0,~~ \M(0) \succeq 0,~~\M(\beta)\succeq0  \\
& \operatorname{Tr} A^{(i)} \M(\tau)= 0 \\
& \operatorname{Tr}\left(D^{(k)}-C^{(k)} \frac{\d}{\d \tau}\right) \M(\tau)=0 \\
& \operatorname{Tr} B^{(j)} \M(0)= b^{(j)} \\
&\operatorname{Tr}\left(E^{(l)}\M(0)\right)-\operatorname{Tr}\left(F^{(l)}\M(\beta)\right)=0.
\end{split}
\end{equation}
The dual problem is  
\begin{equation}\label{thermal_no_init_primaldual}
\boxed{\begin{split}
\text { maximize } & -\lambda_B^{(j)} b^{(j)}\\
\text { subject to } 
&  \widehat{O} + C^{(k)} \left[\lambda_D^{(k)}(T+\epsilon) -\lambda_D^{(k)}(T-\epsilon)\right] \succeq 0\\
& \lambda_A^{(i)}(\tau) A^{(i)}+\left(D^{(k)}+C^{(k)} \frac{\d}{\d \tau}\right) \lambda_D^{(k)}(\tau) \succeq 0 \\
& \lambda_D^{(k)}(0) C^{(k)} + \lambda_B^{(j)} B^{(j)}-\lambda_E^{(l)} E^{(l)} \succeq 0 \\
& -\lambda_D^{(k)}(\beta) C^{(k)} +\lambda_E^{(l)} F^{(l)} \succeq 0.
\end{split}}
\end{equation}
One may be interested in the infeasibility of the primal problem. Farkas' lemma shows that if we consider the following \textit{dual} problem
\begin{equation}\label{eqn:noinitthermalInfeas}
\begin{split}
\text { minimize } & \lambda_B^{(j)} b^{(j)}\\
\text { subject to } 
&  C^{(k)} \left[\lambda_D^{(k)}(T+\epsilon) -\lambda_D^{(k)}(T-\epsilon)\right] \succeq 0\\
& \lambda_A^{(i)}(t) A^{(i)}+\left(D^{(k)}+C^{(k)} \frac{\d}{\d t}\right) \lambda_D^{(k)}(t) \succeq 0 \\
& \lambda_D^{(k)}(0) C^{(k)} + \lambda_B^{(j)} B^{(j)}-\lambda_E^{(l)} E^{(l)} \succeq 0 \\
& -\lambda_D^{(k)}(\beta) C^{(k)} +\lambda_E^{(l)} F^{(l)} \succeq 0
\end{split}
\end{equation}
and the resulting minimal dual objective satisfies $\lambda_B^{(j)}b^{(j)}<0$, then the primal problem (\ref{thermal_no_init_primal}) is infeasible.

\subsection{Thermal state, with inequalities on zero-time input}\label{app:initialIneqThermal}

Here we assume that we are given bounds on the elements of ${\cal M}(0)$, as in (\ref{eq:boundsOnM0}). Then, the primal problem is 
\begin{equation}\label{thermal_init_ineq_primal}
\begin{split}
\operatorname{minimize} & \operatorname{Tr} \widehat{O} \M(T) \\
\text { subject to } & \M(\tau) \succeq 0, ~\M(0)\succeq0,~\M(\beta)\succeq0  \\
& \operatorname{Tr} A^{(i)} \M(\tau)= 0 \\
& \operatorname{Tr}\left(D^{(k)}-C^{(k)} \frac{\d}{\d \tau}\right) \M(\tau)=0 \\
& l^{(p)}\leq\operatorname{Tr} J^{(p)} \M(0)\leq u^{(p)} \\
&\operatorname{Tr}\left(E^{(l)}\M(0)\right)-\operatorname{Tr}\left(F^{(l)}\M(\beta)\right)=0.
\end{split}
\end{equation}
The dual problem is  
\begin{equation}\label{thermal_init_ineq_dual}
\boxed{\begin{split}
\text { maximize } & \lambda^{(p)}_l l^{(p)}-\lambda^{(p)}_u u^{(p)}\\
\text { subject to } 
& \widehat{O} + C^{(k)} \left[\lambda_D^{(k)}(T+\epsilon) -\lambda_D^{(k)}(T-\epsilon)\right] \succeq 0\\
& \lambda_A^{(i)}(t) A^{(i)}+\left(D^{(k)}+C^{(k)} \frac{\d}{\d t}\right) \lambda_D^{(k)}(t) \succeq 0 \\
& \lambda_D^{(k)}(0) C^{(k)} + (\lambda_u^{(p)}-\lambda_l^{(p)})J^{(p)}-\lambda_E^{(l)} E^{(l)} \succeq 0 \\
& -\lambda_D^{(k)}(\beta) C^{(k)} +\lambda_E^{(l)} F^{(l)} \succeq 0\\
&\lambda_l^{(p)}\geq0,~\lambda_u^{(p)}\geq0.
\end{split}}
\end{equation}

\section{Clamped B-spline}\label{app:b-spline}

In this appendix, we review the details of the clamped B-spline basis used to solve the finite-dimensional dual problems, such as~(\ref{SDP:finiteDim1}). This basis was recently employed to study the time evolution of stochastic processes in \cite{Cho:2025dgc}, whose discussion we closely follow. To begin, define the knot vector
\begin{equation*}
    v_{\mathrm{knot}} = (\tau_0, \tau_1, \cdots, \tau_{N+d}) \ ,\ \text{with}\ ,\  \tau_0 = \cdots = \tau_d = 0 \ ,\ \tau_k=\frac{k-d}{N-d}T\ ,\ \text{and}\ ,\ \tau_N = \cdots = \tau_{N+d} = T,
\end{equation*}
used to construct $N$ independent clamped B-splines of degree $d$, denoted by $\phi^{(d)}_{k_{(d)}}(\tau)$ for $k_{(d)} = 1, \cdots, N$. Each spline $\phi^{(d)}_{k_{(d)}}(\tau)$ is determined, up to an overall rescaling, by the requirement that it has nontrivial support only over the interval $[\tau_{k_{(d)} - 1}, \tau_{k_{(d)} + d})$.

The derivative of $\phi^{(d)}_{k_{(d)}}(\tau)$ is a linear combination of the clamped B-spline basis of polynomial degree $d-1$ over the same knot vector $v_{\mathrm{knot}}$. This naturally leads us to consider clamped B-spline bases of polynomial degrees $\mu = 1, 2, \cdots, d$ over the knot vector $v_{\mathrm{knot}}$. Keeping only the splines that have nontrivial support in the interior region $(\tau_d, \tau_N)$, we obtain the following set of clamped B-spline basis functions: 
\begin{equation}
    {\cal B}_{\cal I} = \left\{\phi^{(\mu)}_{k_{(\mu)}}(\tau)\right\}^{\mu=1,\cdots,d}_{k_{(\mu)}=1,\cdots,N+\mu-d} ,    
\end{equation}
where $\phi^{(\mu)}_{k_{(\mu)}}(\tau)$ denotes the $k_{(\mu)}$-th element of the clamped B-spline basis of degree $\mu$ over the knot vector $v_{\mathrm{knot}}$, and is nonzero only on the interval $[\tau_{k_{(\mu)} - 1 + d - \mu}, \tau_{k_{(\mu)} + d})$. The index set $\cal I$ appearing in (\ref{SDP:finiteDim1}) is thus given by ${\cal I} = \{(\mu, k_{(\mu)}) ~|~ \mu = 1, \cdots, d,~ k_{(\mu)} = 1, \cdots, N + \mu - d \}$.

The basis ${\cal B}_{\cal I}$ is linearly independent, although not orthonormal over $\tau \in [0,T]$. Upon appropriate rescaling, the functions satisfy
\begin{equation}
    \phi^{(\mu)}_{k_{(\mu)}}(\tau)\geq0,~~\forall \tau\in[0,T],~~~~\phi^{(\mu)}_{k_{(\mu)}}(0)=\delta_{k_{(\mu)},1},~~~~\phi^{(\mu)}_{k_{(\mu)}}(T)=\delta_{k_{(\mu)},N+\mu-d},  ~~~~~~~\forall\mu,~~\forall k_{(\mu)}.
\end{equation}
Therefore, they form a basis of positive functions that can be used to construct feasible solutions of interest. Conveniently, in \texttt{Mathematica}, the function \texttt{BSplineBasis} generates the basis ${\cal B}_{\cal I}$.

Derivatives of elements in ${\cal B}_{\cal I}$, except for $\mu = 1$, can be expanded again in terms of elements in ${\cal B}_{\cal I}$. Concretely,
\begin{equation}
 \partial_\tau \phi^{(\mu)}_{k_{(\mu)}}(\tau)=\sum_{l_{(\mu-1)}=1}^{N+\mu-1-d}U^{(\mu)}_{k_{(\mu)}l_{(\mu-1)}}\phi^{(\mu-1)}_{l_{(\mu-1)}}(\tau),~~~\mu=2,3,\cdots,d,
\end{equation}
where the expansion coefficients $U^{(\mu)}_{k_{(\mu)}l_{(\mu-1)}}$ can be computed straightforwardly from the following inner products:
\begin{equation}
\begin{split}
    V^{(\mu)}_{k_{(\mu)},l_{(\mu)}}&=\int_0^Td\tau\phi^{(\mu)}_{k_{(\mu)}}(\tau)\phi^{(\mu)}_{l_{(\mu)}}(\tau),~~~~\mu=1,\cdots,d-1,\\
    W^{(\mu)}_{k_{(\mu)},l_{(\mu-1)}}&=\int_0^Td\tau\left({d\phi^{(\mu)}_{k_{(\mu)}}(\tau)\over d\tau}\right)\phi^{(\mu-1)}_{l_{(\mu-1)}}(\tau),~~~~\mu=2,\cdots,d,\\
    \Rightarrow U^{(\mu)}&=W^{(\mu)}\left(V^{(\mu-1)}\right)^{-1},~~~~\mu=2,\cdots,d.
\end{split}
\end{equation}
In contrast, $\phi^{(1)}_{k_{(1)}}(\tau)$ does not have a well-defined first-order derivative. In the notation of (\ref{SDP:finiteDim1}), $\mu=1$ and $k_{(\mu=1)}=1,\cdots,N+1-d$ are the only elements of the set ${\cal I}\backslash\cal J$. In this work, we chose $N=20$ and $d=10$ unless specified otherwise.

\section{Polynomial Formulation}\label{app:PMP}

In this appendix, we present the polynomial formulation of the dual problem discussed in Appendix~\ref{app:implementation}. In this approach, the dual functions are represented as truncated polynomials of degree $d$, which satisfy the dual constraints listed in Appendix~\ref{app:implementation}. Each feasible dual solution provides a rigorous bound on the primal problem. The optimal dual solution is recovered in the limit $d \to \infty$\footnote{Convergence is guaranteed by the Stone–Weierstrass theorem.}.

This formulation can be recast as a \emph{Polynomial Matrix Program} (PMP), which serves as the standard input for \texttt{SDPB}~\cite{Simmons-Duffin:2015qma}:
\begin{equation}\label{eq: SDPB}
\begin{aligned}
\text{maximize} \quad & b \cdot y \quad \text{over } y \in \mathbb{R}^N, \\
\text{subject to} \quad & M^0_j(x) + \sum_{n=1}^N y_n M^n_j(x) \succeq 0,
\quad \forall\, x \ge 0,\; 1 \le j \le J,
\end{aligned}
\end{equation}
where
\begin{equation}
M^n_j(x) =
\begin{pmatrix}
P^n_{j,11}(x) & \cdots & P^n_{j,1m_j}(x) \\
\vdots & \ddots & \vdots \\
P^n_{j,m_j1}(x) & \cdots & P^n_{j,m_jm_j}(x)
\end{pmatrix},
\end{equation}
labeled by $0 \le n \le N$ and $1 \le j \le J$, with each element $P^n_{j,rs}(x)$ being a polynomial.

\vspace{0.5em}
\noindent\textbf{Example: bounding the two-point correlator with zero-time input.}
We illustrate this formulation using a simpler problem: bounding a time-separated operator correlator under a given initial condition (zero-time input),
\begin{equation}\label{eq: toy}
\begin{split}
\operatorname{minimize } \quad & \operatorname{Tr}\big[\widehat{O}\mathcal{M}(T)\big], \\
\text{subject to} \quad & \mathcal{M}(\tau) \succeq 0,\\
& \operatorname{Tr}\big[A^{(i)} \mathcal{M}(\tau)\big]= 0, \\
& \operatorname{Tr}\!\left[\!\left(D^{(k)} - C^{(k)} \frac{\mathrm{d}}{\mathrm{d}\tau}\right)\! \mathcal{M}(\tau)\!\right]=0.
\end{split}
\end{equation}
Note that this represents a strict relaxation of problem~\eqref{eq:probGSInit}, obtained by omitting all conditions involving $\mathcal{N}$. The corresponding dual problem reads:
\begin{equation}\label{eq: toydual}
\begin{split}
\text{maximize} \quad & \lambda_D^{(k)}(0)\, \operatorname{Tr}\!\left[C^{(k)} \mathcal{M}(0)\right], \\
\text{subject to} \quad & \widehat{O}- \lambda_D^{(k)}(T) C^{(k)} = 0, \\
& \lambda_A^{(i)}(\tau) A^{(i)}+\big(D^{(k)}+C^{(k)} \partial_\tau \big) \lambda_D^{(k)}(\tau) \succeq 0.
\end{split}
\end{equation}
We seek feasible solutions where $\lambda_A^{(i)}(\tau)$ and $\lambda_D^{(k)}(\tau)$ are constructed from degree-$d$ polynomials. This yields a suboptimal but still rigorous bound for the primal problem~\eqref{eq: toy}. Since the positivity constraint applies over the finite interval $[0,T]$, we map it to the semi-infinite domain $y \ge 0$ via 
\begin{equation}
    \tau = \tfrac{T y}{1+y},    
\end{equation}
and define
\begin{equation}
    p(y) = (1+y)^d \lambda\!\left(\tfrac{T y}{1+y}\right) = \sum_{j=0}^d q_j y^j.
\end{equation}
The boundary conditions translate to
\begin{equation}
    \lambda(T)=\lim_{y\rightarrow\infty}\frac{p(y)}{(1+y)^d}=q_d,\quad\lambda(0)=\lim_{y\rightarrow 0}\frac{p(y)}{(1+y)^d}=q_0 .
\end{equation}
The derivative transforms as ( $\dot{}$ stands for the derivative in $\tau$ while $'$ stands for that in $y$)
\begin{equation}
    \dot{\lambda}(\tau) = \frac{1+y}{T(1+y)^d}\!\left[(1+y) p'(y) - d\,p(y)\right].
\end{equation}
We introduce the notation $\tilde{\partial}_T$ as
\begin{equation}
    \tilde{\partial}_T p(y) = \frac{1+y}{T}\!\left[(1+y) p'(y) - d\,p(y)\right].
\end{equation}
On the right-hand side, the leading powers cancel between the two terms, leaving a degree-$d$ polynomial. Consequently, the dual problem is reformulated as
\begin{equation}\label{eq: toypol}
\begin{split}
\text{maximize} \quad & q_{D,0}^{(k)}\, \operatorname{Tr}\!\left[C^{(k)} \mathcal{M}(0)\right],\\
\text{subject to} \quad & \widehat{O} - q_{D,d}^{(k)} C^{(k)} = 0,\\
& p_A^{(i)}(y) A^{(i)}+\big(D^{(k)}+C^{(k)} \tilde{\partial}_T \big)p_D^{(k)}(y) \succeq 0,
\end{split}
\end{equation}
where $p_A^{(i)}(y)=\sum_{j=0}^d q_{A,j}^{(i)} y^j$ and $p_D^{(k)}(y)=\sum_{j=0}^d q_{D,j}^{(k)} y^j$. Here, positivity is imposed for all $y \ge 0$, matching the input convention of \texttt{SDPB}.

\subsection{Revisiting the Universal Bound}

To make the construction concrete, consider minimizing the two-point correlator
\begin{equation}
\begin{split}
\operatorname{minimize } \quad & \langle x(T) x(0)\rangle, \\
\text{subject to } & 
\mathcal{M}(\tau) =
\begin{pmatrix}
\langle x(\tau) x(0)\rangle & -i\langle x(\tau)p(0)\rangle \\
-i\langle x(\tau)p(0)\rangle & \langle p(\tau)p(0)\rangle
\end{pmatrix} \succeq 0,
\end{split}
\end{equation}
for a one-dimensional quantum system with Hamiltonian
\begin{equation}
    H=\frac{1}{2} p^2 + V(x).
\end{equation}
The Heisenberg equations yield
$\tfrac{d}{d\tau}\langle x(\tau)x(0)\rangle = -i\langle p(\tau)x(0)\rangle$ and
$\tfrac{d}{d\tau}\langle x(\tau)p(0)\rangle = -i\langle p(\tau)p(0)\rangle$.
As shown in Section~\ref{universalBd}, the exact analytic bound is
\begin{equation}\label{eq: toyexact}
    \langle x(T) x(0)\rangle \ge \langle x^2\rangle\, e^{-T/(2\langle x^2\rangle)}.
\end{equation}
In the notation of~\eqref{eq: toy}, we have
\begin{align}
    \{A^{(i)}\} &= \varnothing,\quad
    \mathcal{O}=\begin{pmatrix}1 & 0 \\ 0 & 0\end{pmatrix},\\
    C^{(1)} &= \begin{pmatrix}1 & 0 \\ 0 & 0\end{pmatrix}, \quad
    C^{(2)} = \begin{pmatrix}0 & 1/2 \\ 1/2 & 0\end{pmatrix},\\
    D^{(1)} &= \begin{pmatrix}0 & -1/2 \\ -1/2 & 0\end{pmatrix}, \quad 
    D^{(2)} = \begin{pmatrix}0 & 0\\ 0 & -1\end{pmatrix},\\
    \mathcal{M}(0) &= \begin{pmatrix}\langle x^2\rangle & 1/2\\ 1/2 & \langle p^2\rangle\end{pmatrix}.
\end{align}
The dual formulation becomes
\begin{equation}
\begin{split}
\text{maximize } \quad & \lambda_D^{(1)}(0)\langle x^2\rangle+\frac{1}{2}\lambda_D^{(2)}(0),\\
\text{subject to } \quad &
\begin{pmatrix}
\lambda_D^{(1)}(T) & \lambda_D^{(2)}(T)/2\\
\lambda_D^{(2)}(T)/2 & 0
\end{pmatrix}
= 
\begin{pmatrix}
1 & 0 \\ 0 & 0
\end{pmatrix},\\
&
\begin{pmatrix}
\lambda_D^{\prime(1)}(\tau) & \lambda_D^{\prime(2)}(\tau)/2\\
\lambda_D^{\prime(2)}(\tau)/2 & 0
\end{pmatrix}
-
\begin{pmatrix}
0 & \lambda_D^{(1)}(\tau)/2\\
\lambda_D^{(1)}(\tau)/2 & \lambda_D^{(2)}(\tau)
\end{pmatrix} \succeq 0.
\end{split}
\end{equation}
Using the polynomial basis introduced above,
\begin{equation}
\begin{split}
\text{maximize } \quad & q_0^1\langle x^2\rangle+\tfrac{1}{2}q_0^2,\\
\text{subject to } \quad & q_d^1=1,\quad q_d^2=0,\\
&
\begin{pmatrix}
\tilde{\partial}_T p_D^{(1)}(y) & \tilde{\partial}_T p_D^{(2)}(y) - p_D^{(1)}(y)/2\\
\tilde{\partial}_T p_D^{(2)}(y) - p_D^{(1)}(y)/2 & -p_D^{(2)}(y)
\end{pmatrix} \succeq 0,
\end{split}
\end{equation}
with $2(d+1)$ variables $q_j^k$ defined by
\begin{equation}
    p_D^{(k)}(y) = (1+y)^d \lambda\!\left(\tfrac{yT}{1+y}\right) = \sum_{j=0}^d q_j^k y^j.
\end{equation}
At the lowest truncation level $d=1$, one can verify the exact analytic solution:
\begin{equation}
    \langle x(T) x(0)\rangle \ge \langle x^2\rangle - \frac{T}{2},
\end{equation}
which is weaker than~\eqref{eq: toyexact}. Convergence is rapid; for example for $\langle x^2\rangle$ with $T = 1,\, d=5$, numerical results from \texttt{SDPB} give
\begin{equation}
    0.60653053 \le e^{-1/2} \simeq 0.60653066.
\end{equation}

\subsection{Polynomial Formulation for Ground State and Thermal Ensembles}

The same construction extends naturally to more complex setups. The final PMP corresponding to the ground-state bootstrap without zero-time data~\eqref{dual0T} reads:
\begin{equation}
\begin{split}
\text{maximize } \quad & -\lambda_B^{(j)}b^{(j)},\\
\text{subject to } \quad &
\widehat{O} - q_{D,d}^{(k)} C^{(k)} = 0,\\
& q_{D,0}^{(k)} C^{(k)} + \lambda_B^{(j)}B^{(j)} + \lambda_E^{(l)}E^{(l)} \succeq 0,\\
& -\lambda_E^{(l)}F^{(l)} \succeq 0,\\
& p_A^{(i)}(y)A^{(i)} + p_G^{(n)}(y)G^{(n)} + \big(D^{(k)}+C^{(k)}\tilde{\partial}_{T}\big)p_D^{(k)}(y) \succeq 0,\\
& -p_G^{(n)}(y)H^{(n)} \succeq 0.
\end{split}
\end{equation}

In the thermal case, the main complication arises from the need to impose matrix positivity over two disjoint closed intervals, $[0, T]$ and $[T, \beta]$. Each interval corresponds to a different analytic continuation of the correlator in Euclidean time. To handle this structure, we introduce a second set of polynomial variables to represent the dual functions in the second interval.

Specifically, in addition to
\begin{equation}
    p(y) = (1+y)^d \lambda\!\left(\tfrac{T y}{1+y}\right) = \sum_{j=0}^d q_j y^j,
\end{equation}
which parameterizes $\lambda(\tau)$ on $[0,T]$, we define
\begin{equation}
    r(y) = (1+y)^d \lambda\!\left(\tfrac{(\beta-T) y}{1+y}\right) = \sum_{j=0}^d s_j y^j,
\end{equation}
to encode its continuation on the second interval $[T, \beta]$. These two sets of variables, $(q_j)$ and $(s_j)$, capture the polynomial structure of the dual functions on the respective domains and allow us to impose the required positivity conditions independently.

The resulting polynomial matrix program for the thermal state without zero-time data~\eqref{thermal_no_init_primaldual} becomes:
\begin{equation}
\begin{split}
\text{maximize } \quad & -\lambda_B^{(j)}b^{(j)},\\
\text{subject to } \quad &
\widehat{O} - q_{D,d}^{(k)} C^{(k)} + s_{D,0}^{(k)} C^{(k)} \succeq 0,\\
& p_A^{(i)}(y)A^{(i)}+\big(D^{(k)}+C^{(k)}\tilde{\partial}_{T}\big)p_D^{(k)}(y) \succeq 0,\\
& r_A^{(i)}(y)A^{(i)}+\big(D^{(k)}+C^{(k)}\tilde{\partial}_{\beta-T}\big)r_D^{(k)}(y) \succeq 0,\\
& q_{D,0}^{(k)} C^{(k)} + \lambda_B^{(j)}B^{(j)} - \lambda_E^{(l)}E^{(l)} \succeq 0,\\
& -s_{D,d}^{(k)} C^{(k)} + \lambda_E^{(l)}F^{(l)} \succeq 0.
\end{split}
\end{equation}
The introduction of two polynomial bases, $p(y)$ and $r(y)$, ensures that the positivity constraints on both segments of the Euclidean time circle are treated on an equal footing, rendering the thermal bootstrap problem fully compatible with the \texttt{SDPB} input format.

\section{Improved LMN bootstrap \label{app:lawrence}}
Here we summarize the method developed in \cite{Lawrence:2024mnj}.
One defines a matrix of one-point functions: $\M_{ij}(t) = \ev{\O_i^\dagger(t) \O_j(t)}$.
We imagine that $\M_{ij}(0)$ is ``known,'' which allows one to (partially) specify the initial density matrix $\rho_0$.
The primal problem is:
\begin{align}
\operatorname{minimize} & \operatorname{Tr} \widehat{O} \M(T) \\
\text { subject to } & \M(t) \succeq 0 \label{posM} \\
& \operatorname{Tr} A^{(i)} \M(t)= 0 \label{constrA}\\
& \operatorname{Tr}\left(D^{(k)}-C^{(k)} \frac{\d}{\d t}\right) \M(t)=0 \label{constrB}%
\end{align}
Following \cite{Lawrence:2024mnj}, one should derive the dual version of this semidefinite problem. This is accomplished by writing a action:
\begin{align}
	S[M, \lambda_A,\lambda_D,\Lambda] =& \operatorname{Tr} \widehat{O} \M(T)\nonumber\\
    &+\int_0^T\, \d t  \Tr \left[\lambda_A^{(i)}(t) A^{(i)}+\lambda_D^{(k)}(t) \left(D^{(k)}-C^{(k)} \frac{\d}{\d t}\right) -\Lambda(t) \right] \M(t) .
\end{align}
Here we have written two sets of (scalar) Lagrange multipliers that enforce the constraints \eqref{constrA} and \eqref{constrB} and a positive semidefinite matrix function $\Lambda(t) \succeq 0$ that enforces \eqref{posM}. Then integrating by parts,
\begin{align}
S&= \Tr \Big\{\widehat{O} \M(T) +\lambda^{(k)}_D(0) C^{(k)} \M(0)-\lambda^{(k)}_D(T) C^{(k)} \M(T) \nonumber\\
&\qquad\qquad+ \int_0^T \d t \,  \left[ \lambda_A^{(i)} A^{(i)}+ \left(D^{(k)}+C^{(k)} \frac{\d}{\d t}\right) \lambda_D^{(k)} -\Lambda\right] \M(t) \Big \}\\
&=\Tr \left\{ \lambda^{(k)}_D(0)  C^{(k)} \M(0) + \int_0^T \d t \,  \left[ \lambda_A^{(i)} A^{(i)}+ \left(D^{(k)}+C^{(k)} \frac{\d}{\d t}\right) \lambda_D^{(k)} -\Lambda\right] \M(t) \right \} \nonumber
\end{align}
In the last line, %
{we integrated out $\M(T)$ resulting in the boundary condition for $\lambda_D$,}
\begin{align}
	\sum_k \lambda_D^{(k)}(T) C^{(k)} =\widehat{O}. \label{bdCond}
\end{align}
In the primal problem, we imagine varying this functional with respect to $\lambda_A, \lambda_D, \Lambda$ first. This constrains $\M$. In the dual problem, we first vary with respect to $\M$ before varying with respect to $\Lambda$. The equations of motion for $\M$ set
\begin{align}\label{Lsol}
\Lambda =	 \lambda_A^{(i)} A^{(i)}+ \left(D^{(k)}+C^{(k)} \frac{\d}{\d t}\right) \lambda_D^{(k)}  
\end{align}
So we can eliminate $\Lambda$ as long as we remember to impose positivity, which comes from $\Lambda \succeq 0 $:
\begin{align}\label{posL}
	\lambda_A^{(i)} A^{(i)}+ \left(D^{(k)}+C^{(k)} \frac{\d}{\d t}\right) \lambda_D^{(k)}   \succeq 0. 
\end{align}
The resulting functional is independent of $\M(t)$. Therefore, we optimize $  \lambda^{(k)}_D(0)  \Tr C^{(k)} \M(0) $ over possible $\lambda_A, \lambda_D$, subject to the boundary condition \eqref{bdCond} and the positivity constraint \eqref{posL}:
\begin{align}
\label{ineqLMND}
\text { maximize } & \lambda_D^{(k)}(0) \operatorname{Tr} C^{(k)} \M(0) \nonumber\\
\text { subject to } & \lambda_D^{(k)}(T) C^{(k)}= \widehat{O} \\
& \lambda_A^{(i)}(t) A^{(i)}+\left(D^{(k)}+C^{(k)} \frac{\d}{\d t}\right) \lambda_D^{(k)} \succeq 0 .\nonumber 
\end{align}
Up to this point, we have just reviewed the work of LMN \cite{Lawrence:2024mnj}.

LMN then propose to use a barrier function method to ensure that the positive semidefinite inequality in \eqref{ineqLMND} holds for all time $0< t < T$. 
The barrier function $\int \d \tau \mathcal{F}(\Lambda)$ is engineered to prevent $\Lambda$ from having negative eigenvalues. Numerical implementation requires evaluating this integral. A naive fixed-grid quadrature does not by itself guarantee that the proposed $\vec{\lambda}(t)$ satisfies the positivity inequality at every time.\footnote{This is a limitation of naive fixed-grid quadrature, not of the barrier method itself. Adaptive integration with controlled errors, together with derivative bounds or another certified separation procedure, can make the continuous positivity check rigorous \cite{Lawrence:2024spectral}.}

\begin{figure}[H]
    \centering
    \includegraphics[width=0.7\linewidth]{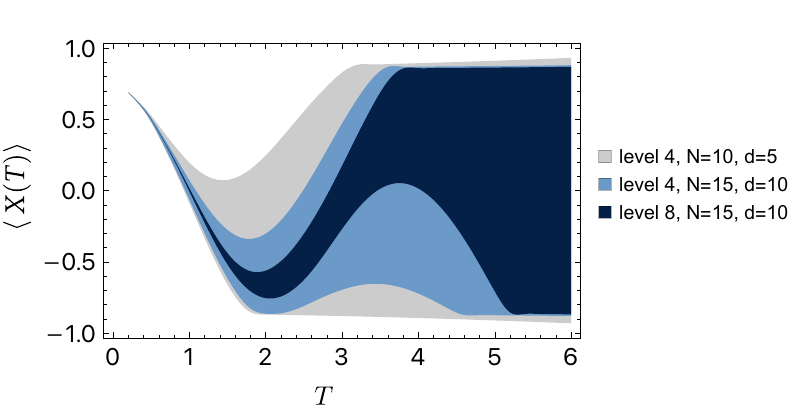}
    \caption{LMN bootstrap for the anharmonic oscillator with the rigorous spline approach \eqref{ineqLMND}. Here we bootstrap $\ev{x(t)}$ where the initial state at $t=0$ is given by $|\psi\rangle = \frac{6}{7} \left(|0\rangle_0 +\hf |1\rangle_0 + \qrt |2\rangle_0 \right)$. The results are qualitatively similar to the results reported in LMN \cite{Lawrence:2024mnj} (their Figure 1).}
    \label{fig:LMNbootstrapImproved}
\end{figure}

Based on the method introduced around (\ref{finiteBasis}), it is clear how to treat the dual problem in a rigorous way. One introduces a basis of positive functions:
\begin{equation}
    \lambda_A^{(i)}(t)=\sum_{I\in\cal I}a_{i}^I \phi_I(t),~~\lambda_D^{(k)}(t)=\sum_{I\in\cal I}d^{I}_{k}\phi_I(t).%
\end{equation}
We again assume $\phi_I$ are closed under derivatives \eqref{dDJI}. Then the dual problem becomes
\begin{align}\label{ineqLMNDfinite}
\text { maximize } & d^I_k \phi_I(0) \operatorname{Tr} C^{(k)} \M(0) \\
\text { subject to } & d_k^I \phi_I(T) C^{(k)}= \widehat{O} \\
&  a^I_i  A^{(i)}+\left(D^{(k)} d^I_k +C^{(k)} \mathcal{D}_{JI} d^J_k \right)   \succeq 0, \quad \forall I \in \mathcal{I} . 
\end{align}
In practice, we choose the B-spline basis. We repeat the exercise of LMN and bootstrap the time evolution of the following initial state%
\begin{equation}
    |\psi\rangle = \frac{6}{7} \left(|0\rangle_0 +\hf |1\rangle_0 + \qrt |2\rangle_0 \right),
\end{equation}
where $|n\rangle_0$ stands for the $n$-th energy eigenstate of the harmonic oscillator for which all zero time expectation values $\M(0)$ can be analytically evaluated, under the anharmonic oscillator Hamiltonian,
\begin{equation}
    H = \frac{1}{2}p^2 + \frac{1}{2} x^2 + \frac{1}{4} x^4\ .
\end{equation}
We report the results in Figure \ref{fig:LMNbootstrapImproved}, where we derive upper and lower bounds on the one-point function $\ev{x(T)}$. Our results are qualitatively similar to the results reported in LMN. We believe that the reason the allowed region for $\ev{x(T)}$ grows is simply that, at a finite level, the initial state is not fully specified. E.g., since the Hilbert space of the anharmonic oscillator is infinite-dimensional, it is impossible to fully specify the initial state with finitely many moments.  

\section{Testing later time bootstrap constraints for the ground state \label{app:extension}}

In this appendix, we consider the effect of imposing the positivity constraints for  $\tau > T$ (with the objective being $\ev{\O(T)\O(0)}$) in the ground state bootstrap problem \eqref{primalMN}. This is motivated by the thermal state bootstrap problem, where we consider the positivity of $\M(\tau)$ in the full thermal circle~$[0,\beta]$.

First, we derive the dual problem where we consider imposing the primal constraints for $\tau = [0,T_2]$ where $T_2 > T$. The action is 
\begin{align}
	I &= \Tr  \bigg\{\widehat{O} \M(T)\nonumber\\ &+ \int_0^{T_2}    \left[ \lambda_A D^\dagger - D \lambda_A  +\hf (D \lambda_G + \lambda_G D^\dagger )+   \hf (\lambda_D D^\dagger+ D \lambda_D) - \lambda_D \partial_\tau   -\Lambda_\M \right] \mathcal{M}(\tau) \d \tau \nonumber\\
    & + \int_0^{T_2}  \left[ \lambda_G  - \Lambda_\N\right] \mathcal{N}(\tau)  \, \d \tau \bigg \}\\
    &= \Tr  \bigg\{ \int_0^{T_2}    \left[ \lambda_A D^\dagger   - D \lambda_A+\hf (D \lambda_G + \lambda_G D^\dagger )  + \hf (\lambda_D D^\dagger+ D \lambda_D)   + \partial_\tau\lambda_D     -\Lambda_\M \right] \mathcal{M}(\tau)  \, \d \tau \nonumber\\
    &\quad + \int_0^{T_2} \left[ -\lambda_G - \Lambda_\N\right] \mathcal{N}(\tau) \, \d \tau + \lambda_D(0)  \M(0) + \left[ \widehat{O}  \M(T)  -\lambda_D(T_2)  \M(T_2)  \right] \bigg \}\nonumber
\end{align}
We arrive at the {dual problem}:
\begin{equation}\label{SDP:MNbootT2}
\begin{split}
        \text { maximize } &  \operatorname{Tr} \lambda_D(0) \M(0) \\
\text { subject to } & \widehat{O} + \lambda_D(T+\epsilon)-\lambda_D(T-\epsilon) \succeq 0   \\
&- \lambda_D(T_2) \succeq 0\\
&  \lambda_A D^\dagger - D \lambda_A    + \hf (D \lambda_G +\lambda_G D^\dagger + D \lambda_D + \lambda_D D^\dagger )  + \partial_\tau \lambda_D     \succeq 0\\
&-\lambda_G(\tau)  \succeq 0, \quad \forall \tau \in [0,T_2]
\end{split}
\end{equation}
To make direct comparison with the previous dual problem \eqref{SDP:MNboot}, we introduce some notation where the Lagrange multipliers for $\tau \in [0,T]$ continue to be denoted by $\lambda$ whereas the Lagrange multipliers for $\tau' \in [T,T_2]$ are denoted by $\widehat\lambda$. Then, the dual problem can be rewritten as 
\begin{equation}\label{SDP:MNbootT2tilde}
\begin{split}
        \text { maximize } &  \operatorname{Tr} \lambda_D(0) \M(0) \\
\text { subject to } & \widehat{O}  \succeq \lambda_D(T)  -\widetilde\lambda_D(T)    \\
&  \lambda_A D^\dagger - D \lambda_A    + \hf (D \lambda_G +\lambda_G D^\dagger + D \lambda_D + \lambda_D D^\dagger )  + \partial_\tau \lambda_D     \succeq 0\\
&-\lambda_G(\tau)  \succeq 0, \quad \forall \tau \in [0,T]\\
&-\widetilde\lambda_D(T_2) \succeq 0\\
&  \widetilde\lambda_A D^\dagger - D \widetilde\lambda_A    + \hf (D \widetilde\lambda_G +\widetilde\lambda_G D^\dagger + D \widetilde\lambda_D +\widetilde \lambda_D D^\dagger )  + \partial_\tau \widetilde\lambda_D     \succeq 0\\
&-\widetilde\lambda_G(\tau')  \succeq 0, \quad \forall \tau' \in [T,T_2]
\end{split}
\end{equation}

Notice that if we restrict to $\widetilde \lambda_D (T) = 0$, the last three constraints in \eqref{SDP:MNbootT2tilde} are decoupled from the rest, and this problem factorizes into a problem that involves $\lambda$ and a problem that involves $\tilde\lambda$. In this case, there will be no improvement in the bounds. {Moreover, if we set to any value $-\widetilde \lambda_D (T) \succeq 0$ the dual problem in the region $[0,T]$ becomes more restrictive and the bounds cannot improve either.}

To this end, we conjecture that the maximum of \eqref{SDP:MNbootT2tilde} is indeed achieved at  $\widetilde \lambda_D (T) = 0$, and provide strong numerical evidence that this new dual formulation does not lead to further improvement. This is demonstrated in Table \ref{tab:extension-lemma}, where we compare the bootstrap results of $\langle x(\tau=1)x(0)\rangle$ in an anharmonic oscillator with and without the later time positivity. We found no improvement  up to a precision of $10^{-10}$ as the degree of the polynomial in the Lagrange multipliers increases. 
It would be interesting to prove an ``extension lemma'' that shows that \eqref{SDP:MNbootT2tilde} yields the same bounds for all values of $T_2 \ge T$.

\begin{table}[]
    \centering
    \begin{tabular}{c|c|c|c}
         &level 4& level 6 & level 8\\
         \hline 
         $d=2$ & \makecell{\{0.0615063911,0.3548402511\} \\ \{0.0615063911,0.3548402511\}}& \makecell{\{0.0617578682,0.0965520056\} \\ \{0.0617968775,0.0965520056\}} & \makecell{\{0.0617582006,0.0957029026\} \\ \{0.0619257739,0.0957029026\}}\\
         \hline 
         $d=4$ & \makecell{\{0.0865444547,0.3548402511\} \\ \{0.0865444547,0.3548402511\}}& \makecell{\{0.0867969987,0.0878682569\} \\ \{0.0867975182,0.0878682569\}} & \makecell{\{0.0867974812,0.0870736932\} \\ \{0.0867981954,0.0870736932\}}\\
         \hline 
         $d=6$ & \makecell{\{0.0867105637,0.3548402511\} \\ \{0.0867105637,0.3548402511\}}& \makecell{\{0.0869647943,0.0877609256\} \\ \{0.0869647943,0.0877609256\}} & \makecell{\{0.0869652878,0.0869683882\} \\ \{0.0869652879,0.0869683882\}}\\
         \hline 
         $d=8$ & \makecell{\{0.0867111854,0.3548402511\} \\ \{0.0867111854,0.3548402511\}}& \makecell{\{0.0869655231,0.0877604498\} \\ \{0.0869655231,0.0877604498\}} & \makecell{\{0.0869660166,0.0869678669\} \\ \{0.0869660166,0.0869678669\}}\\

    \end{tabular}
    \caption{We compare the two-point correlators of $\ev{x(T)x(0)}$ in anharmonic oscillator with and without imposing the positivity after $T$ ($T=1$ here). The table shows the results of bootstrap levels-$\{4,6,8\}$ with the degrees of the polynomial in the Lagrange multipliers $d=\{2,4,6,8\}$. For each value of $d$, the first row shows the \{upper, lower\}  bounds for the problem without inputting positivity after $T$ \eqref{SDP:MNboot}, and the second row (underneath) shows the results with positivity after $T$ \eqref{SDP:MNbootT2}. We found no improvement up to a precision of $10^{-10}$ as the degree $d$ increases in all bootstrap levels we consider here. }
    \label{tab:extension-lemma}
\end{table}

\section{One-matrix quantum mechanics}\label{app: analytic}
\subsection{Ground state}
The 1-MQM restricted to the gauge singlet states reduces to $N$ non-interacting fermions in the potential $V(\lambda)$ \cite{Brezin:1977sv}. At large $N$ we can use the semi-classical statistical mechanics approximation for Fermi-Dirac statistics \cite{Brezin:1977sv}:
\begin{align}
E & =N \mu_F-\int \frac{\d \lambda \, \d p}{2 \pi} \Theta\left(\mu_F-\frac{p^2}{2}-V(\lambda) \right)\left(\mu_F-\frac{p^2}{2}-V(\lambda)  \right) \\
N & =\int \frac{\d \lambda \, \d p}{2 \pi} \Theta\left(\mu_F-\frac{p^2}{2}-V(\lambda) \right) .
\end{align}
We may think of the Fermi level $\mu_F$ as defined by the second equation.

The eigenvalue density can be obtained by integrating out $p$:
\begin{align}
\rho(\lambda) = \frac{1}{\pi}\sqrt{2 (\mu_F-V)} \Theta(\mu_F-V(\lambda)  ) %
\end{align}
Let's now specialize to the potential $V = \hf x^2 + g x^4$. Then the even moments are given by
\begin{align}
    \int \rho(\lambda) \lambda^{2k} = 
    \frac{\left(4 g \lambda_0^3+\lambda_0\right)^2 \lambda_0^{2 k} \Gamma \left(k+\frac{1}{2}\right) \Gamma (k+2) \, _2F_1\left(\frac{3}{2},k+\frac{5}{2};k+2;-\frac{2 g \lambda_0^2}{2 g \lambda_0^2+1}\right)}{2 \sqrt{\pi } \left(2 g \lambda_0^2+1\right)^{3/2}}
\end{align}
We can solve for $\lambda_0$ by demanding that $\tr 1 = 1$, e.g., that the above expression at $k=0$ gives 1. Then the critical point is where
\begin{align} \label{muf}
    \mu_F(g_c) = \hf \lambda^2_0 + g_c \lambda^4_0 = V_\text{max} = -\frac{1}{16 g_c}.
\end{align}

\subsection{Adjoint sector}\label{app:longstring}
The Schr{\"o}dinger equation for states in the adjoint representation reduces, in the large $N$ limit, to the following singular integral eigenvalue problem, known as the Marchesini-Onofri equation \cite{Marchesini:1979yq,Casartelli:1979iz,Gross:1990md,Maldacena:2005hi,Balthazar:2018qdv}:
\begin{align}\label{eqn:longstringwaveEq}
    \int_{-a}^{+a} \frac{\mathrm{~d} y}{\pi} \sqrt{2(\mu_F - V(y))} \frac{\psi_n(x)-\psi_n(y)}{(x-y)^2}=\Delta_n \psi_n(x),
\end{align}
where $\pm a$ are the turning points of the potential $V(y)$. $\psi_n(x)$ and $\Delta_n$ for $n=1,2,\cdots$ denote, respectively, the adjoint wavefunctions and the adjoint energy levels above the ground-state energy, satisfying the orthonormality conditions,
\begin{equation}
    \int_{-a}^{+a} \frac{\mathrm{~d} y}{\pi} \sqrt{2(\mu_F - V(y))}\psi_n(y)=0 \quad,\quad\text{and} \quad,\quad\int_{-a}^{+a} \frac{\mathrm{~d} y}{\pi} \sqrt{2(\mu_F - V(y))}|\psi_n(y)|^2=1.
\end{equation}
We order them as $\Delta_1\leq\Delta_2\leq\cdots$, so that $\Delta_1$ represents the gap in the adjoint sector.

In practice, (\ref{eqn:longstringwaveEq}) can be solved numerically by discretizing the interval $y\in[-a,a]$, reducing it to an ordinary finite-dimensional matrix eigenvalue problem. We discretize the interval into a uniform lattice with $d+1$ sites $y_r=-a+{2ar\over d}$ for $r=0,1,\cdots,d$, which yields estimates $\Delta_n^{(d)}$ for $\Delta_n$. After obtaining such estimates for various values of $d$ up to a maximum $d_{max}$, we extrapolate to $d\rightarrow\infty$ by fitting $\Delta_n^{(d)}=\sum_{u=0}^\gamma a_ud^{-u}$, where $\gamma-1$ is the number of $d$ values used in the fit. The extrapolated coefficients $a_0$ obtained in this work remain stable up to the 6th digit as $d_{max}$ varies from $\sim 200$ to $\sim 400$, indicating that they provide reliable estimates for $\Delta_n$. %

For every such adjoint wavefunction, there are $N^2$ energy eigenstates $|n,ab\rangle$, $a,b=1,\cdots,N$, with energy $\Delta_n$, where $ab$ label the adjoint indices. 
With the normalization $\langle n,ab|m,cd\rangle=\delta_{nm}\delta_{ad}\delta_{bc}$, the matrix elements of $X^k_{ab}$ between the ground state $|\Omega\rangle$ and $|n,cd\rangle$ are given by
\begin{equation}
    \langle\Omega|X^k_{ab}|n,cd\rangle= \frac{\delta_{ad}\delta_{bc}}{N^2}\int_{-a}^{+a} \frac{\mathrm{~d} y}{\pi} \sqrt{2(\mu_F - V(y))}y^k\psi_n(y).
\end{equation}
Numerical estimates for $\langle\Omega|X^k_{ab}|n,cd\rangle$ are obtained by evaluating the expression with the discretized wavefunction found above and then extrapolating the results, following exactly the same procedure described above for determining $\Delta_n$.

\section{Bootstrapping One-Point Functions of the MQM}\label{app:1ptMQM}

In this appendix, we summarize the bootstrap ingredients for the 1-MQM with Hamiltonian\footnote{Here $X$ and $P$ are the rescaled variables as in the main text.} given by \eqref{eq:HamiltonianMQM}.
The basic setup closely follows the formulation in~\cite{Lin:2025srf}. In summary, for all operators $\O$, $\O_1$ and $\O_2$:
\begin{itemize}
    \item \textbf{Equality conditions:}  
    We impose the Heisenberg equation of motion $\langle [H,\mathcal{O}] \rangle = 0$, the cyclicity condition $\langle \operatorname{tr} \mathcal{O}_1 \mathcal{O}_2 \rangle = \langle \operatorname{tr} \mathcal{O}_2 \mathcal{O}_1 \rangle + \langle \operatorname{tr} [\mathcal{O}_1,\mathcal{O}_2] \rangle$, and the gauge condition $\langle \operatorname{tr} \mathcal{O} C \rangle = 0$.  
    We also note the presence of discrete symmetries, as discussed in~\cite{Lin:2025srf}.
    \item \textbf{Inequality conditions:}  
    We require inner-product positivity $\langle \operatorname{tr} \mathcal{O}^\dagger \mathcal{O} \rangle \ge 0$ and ground-state positivity $\langle \operatorname{tr} \mathcal{O}^\dagger [H,\mathcal{O}] \rangle \ge 0$.
\end{itemize}

\begin{figure}[t]
    \centering
    \includegraphics[width=1.05\linewidth]{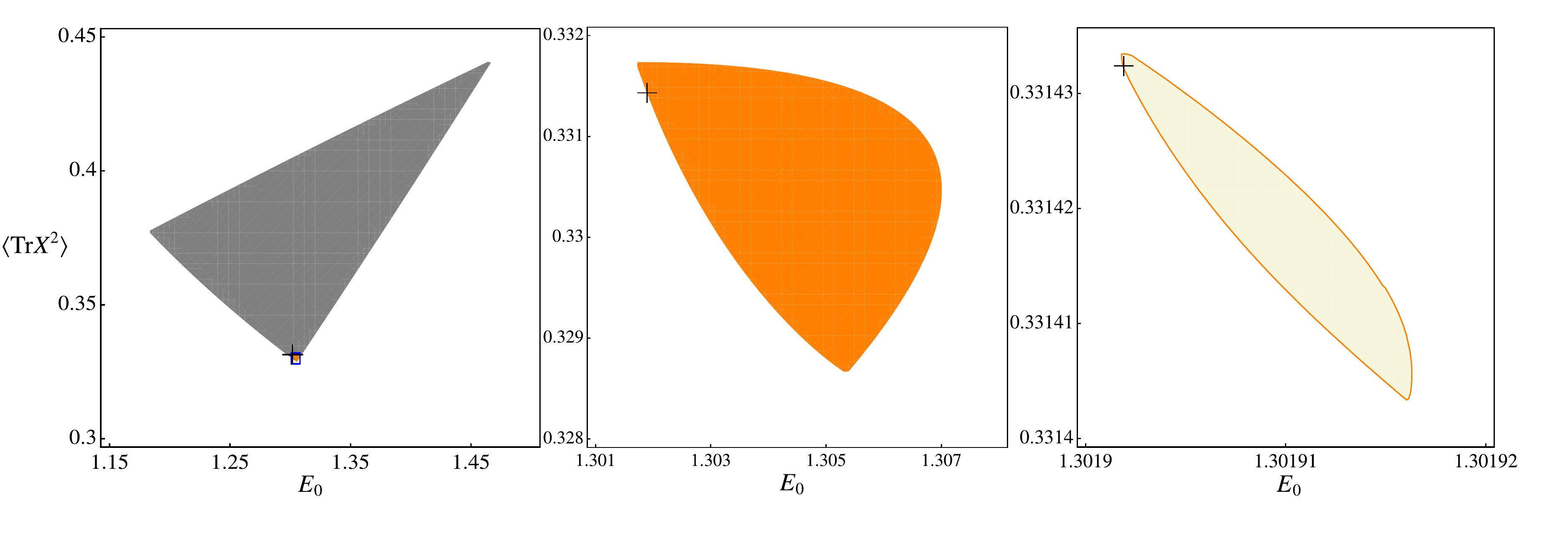}
    \caption{Bootstrap results for the ground state of the 1-MQM with $V(x)=x^2+x^4$.  
    From left to right: level-4, level-6, and level-8 bootstraps, corresponding to moments involving at most $4$, $6$, and $8$ powers of $X$, respectively.  
    The black cross denotes the exact values of $\langle \operatorname{tr} X^2 \rangle$ and $E_0$ from Appendix~\ref{app: analytic}.}
    \label{fig:1ptMQM}
\end{figure}

Figure~\ref{fig:1ptMQM} shows the allowed bootstrap region for $(E_0,\, \langle \operatorname{tr} X^2 \rangle)$ at different truncation levels.  
The convergence is remarkably fast: at level~8 the result already achieves five-digit precision.  
The semidefinite program (SDP) at this level involves only four blocks of sizes $(5,\,4,\,4,\,4)$, with four primal variables.

As discussed in Section~\ref{app: analytic}, the model defined by~\eqref{eq:HamiltonianMQM} is exactly solvable, and this solvability is implicitly reflected in the equality and inequality constraints above.  
In particular, the size of the semidefinite matrices grows quadratically, while the number of variables increases only linearly with the cutoff level.

\vspace{0.5em}
\noindent\textbf{Gauge symmetry.}  
For the ground state, the following operator relation holds:
\begin{equation}
    \mathcal{O}_{1,ab} C_{bc} \mathcal{O}_{2,cd} |\Omega\rangle
    = \mathcal{O}_{1,ab} [C_{bc},\, \mathcal{O}_{2,cd}] |\Omega\rangle,
\end{equation}
where $C_{bc}$ is the gauge generator \eqref{gaugeGen}, %
which satisfies%
\begin{equation}
    [C_{ab},\, \mathcal{O}_{cd}] = \frac{1}{N}
    \left( \mathcal{O}_{ad}\delta_{bc} - \mathcal{O}_{cb}\delta_{ad} \right),
\end{equation}
manifesting the $U(N)$ gauge transformation.  
Because $C_{ab}|\Omega\rangle = 0$, we have
\begin{equation}
    \mathcal{O}_{1,ab} C_{bc} \mathcal{O}_{2,cd} |\Omega\rangle
    = (\mathcal{O}_1 \mathcal{O}_2)_{ad} |\Omega\rangle
    - \langle \operatorname{tr} \mathcal{O}_2 \rangle (\mathcal{O}_1)_{ad} |\Omega\rangle,
\end{equation}
or equivalently,
\begin{equation}
    (\mathcal{O}_1 P X \mathcal{O}_2)_{ad} |\Omega\rangle
    = (\mathcal{O}_1 X P \mathcal{O}_2)_{ad} |\Omega\rangle
    - \i \, \langle \operatorname{tr} \mathcal{O}_2 \rangle (\mathcal{O}_1)_{ad} |\Omega\rangle.
\end{equation}
Thus, the state vectors
$(\mathcal{O}_1 P X \mathcal{O}_2)_{ad}|\Omega\rangle$,
$(\mathcal{O}_1 X P \mathcal{O}_2)_{ad}|\Omega\rangle$, and
$(\mathcal{O}_1)_{ad}|\Omega\rangle$
are not linearly independent.  
Consequently, when constructing the positivity matrices, it suffices to consider vectors of the form $(X^m P^n)_{ad}|\Omega\rangle$; other words containing the same numbers of $X$ and $P$ do not provide independent constraints.

\vspace{0.5em}
\noindent\textbf{Reduction of mixed expectation values.}  
The above argument generalizes to operator expectation values.  
Any mixed word composed of $m$ factors of $X$ and $n$ factors of $P$ can be reduced to
\begin{equation} %
\label{reduce-gs-moments}
    \langle \operatorname{tr} X^m P^n \rangle
    + \sum 
    a_{m_1,n_1,m_2,n_2}\,
    \langle \operatorname{tr} X^{m_1} P^{n_1} \rangle
    \langle \operatorname{tr} X^{m_2} P^{n_2} \rangle,
\end{equation}
where the coefficients $a_{m_1,n_1,m_2,n_2}$ are numerical constants satisfying
\[
m_1 + m_2 < m, \qquad n_1 + n_2 < n.
\]
This recursive structure implies that all operator moments built from $X$ and $P$ can be expressed as polynomials of the purely positional moments $\langle \operatorname{tr} X^m \rangle$.  
We demonstrate this statement via induction on the number of $P$ insertions.

\vspace{0.5em}
\noindent\textbf{Inductive step.}  
The base case with no $P$’s is trivial.  
Assume that all expectation values containing at most $(k-1)$ powers of $P$ can be expressed as polynomials in $\langle \operatorname{tr} X^m \rangle$.  
For expectation values with exactly $k$ powers of $P$, the Heisenberg equation of motion gives
\begin{equation}
    \langle [H,\, \operatorname{tr} X^m P^{k-1}] \rangle = 0,
\end{equation}
which implies
\begin{equation}
    \left\langle
    \sum_{i=0}^{m-1}
    X^i P X^{m-i-1} P^{k-1}
    \right\rangle
    =
    \left\langle
    \sum_{i=0}^{k-2}
    X^m P^i V'(X) P^{k-2-i}
    \right\rangle.
\end{equation}
The left-hand side contains a leading term
$m\, \langle \operatorname{tr} X^{m-1} P^k \rangle$,
while all remaining terms are polynomials in moments involving at most $(k-1)$ $P$’s by~\eqref{reduce-gs-moments}.  
The right-hand side contains at most $(k-2)$ powers of $P$, since $V'(X)$ is a polynomial in $X$.  
Hence $\langle \operatorname{tr} X^{m-1} P^k \rangle$, and therefore all moments with $k$ $P$’s, can be expressed as polynomials in $\langle \operatorname{tr} X^m \rangle$.  
By induction, every mixed moment composed of $X$ and $P$ reduces to a polynomial in the purely positional moments $\langle \operatorname{tr} X^m \rangle$.

\section{Monte Carlo simulation of the ungauged model \label{app:montecarlo}}

\begin{figure}[H]
    \centering
\includegraphics[width=1\linewidth]{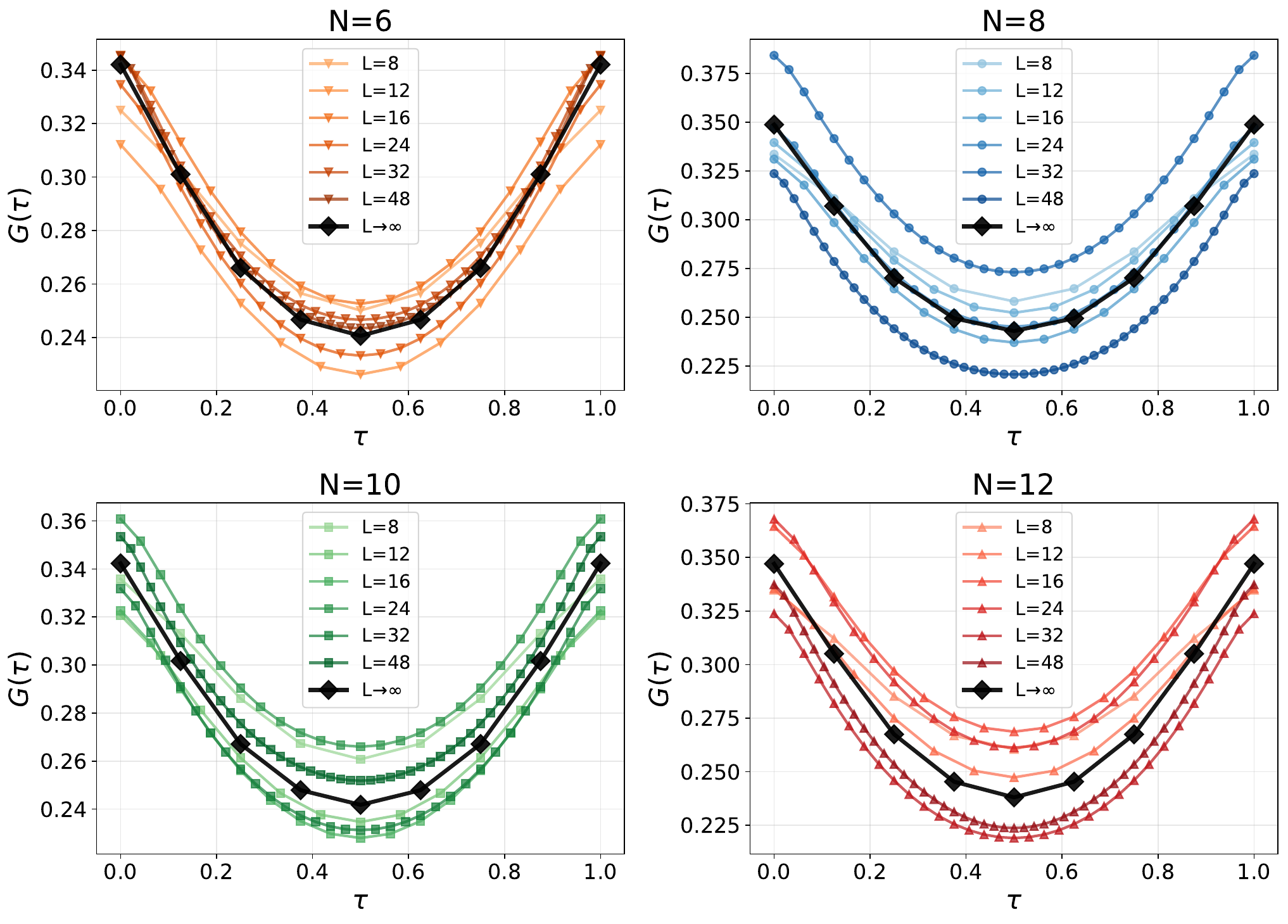}
    \caption{Monte Carlo for various values of $N$ and $L$ results with 500k burn in + 100k measurements. For each value of $N$, we show the continuum extrapolation to $L\to \infty$.}
    \label{fig:N_comparison}
\end{figure}

\begin{figure}[H]
    \centering
\includegraphics[width=1\linewidth]{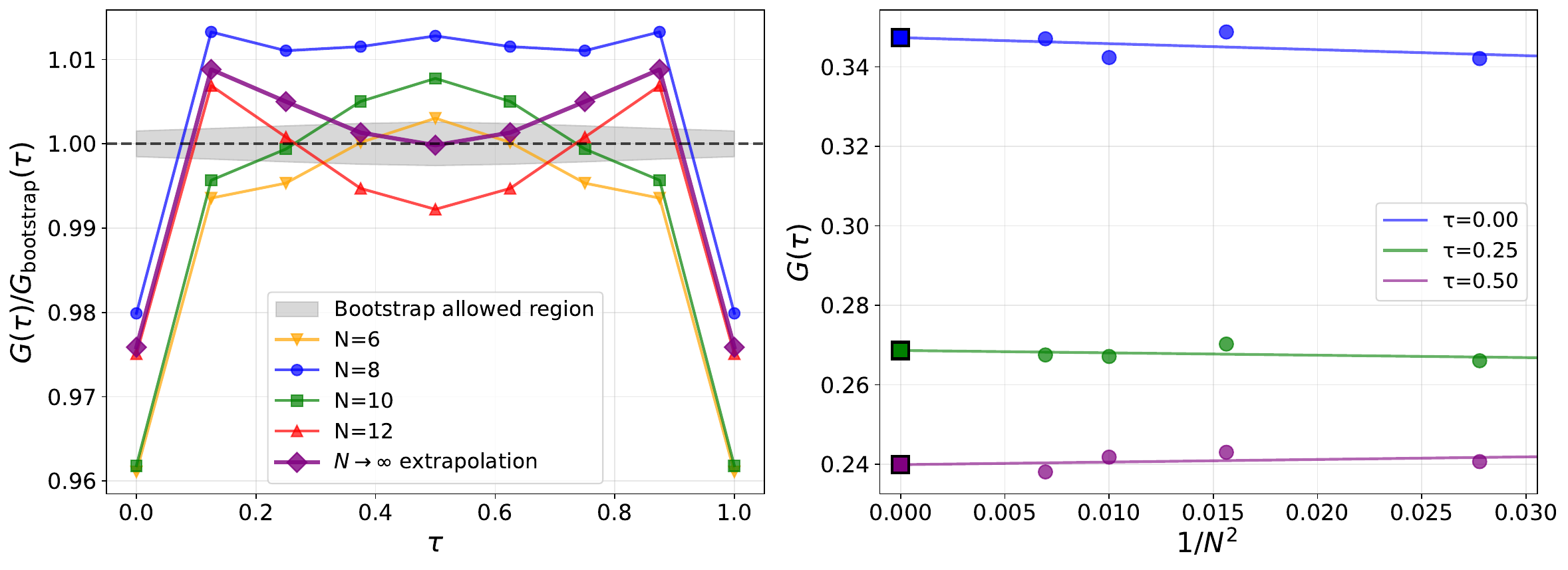}
    \caption{Extrapolation to infinite $N$. Here we take the large $L$ extrapolation of the finite $N$ data points and extrapolate to $N \to \infty$ by fitting $G(\tau) = G_{\infty}(\tau) + \frac{1}{N^2} g(\tau)$. We also show the normalized data using the average of the level 8 bootstrap upper/lower bounds.}
    \label{fig:monteCarloNextrap}
\end{figure}

The Euclidean action for the MQM with a quartic potential is:
\begin{equation}
S = N^2 \int_0^\beta d\tau \, \text{tr}\left[\frac{1}{2}\left(\frac{\d X}{\d\tau}\right)^2 + \frac{1}{2}X^2 +g  X^4\right]
\end{equation}
where $X(\tau)$ is an $N \times N$ Hermitian matrix, $\beta = 1/T$ is the inverse temperature, and $g$ is the quartic coupling. Our simulations were performed at $\beta = 1$ and $g=1$. Since we are interested in the ungauged model, we did not add any gauge field.

We discretize Euclidean time into $L$ sites with lattice spacing $a = \beta/L$. The standard naive discretization leads to $\mathcal{O}(a)$ errors in observables. We use the ``improved'' lattice action \cite{Berkowitz:2016jlq} for the ungauged matrix model:
\begin{align}
    S &=  \frac{N}{L}\sum_{i=1}^L \frac{1}{2a}  \tr (\Delta X_i)^2  + a \, \tr V(X_i), \quad V=\hf X^2 + X^4\\
    \Delta X_i &= -\hf X_{i+2}+ 2 X_{i+1}-\tfrac{3}{2} X_{i}
\end{align}
We use a local update Metropolis-Hastings rule, where we propose a change $X_i \to X_i + \epsilon \, \delta H$, where $\delta H$ is a Hermitian matrix drawn from the Gaussian unitary ensemble (GUE), normalized so that the diagonal elements of $\delta H$ have unit variance. We then compute $\Delta S$ which is ``5-site'' local, e.g., depends on $X_{i-2}, X_{i-1}, \cdots,  X_{i+2}$. We accept/reject with probability $P_{\text{accept}} = \min\{1, e^{-\Delta S}\}$. We define a ``sweep'' to be a sequential update on all $L$ lattice sites. For each value of $N$ and $L$, we then ran $N_{\text{therm}} = 5\times 10^5$ sweeps to equilibrate the system. We dynamically tune the parameter $\epsilon$ to keep the acceptance rate $0.3 < P_\text{accept} < 0.7$. After these thermalization runs, we ran additional $N_{\text{sweeps}} = 10^5$ sweeps, measuring observables every sweep. This yields a statistical error that is completely negligible compared to the systematic errors we will encounter due to extrapolation.

\subsection{Extrapolation Ansatz}
We performed a simulation at finite $N$ for multiple $L$ values $L \in \{8, 12, 16, 24, 32, 48\}$. 
Since different $L$ values have different time grids $\{\tau_i\}$, we choose the $L=8$ grid as a reference grid. We then perform a cubic interpolation to map all data to this grid. Then at each reference grid point, we perform a linear fit. Based on the $\mathcal{O}(a^2)$ discretization error analysis, we use:
\begin{equation}
G(\tau; N, L) = G_{\text{continuum}}(\tau; N) + c_2(\tau) \left(\frac{1}{L}\right)^2.
\end{equation}
We show these fits in Figure \ref{fig:N_comparison}. In the 't Hooft limit, the leading large $N$ corrections scale as $1/N^2$ for matrix models. Hence
\begin{equation}
G_{\text{continuum}}(\tau; N) = G_\infty(\tau) + \frac{g(\tau)}{N^2}
\end{equation}
For each $\tau$, we fit data from $N = 6, 8, 10, 12$ and extract the $N \to \infty$ limit as the intercept $G_\infty(\tau)$.

\bibliography{main}
\bibliographystyle{JHEP}

\end{document}